\documentclass[a4paper,fleqn,usenatbib]{mnras}

\usepackage{Times}

\usepackage[T1]{fontenc}
\usepackage{ae,aecompl}

\usepackage{graphicx}	
\usepackage{amsmath}	
\usepackage{amssymb}	

\title[LyC photon production and escape from SFGs at $z\sim2$]{The production and escape of Lyman-Continuum radiation from star-forming galaxies at $\bf z\sim2$ and their redshift evolution}

\author[J. Matthee et al.]{Jorryt Matthee$^{1}$\thanks{E-mail: matthee@strw.leidenuniv.nl}, David Sobral$^{1,2}$, Philip Best$^{3}$, Ali Ahmad Khostovan$^{4}$,
\newauthor Iv\'an Oteo$^{3,5}$, Rychard Bouwens$^{1}$, Huub R\"ottgering$^{1}$ \\
$^{1}$ Leiden Observatory, Leiden University, P.O.\ Box 9513, NL-2300 RA Leiden, The Netherlands\\
$^{2}$ Department of Physics, Lancaster University, Lancaster, LA1 4YB, UK \\ 
$^{3}$ Institute for Astronomy, University of Edinburgh, Royal Observatory, Blackford Hill, Edinburgh EH9 3HJ UK\\
$^{4}$ University of California, Riverside, 900 University Ave, Riverside, CA, 92521, USA\\
$^{5}$ European Southern Observatory, Karl-Schwarzschild-Str. 2, 85748 Garching, Germany \\
 }

\pubyear{2016}

\begin{document}
\label{firstpage}
\pagerange{\pageref{firstpage}--\pageref{lastpage}}
\maketitle

\begin{abstract}
We study the production rate of ionizing photons of a sample of 588 H$\alpha$ emitters (HAEs) and 160 Lyman-$\alpha$ emitters (LAEs) at $z=2.2$ in the COSMOS field in order to assess the implied emissivity from galaxies, based on their UV luminosity. By exploring the rest-frame Lyman Continuum (LyC) with {\it GALEX}/$NUV$ data, we find f$_{\rm esc} < 2.8\, (6.4)$\% through median (mean) stacking. By combining the H$\alpha$ luminosity density with IGM emissivity measurements from absorption studies, we find a globally averaged $\langle$f$_{\rm esc}\rangle$ of $5.9^{+14.5}_{-4.2}$ \% at $z=2.2$ if we assume HAEs are the only source of ionizing photons. We find similarly low values of the global $\langle$f$_{\rm esc}\rangle$ at $z\approx3-5$, also ruling out a high $\langle$f$_{\rm esc}\rangle$ at $z<5$. These low escape fractions allow us to measure $\xi_{ion}$, the number of produced ionizing photons per unit UV luminosity, and investigate how this depends on galaxy properties. We find a typical $\xi_{ion} \approx 10^{24.77\pm0.04}$ Hz erg$^{-1}$ for HAEs and $\xi_{ion} \approx 10^{25.14\pm0.09}$ Hz erg$^{-1}$ for LAEs. LAEs and low mass HAEs at $z=2.2$ show similar values of $\xi_{ion}$ as typically assumed in the reionization era, while the typical HAE is three times less ionizing. Due to an increasing $\xi_{ion}$ with increasing EW(H$\alpha$), $\xi_{ion}$ likely increases with redshift. This evolution alone is fully in line with the observed evolution of $\xi_{ion}$ between $z\approx2-5$, indicating a typical value of $\xi_{ion} \approx 10^{25.4}$ Hz erg$^{-1}$ in the reionization era.
\end{abstract}

\begin{keywords}
galaxies: high-redshift -- galaxies: evolution -- cosmology:observations -- cosmology: dark ages, re-ionisation, first stars. \end{keywords}



\section{Introduction}
One of the most important questions in galaxy formation is whether galaxies alone have been able to provide the ionizing photons which reionized the Universe. Optical depth measurements from the Planck satellite place the mean reionization redshift between $z\approx7.8-8.8$ \citep{Planck2016}. The end-point of reionization has been marked by the Gun-Peterson trough in high-redshift quasars at $z\approx5-6$, with a typical neutral fraction of $\sim10^{-4}$ \citep[e.g.][]{Fan2006,McGreer2015}. Moreover, recent observations indicate that there are large opacity fluctuations among various sight-lines, indicating an inhomogeneous nature of reionization \citep{Becker2015}.
 
Assessing whether galaxies have been the main provider of ionizing photons at $z\gtrsim5$ (alternatively to Active Galactic Nucleii, AGN; e.g. \citealt{MadauHaardt2015,Giallongo2015,Weigel2015}) crucially depends on i) precise measurements of the number of galaxies at early cosmic times, ii) the clumping factor of the IGM \citep[e.g.][]{Pawlik2015}, iii) the amount of ionizing photons that is produced (Lyman-Continuum photons, LyC, $\lambda<912${\AA}) and iv) the fraction of ionizing photons that escapes into the inter galactic medium (IGM). All these numbers are currently uncertain, with the relative uncertainty greatly rising from i) to iv).

Many studies so far have focussed on counting the number of galaxies as a function of their UV luminosity (luminosity functions) at $z>7$ \citep[e.g.][]{McLure2013,Bowler2014,Atek2015,Bouwens2015,Finkelstein2015,Ishigaki2015,McLeod2015,Castellano2016,Livermore2016}. These studies typically infer luminosity functions with steep faint-end slopes, and a steepening of the faint-end slope with increasing redshift (see for example the recent review from \citealt{FinkelsteinReview}), leading to a high number of faint galaxies. Assuming ``standard'' values for the other parameters such as the escape fraction, simplistic models indicate that galaxies may indeed have provided the ionizing photons to reionize the Universe \citep[e.g.][]{Madau1999,Robertson2015}, and that the ionizing background at $z\sim5$ is consistent with the derived emissivity from galaxies \citep{Choudhury2015,Bouwens2015reion}. However, without validation of input assumptions regarding the production and escape of ionizing photons (for example, these simplistic models assume that the escape fraction does not depend on UV luminosity), the usability of these models remains to be evaluated. 

The most commonly adopted escape fraction of ionizing photons, f$_{\rm esc}$, is 10-20 \%, independent of mass or luminosity \citep[e.g.][]{Mitra2015,Robertson2015}. However, hydrodynamical simulations indicate that f$_{\rm esc}$ is likely very anisotropic and time dependent \citep{Cen2015,Ma2015}. An escape fraction which depends on galaxy properties (for example a higher f$_{\rm esc}$ for lower mass galaxies, e.g. \citealt{Paardekooper2015}) would influence the way reionization happened \citep[e.g.][]{Sharma2016}. Most importantly, it is impossible to measure f$_{\rm esc}$ directly at high-redshift ($z>6$) because of the high opacity of the IGM for ionizing photons \citep[e.g.][]{Inoue2014}. Furthermore, to estimate f$_{\rm esc}$ it is required that the intrinsic amount of ionizing photons is measured accurately, which requires accurate understanding of the stellar populations, SFR and dust attenuation \citep[i.e.][]{deBarros2016}.  

Nevertheless, several attempts have been made to measure f$_{\rm esc}$, both in the local Universe \citep[e.g.][]{Leitherer1995,Deharveng2001,Leitet2013,Alexandroff2015} and at intermediate redshift, $z\sim3$, where it is possible to observe redshifted LyC radiation with optical CCDs \citep[e.g.][]{Inoue2006,Boutsia2011,Vanzella2012,Bergvall2013,Mostardi2015}. However, the number of reliable direct detections is limited to a handful, both in the local Universe and at intermediate redshift \citep[e.g.][]{Borthakur2014,Izotov2016,Izotov2016b,deBarros2016,Leitherer2016}, and strong limits of f$_{\rm esc} \lesssim 5-10$ \% exist for the majority \citep[e.g.][]{Grazian2016,Guaita2016,Rutkowski2015}. An important reason is that contamination from sources in the foreground may mimic escaping LyC, and high resolution UV imaging is thus required \citep[e.g.][]{Mostardi2015,Siana2015}. Even for sources with established LyC leakage, estimating f$_{\rm esc}$ reliably depends on the ability to accurately estimate the intrinsically produced amount of LyC photons and precisely model the transmission of the IGM \citep[e.g.][]{Vanzella2016}.

The amount of ionizing photons that are produced per unit UV (rest-frame $\approx1500$ {\AA}) luminosity ($\xi_{ion}$) is generally calculated using SED modelling \citep[e.g.][]{Madau1999,Bouwens2012,Kuhlen2012} or (in a related method) estimated from the observed values of the UV slopes of high-redshift galaxies \citep[e.g.][]{Robertson2013,Duncan2015}. Most of these studies find values around $\xi_{ion} \approx10^{25.2-25.3}$ Hz erg$^{-1}$ at $z\sim8$. More recently, \cite{Bouwens2015xi} estimated the number of ionizing photons in a sample of Lyman break galaxies (LBGs) at $z\sim4$ to be $\xi_{ion} \approx10^{25.3}$ Hz erg$^{-1}$ by estimating H$\alpha$ luminosities with {\it Spitzer}/IRAC photometry.  

Progress in the understanding of f$_{\rm esc}$ and $\xi_{ion}$ can be made by expanding the searched parameter space to lower redshifts, where rest-frame optical emission lines (e.g. H$\alpha$) can provide valuable information on the production rate of LyC photons and where it is possible to obtain a complete selection of star-forming galaxies. 

In this paper, we use a large sample of H$\alpha$ emitters (HAEs) and Ly$\alpha$ emitters (LAEs) at $z=2.2$ to constrain f$_{\rm esc}$ and measure $\xi_{ion}$ and how this may depend on galaxy properties. Our measurements of $\xi_{ion}$ rely on the assumption that f$_{\rm esc}$ is negligible ($<10$ \%), which we validate by constraining f$_{\rm esc}$ with archival {\it GALEX} $NUV$ imaging and by comparing the estimated emissivity of HAEs with IGM emissivity measurements from quasar absorption lines \citep[e.g.][]{BeckerBolton2013}. Combined with rest-frame UV photometry, accurate measurements of $\xi_{ion}$ are possible on a source by source basis for HAEs, allowing us to explore correlations with galaxy properties. Since only a  handful of LAEs are detected in H$\alpha$ (see \citealt{Matthee2016}), we measure the median $\xi_{ion}$ from stacks of Lyman-$\alpha$ emitters from \cite{Sobral2015survey}.

We describe the galaxy sample and definitions of galaxy properties in \S \ref{sec:2}. \S \ref{sec:3} presents the {\it GALEX} imaging. We present upper limits on f$_{\rm esc}$ in \S \ref{sec:4}. We indirectly estimate f$_{\rm esc}$ from the H$\alpha$ luminosity function and the IGM emissivity in \S \ref{sec:5} and measure the ionizing properties of galaxies and its redshift evolution in \S \ref{sec:6}. \S \ref{sec:7} discusses the implications for reionization. Finally, our results are summarised in \S \ref{sec:8}.
We adopt a $\Lambda$CDM cosmology with $H_0$ = 70 km s$^{-1} $Mpc$^{-1}$, $\Omega_{\rm M} = 0.3$ and $\Omega_{\Lambda} = 0.7$. Magnitudes are in the AB system. At $z=2.2$, 1$''$ corresponds to a physical scale of 8.2 kpc.

\begin{figure*}
\begin{tabular}{ccc}
\includegraphics[width=5.6cm]{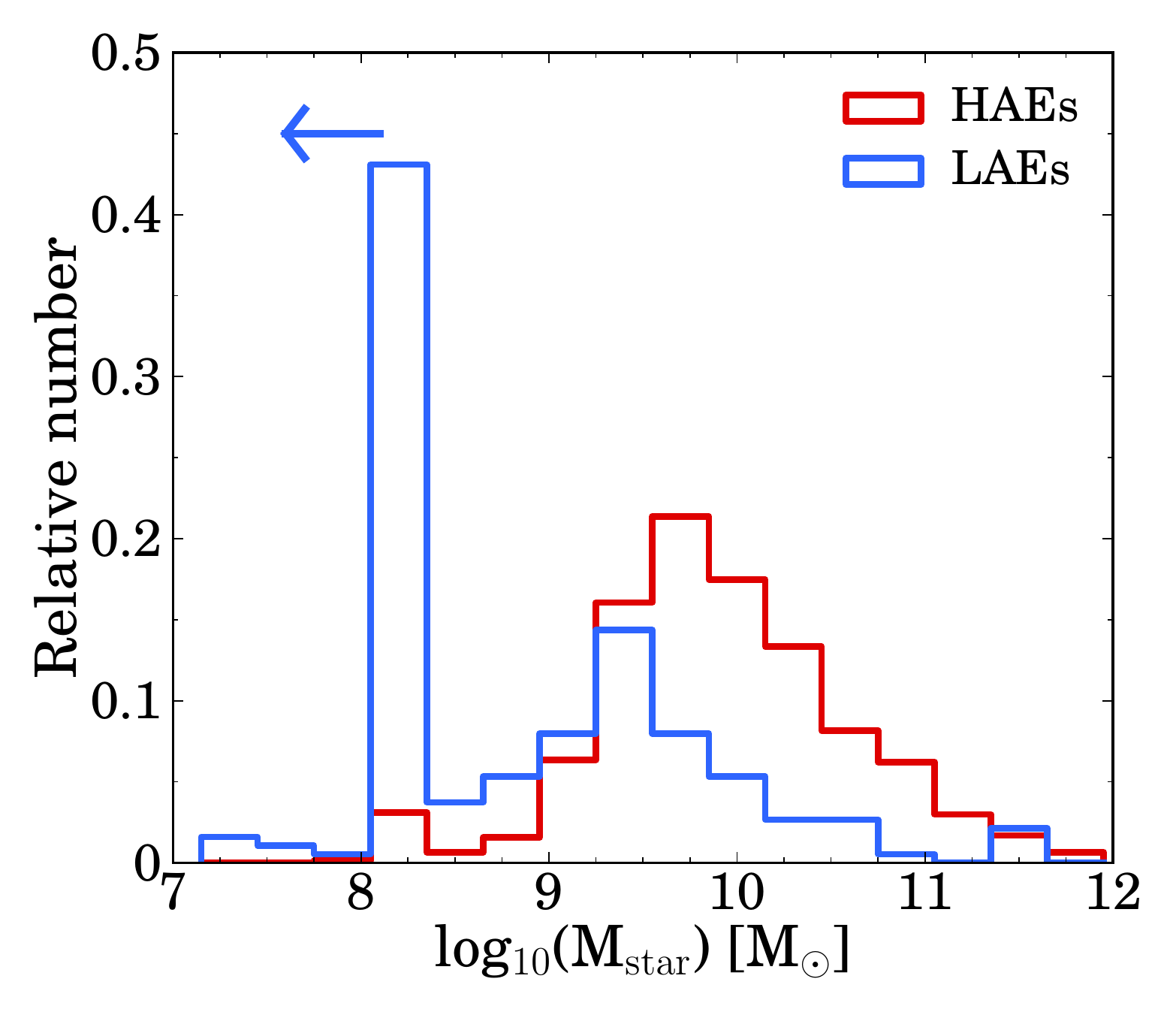}&
\includegraphics[width=5.6cm]{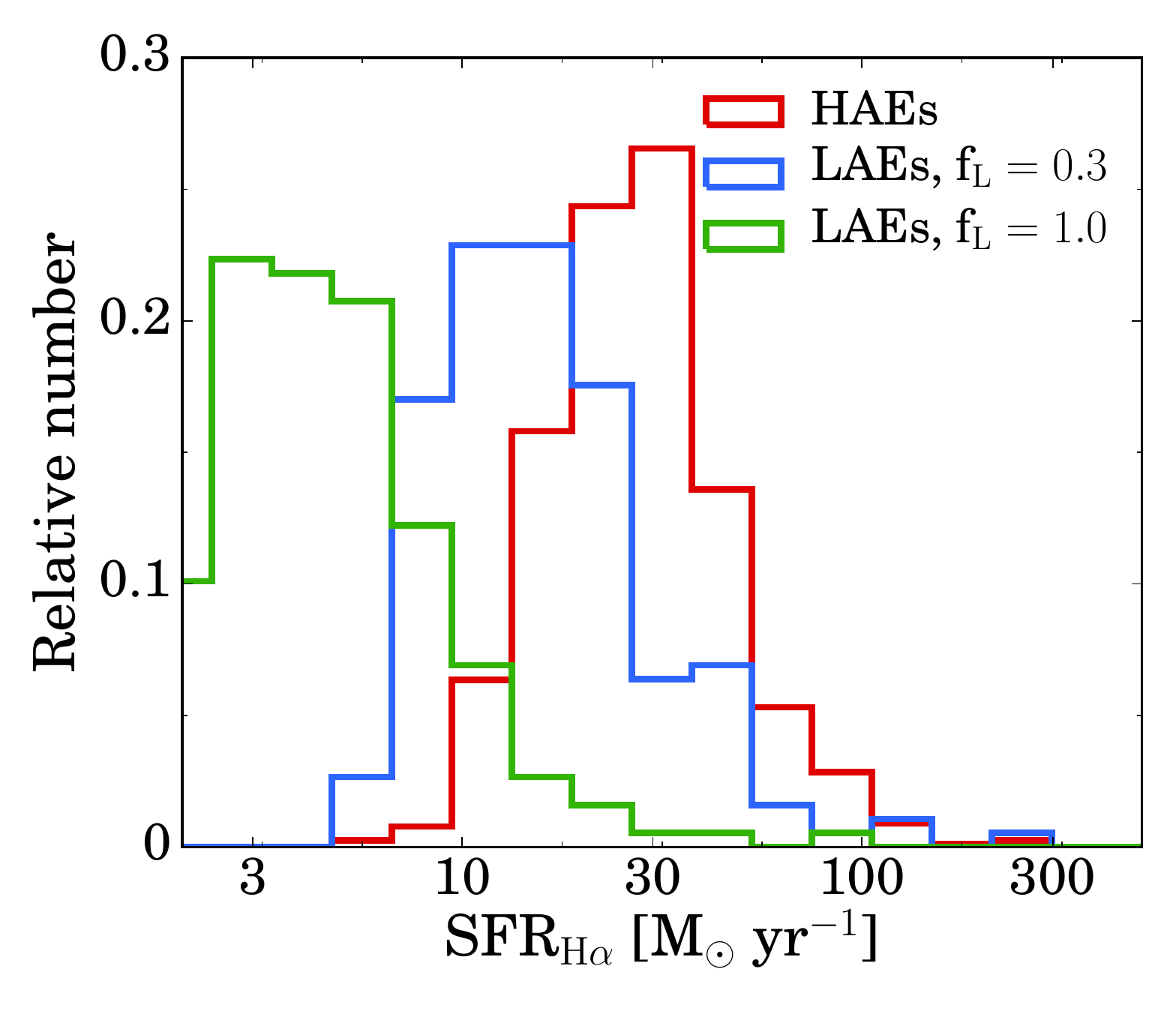}&
\includegraphics[width=5.6cm]{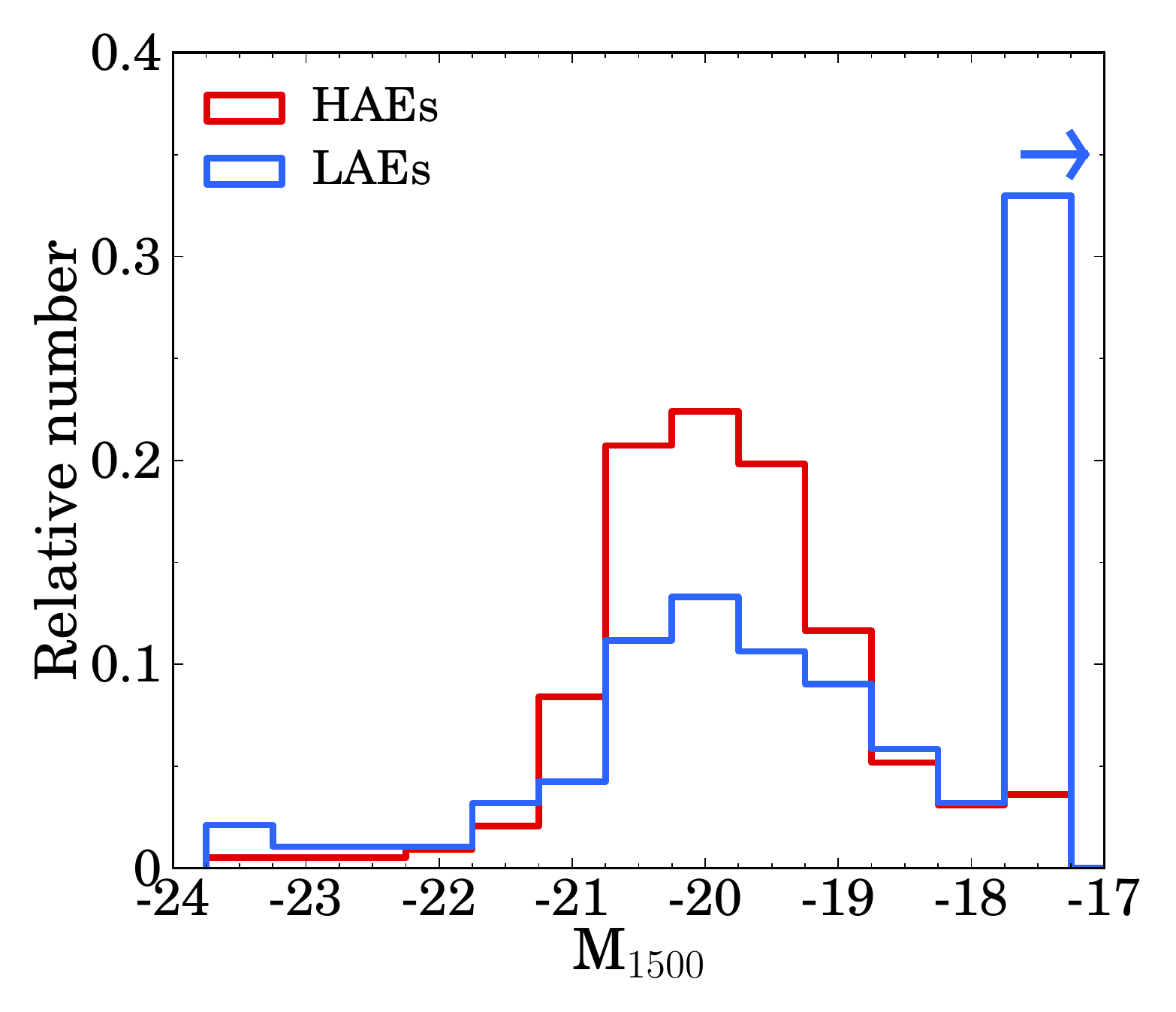}\\
\end{tabular}
\caption{\small{Histogram of the properties of HAEs and LAEs. Stellar mass is obtained through SED fitting (see \S 2.1.1). For HAEs, SFR(H$\alpha$) is obtained from dust-corrected H$\alpha$ (see \S 2.1.2). LAEs that are undetected in broad-bands (and thus without SED fits) are assigned M$_{\rm star} = 10^{8}$ M$_{\odot}$ and M$_{1500} = -17$, corresponding to a $V$ band magnitude of 27 and we assumed those galaxies have no dust in computing SFR(H$\alpha$). For LAEs, we use the observed Ly$\alpha$ luminosity and convert this to H$\alpha$ for two different Ly$\alpha$ escape fractions (f$_L$, the typical escape fraction for LAEs (30 \%) and the maximum of 100 \%, see \citealt{Sobral2015survey}). M$_{1500}$ is obtained by converting the observed $V$ magnitude to absolute magnitude. In general, LAEs trace a galaxy population with lower stellar masses and SFR and fainter UV magnitudes. }} 
\label{fig:galaxy_properties}
\end{figure*}

\section{Galaxy sample}
\label{sec:2}
We use a sample of H$\alpha$ selected star-forming galaxies from the High-$z$ Emission Line Survey (HiZELS; \citealt{Geach2008,Sobral2009}) at $z=2.2$ in the COSMOS field. These galaxies were selected using narrow-band (NB) imaging in the $K$ band with the United Kingdom InfraRed Telescope. H$\alpha$ emitters (HAEs) were identified among the line-emitters using $BzK$ and $BRU$ colours and photometric redshifts, as described in \cite{Sobral2013}, and thus have a photometric redshift of $z=2.22\pm0.02$ where the error is due to the width of the narrow-band filter. In total, there are 588 H$\alpha$ emitters at $z=2.2$ in COSMOS.\footnote{The sample of H$\alpha$ emitters from \cite{Sobral2013} is publicly available through e.g. VizieR, http://vizier.cfa.harvard.edu.} 

HAEs are selected to have EW$_{0, \rm H\alpha+[NII]} > 25$ {\AA}. Since the COSMOS field has been covered by multiple narrow-band filters, a fraction of $z=2.2$ sources are detected with multiple major emission lines in addition to H$\alpha$: {\sc [Oiii]}, {\sc [Oii]} \citep[e.g.][]{Sobral2012,Nakajima2012,Sobral2013} or Ly$\alpha$ \citep[e.g.][]{Oteo2015,Matthee2016}. Multi-wavelength photometry from the observed UV to mid-IR is widely available in COSMOS. In this paper, we make explicit use of $V$ and $R$ band in order to measure the UV luminosity and UV slope $\beta$ (see \S 2.1.3), but all bands have been used for photometric redshifts (see \citealt{Sobral2013}, and e.g. \citealt{Ilbert2009}) and SED fitting \citep{Sobral2014,Oteo2015,Khostovan2016}. 

We also include 160 Lyman-$\alpha$ emitters (LAEs) at $z=2.2$ from the CAlibrating LYMan-$\alpha$ with H$\alpha$ survey (CALYMHA; \citealt{Matthee2016,Sobral2015survey}). For completeness at bright luminosities, LAEs were selected with EW$_{0, \rm Ly\alpha} > 5$ {\AA}, while LAEs are typically selected with a higher EW$_0$ cut of 25 {\AA} (see e.g. \citealt{Matthee2015} and references therein). Only 15 \% of our LAEs have EW$_{0, \rm Ly\alpha} < 25$ {\AA} and these are typically AGN, see \cite{Sobral2015survey}, but they represent some of the brightest. We note that 40 \% of LAEs are too faint to be detected in broad-bands, and we thus have only upper limits on their stellar mass and UV magnitude (see Fig. $\ref{fig:galaxy_properties}$). By design, CALYMHA observes both Ly$\alpha$ and H$\alpha$ for H$\alpha$ selected galaxies. As presented in \cite{Matthee2016}, 17 HAEs are also detected in Ly$\alpha$ with the current depth of Ly$\alpha$ narrow-band imaging. These are considered as HAEs in the remainder of the paper.

We show the general properties of our sample of galaxies in Fig. $\ref{fig:galaxy_properties}$. It can be seen that compared to HAEs, LAEs are typically somewhat fainter in the UV, have a lower mass and lower SFR, although they are also some of the brightest UV objects. 

Our sample of HAEs and LAEs was chosen for the following reasons: i) all are at the same redshift slice where the LyC can be observed with the {\it GALEX} $NUV$ filter and H$\alpha$ with the NB$_K$ filter, ii) the sample spans a large range in mass, star formation rate (SFR) and environments (Fig. $\ref{fig:galaxy_properties}$ and \citealt{Geach2012,Sobral2014}) and iii) as discussed in \cite{Oteo2015}, H$\alpha$ selected galaxies span the entire range of star-forming galaxies, from dust-free to relatively dust-rich (unlike e.g. Lyman-break galaxies).

\subsection{Definition of galaxy properties}
We define the galaxy properties that are used in the analysis in this subsection. These properties are either obtained from: (1) SED fitting of the multi-wavelength photometry, (2) observed H$\alpha$ flux, or (3) observed rest-frame UV photometry.

\subsubsection{SED fitting}
For HAEs, stellar masses (M$_{\rm star}$) and stellar dust attenuations (E$(B-V)$) are taken from \cite{Sobral2014}. In this study, synthetic galaxy SEDs are simulated with \cite{BC2003} stellar templates with metallicities ranging from $Z= 0.0001 - 0.05$, following a \cite{Chabrier2003} initial mass function (IMF) and with exponentially declining star formation histories (with e-folding times ranging from 0.1 to 10 Gyr). The dust attenuation is described by a \cite{Calzetti2000} law. The observed UV to IR photometry is then fitted to these synthetic SEDs. The values of M$_{\rm star}$ and E$(B-V)$ that we use are the median values of all synthetic models which have a $\chi^2$ within $1\sigma$ of the best fitted model. The 1$\sigma$ uncertainties are typically $0.1-0.2$ dex for M$_{\rm star}$ and 0.05-0.1 dex for E$(B-V)$. The smallest errors are found at high masses and high extinctions. The same SED fitting method is applied to the photometry of LAEs. 

We note that the SED fitting from \cite{Sobral2014} uses SED models which do not take contribution from nebular emission lines into account. This means that some stellar masses could be over-estimated. However, the SED fits have been performed on over $>20$ different filters, such that even if a few filters are contaminated by emission lines, the $\chi^2$ values are not strongly affected. Importantly, the {\it Spitzer}/IRAC bands (included in SED fitting and most important for measuring stellar mass at $z=2.2$) are unaffected by strong nebular emission lines at $z=2.2$.

We still investigate the importance of emission lines further by comparing the SED results with those from \cite{Oteo2015}, who performed SED fits for a subsample ($\approx 60$\%) of the HAEs and LAEs, including emission lines. We find that the stellar masses and dust attenuations correlate very well, although stellar masses from \cite{Oteo2015} are on average lower by 0.15 dex. We look at the galaxies with the strongest lines (highest observed EWs) and find that the difference in the stellar mass is actually smaller than for galaxies with low H$\alpha$ EW. This indicates that the different mass estimates are not due to the inclusion of emission lines, but rather due to the details of the SED fitting implementation, such as the age-grid ages and allowed range of metallicities. We therefore use the stellar masses from \cite{Sobral2014}. Our sample spans galaxies with masses M$_{\rm star} \approx 10^{7.5-12}$ M$_{\odot}$, see Fig. $\ref{fig:galaxy_properties}$.

\subsubsection{Intrinsic H$\alpha$ luminosity}
The intrinsic H$\alpha$ luminosity is used to compute instantaneous star formation rates (SFRs) and the number of produced ionizing photons. To measure the intrinsic H$\alpha$ luminosity, we first correct the observed line-flux in the NB$_K$ filter for the contribution of the adjacent {\sc [Nii]} emission-line doublet. We also correct the observed line-flux for attenuation due to dust.

We correct for the contribution from {\sc [Nii]} using the relation between {\sc [Nii]}/H$\alpha$ and EW$_{0, \rm [NII]+ H\alpha}$ from \cite{Sobral2012}. This relation is confirmed to hold up to at least $z\sim 1$ \citep{Sobral2015} and the median ratio of {\sc [Nii]}/(H$\alpha$+ {\sc [Nii]}) = $0.19\pm0.06$ is consistent with spectroscopic follow-up at $z\approx2$ \citep[e.g.][]{Swinbank2012,Sanders2015}, such that we do not expect that metallicity evolution between $z=1-2$ has a strong effect on the applied correction. For 1 out of the 588 HAEs we do not detect the continuum in the $K$ band, such that we use the 1$\sigma$ detection limit in $K$ to estimate the EW and the contribution from {\sc [Nii]}. We apply the same correction to HAEs that are detected as X-ray AGN (see \citealt{Matthee2016} for details on the AGN identification).

If we alternatively use the relation between stellar mass and {\sc [Nii]}/H$\alpha$ from \cite{Erb2006} at $z\sim2$, we find {\sc [Nii]}/(H$\alpha$+ {\sc [Nii]}) = $0.10\pm0.03$. This different {\sc [Nii]} estimate is likely caused by the lower metallicity of the \cite{Erb2006} sample, which may be a selection effect (UV selected galaxies typically have less dust than H$\alpha$ selected galaxies, and are thus also expected to be more metal poor, i.e. \citealt{Oteo2015}). The difference in {\sc [Nii]} contributions estimated either from the EW or mass is smaller for higher mass HAEs, which have a higher metallicity. Due to the uncertainties in the {\sc [Nii]} correction we add 50 \% of the correction to the uncertainty in the H$\alpha$ luminosity in quadrature.

Attenuation due to dust is estimated with a \cite{Calzetti2000} attenuation curve and by assuming that the nebular attenuation equals the stellar attenuation, E$(B-V)_{\rm gas} = $ E$(B-V)_{\rm stars}$. This is in agreement with the average results from the H$\alpha$ sample from MOSDEF \citep{Shivaei2015}, although we note that there are indications that the nebular attenuation is stronger for galaxies with higher SFRs and masses \citep[e.g.][]{Reddy2015,Puglisi2016} and other studies indicate slightly higher nebular attenuations \citep[e.g.][]{ForsterSchreiber2009,Wuyts2011,Kashino2013}. We note that we vary the method to correct for dust in the relevant sections (e.g. \S 6.3) in two ways: either based on the UV slope \citep{Meurer1999}, or from the local relation between dust attenuation and stellar mass \citep{GarnBest2010}.

Star formation rates are obtained from dust-corrected L(H$\alpha$) and using a \cite{Chabrier2003} initial mass function: SFR = $4.4\times10^{-42} $ L(H$\alpha$)  \citep[e.g.][]{Kennicutt1998}, where the SFR is in M$_{\odot}$ yr$^{-1}$ and L(H$\alpha$) in erg s$^{-1}$. The SFRs of galaxies in our sample range from $3-300$ M$_{\odot}$ yr$^{-1}$, with a typical SFR of $\approx30$ M$_{\odot}$ yr$^{-1}$, see Fig. $\ref{fig:galaxy_properties}$. 

\subsubsection{Rest-frame UV photometry and UV slopes} 
For our galaxy sample at $z=2.2$, the rest-frame UV ($\sim1500$\AA) is traced by the $V$ band, which is not contaminated by (possibly) strong Ly$\alpha$ emission. Our full sample of galaxies is imaged in the optical $V$ and $R$ filters with Subaru Suprime-Cam as part of the COSMOS survey \citep{Taniguchi2007}. The 5$\sigma$ depths of $V$ and $R$ are 26.2-26.4 AB magnitude (see e.g. \citealt{Muzzin2013}) and have a FWHM of $\sim0.8''$. The typical HAE in our sample has a $V$ band magnitude of $\approx25$ and is thus significantly detected. 5-7 \% of the HAEs in our sample are not detected in either the $V$ or $R$ band.

We correct the UV luminosities from the $V$ band for dust with the \cite{Calzetti2000} attenuation curve and the fitted E$(B-V)$ values. The absolute magnitude, M$_{1500}$, is obtained by subtracting a distance modulus of $\mu = 44.97$ (obtained from the luminosity distance and corrected for bandwidth stretching with 2.5log$_{10}$($1+z$), $z=2.23$)  from the observed $V$ band magnitudes. The UV slope $\beta$ is measured with observed $V$ and $R$ magnitudes following:
\begin{equation}
\beta = -\frac{V-R}{2.5 \rm log_{10}(\lambda_V/\lambda_R)} - 2 
\end{equation}
Here, $\lambda_V = 5477.83$ {\AA}, the effective wavelength of the $V$ filter and $\lambda_R = 6288.71$ {\AA}, the effective wavelength of the $R$ filter. With this combination of filters, $\beta$ is measured around a rest-frame wavelength of $\sim 1800$ {\AA}. 

\section{{\it GALEX} UV data}
\label{sec:3}
For galaxies observed at $z=2.2$, rest-frame LyC photons can be observed with the $NUV$ filter on the {\it GALEX} space telescope. In COSMOS there is deep {\it GALEX} data (3$\sigma$ AB magnitude limit $\sim 25.2$, see e.g. \citealt{Martin2005,Muzzin2013}) available from the public Deep Imaging Survey. We stress that the full width half maximum (FWHM) of the point spread function (PSF) of the $NUV$ imaging is 5.4$''$\citep{Martin2003} and that the pixel scale is 1.5$''$ pix$^{-1}$. We have acquired $NUV$ images in COSMOS from the Mikulski Archive at the Space Telescope Science Institute (MAST)\footnote{https://mast.stsci.edu/}. All HAEs and LAEs in COSMOS are covered by {\it GALEX} observations, due to the large circular field of view with 1.25 degree diameter. Five pointings in the COSMOS field overlap in the center, which results in a total median exposure time of 91.4 ks and a maximum exposure time of 236.8 ks. 

\begin{figure}
\includegraphics[width=8.8cm]{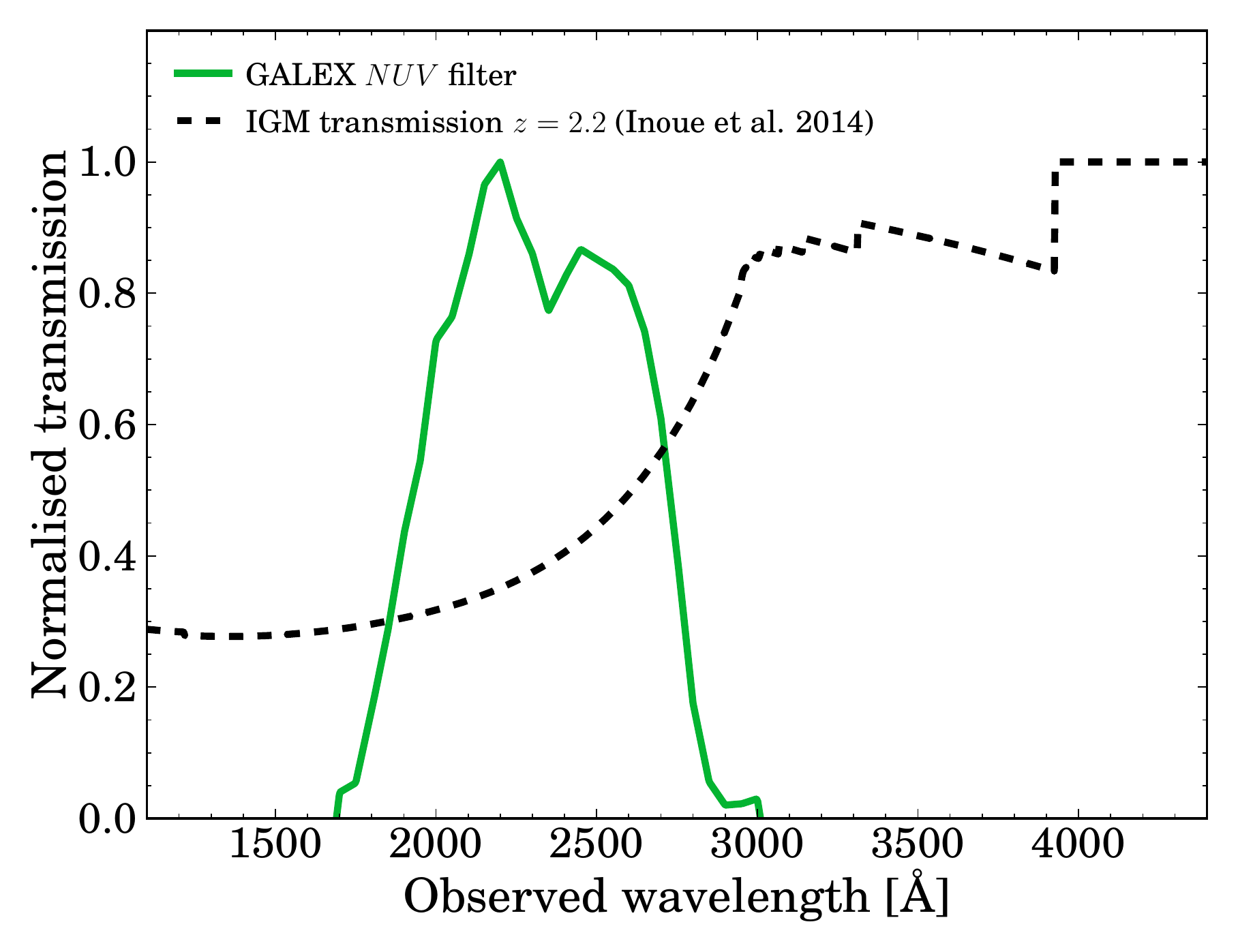}
\caption{\small{Filter transmission of the {\it GALEX} $NUV$ filter (green line) and mean IGM transmission versus observed wavelength (dashed black line). We compute the IGM transmission at $z=2.2$ using the models from \citet{Inoue2014}. The bandpass-averaged IGM transmission is 40.4 \%. As highlighted by a simulation from \citealt{Vasei2016}, the mean value of $T_{IGM}$ is not the most common value. The distribution is bimodal, with a narrow peak at $T_{IGM} \approx 0.0$ and a broad peak around $T_{IGM} = 0.7$. }} 
\label{fig:filter}
\end{figure}

\subsection{Removing foreground/neighbouring contamination}
The large PSF-FWHM of {\it GALEX} $NUV$ imaging leads to a major limitation in measuring escaping LyC photons from galaxies at $z=2.2$. This is because the observed flux in the $NUV$ filter could (partly) be coming from a neighbouring foreground source at lower redshift.  In order to overcome this limitation, we use available high resolution deep optical {\it HST}/ACS F814W (rest-frame $\approx 2500$ {\AA}, \citealt{Koekemoer2007}) imaging to identify sources for which the $NUV$ flux might be confused due to possible foreground or neighbouring sources and remove these sources from the sample. In addition, we use visual inspections of deep ground-based $U$ band imaging as a cross-check for the bluest sources which may be missed with the {\it HST} imaging. These data are available through the COSMOS archive.\footnote{http://irsa.ipac.caltech.edu/data/COSMOS/}

Neighbours are identified using the photometric catalog from \cite{Ilbert2009}, which is selected on deep {\it HST}/ACS F814W data. We find that 195 out of the 588 HAEs in COSMOS have no neighbour inside a radius of 2.7$''$. We refer to this subsample as our {\sc Clean} sample of galaxies in the remainder of the text. The average properties (dust attenuation, UV magnitude mass and SFR) of this sample is similar to the full sample of SFGs.

\subsection{Transmission redward of 912 \AA}
For sources at $z=2.22$, the $NUV$ filter has non-negligible transmission from $\lambda_0= 912-933$ {\AA} of $\approx 1.5$\%. This limits the search for escaping LyC radiation. The fraction of the observed flux in the $NUV$ filter that originates from $\lambda_0>912$ {\AA} depends on the galaxy's SED, the IGM transmission and the filter transmission. In order to estimate this contribution, we first use a set of {\sc Starburst99} \citep{Starburst99} SED models to estimate the shape of the galaxy's SED in the far UV. We assume a single burst of star formation with a Salpeter IMF with upper mass limit 100 M$_{\odot}$, Geneva stellar templates without rotation \citep{Mowlavi2012} and metallicity $Z = 0.02$. Then, we convolve this SED with the filter and IGM transmission curves, to obtain the fraction of the flux in the $NUV$ filter that is non-ionizing at $z=2.2$ (compared to the flux in the $NUV$ filter that is ionizing). By using the SED models with H$\alpha$ EWs within our measured range, we find that $2.6\pm0.4$ \% of the flux observed in the $NUV$ filter is not-ionizing. This means that upper limits from non-detections are slightly over-estimated. For individually detected sources it is theoretically possible that the $NUV$ detection is completely due to non-ionizing flux, depending on the SED shape and normalisation. This is analysed in detail on a source-by-source basis in Appendix A.

\section{The escape fraction of ionizing photons}
\label{sec:4}
\subsection{How to measure f$_{\rm esc}$?}
Assuming that LyC photons escape through holes in the ISM (and hence that H{\sc ii} regions are ionization bounded from which no ionizing photons escape), the escape fraction, f$_{\rm esc}$, can be measured directly from the ratio of observed to produced LyC luminosity (averaged over the solid angle of the measured aperture). 

In this framework, produced LyC photons either escape the ISM, ionise neutral gas (leading to recombination radiation) or are absorbed by dust \citep[e.g.][]{Bergvall2006}. The number of produced ionizing photons per second, Q$_{\rm ion}$, can be estimated from the strength of the (dust corrected) H$\alpha$ emission line as follows:
\begin{equation}
L_{\rm H\alpha}  = Q_{\rm ion} \, c_{\rm H\alpha} \, (1-f_{\rm esc}-f_{\rm dust})
\end{equation}
where Q$_{\rm ion}$ is in s$^{-1}$, L$_{\rm H\alpha}$ is in erg s$^{-1}$, f$_{\rm esc}$ is the fraction of produced ionizing photons that escapes the galaxy and f$_{\rm dust}$ is the fraction of produced ionizing photons that is absorbed by dust. For case B recombinations with a temperature of $T=10\ 000$ K, $c_{\rm H\alpha} = 1.36\times10^{-12}$ erg \citep[e.g.][]{Kennicutt1998,Schaerer2003}. Since the dust attenuation curve at wavelengths below 912 {\AA} is highly uncertain, we follow the approach of \cite{Rutkowski2015}, who use f$_{\rm dust} = 0.5$, which is based on the mean value derived by \cite{Inoue2002} in local galaxies.

Rest-frame LyC photons are redshifted into the $NUV$ filter at $z=2.2$. However, the IGM between $z=2.2$ and our telescopes is not transparent to LyC photons (see Fig. $\ref{fig:filter}$), such that we need to correct the observed LyC luminosity for IGM absorption. The observed luminosity in the $NUV$ filter ($L_{NUV}$) is then related to the produced number of ionizing photons as:

\begin{equation}
L_{NUV} = Q_{\rm ion} \, \epsilon \, f_{\rm esc} \, T_{\rm IGM, NUV}
\end{equation}
Here, $\epsilon$ is the average energy of an ionizing photon observed in the $NUV$ filter (which traces rest-frame wavelengths from 550 to 880 {\AA}, see Fig. $\ref{fig:filter}$). Using the {\sc Starburst99} models as described in \S 3.2, we find that $\epsilon$ is a strong function of age, but that it is strongly correlated with the EW of the H$\alpha$ line (which itself is also a strong function of age). For the range of H$\alpha$ EWs in our sample, $\epsilon = 17.04^{+0.45}_{-0.26}$ eV. We therefore take $\epsilon = 17.0$ eV. 

$T_{\rm IGM, NUV}$ is the absorption of LyC photons due to the intervening IGM, convolved with the $NUV$ filter. Note that $T_{\rm IGM} =e^{-\tau_{\rm IGM}}$, where $\tau_{IGM}$ is the optical depth to LyC photons in the IGM, see e.g \cite{Vanzella2012}. The IGM transmission depends on the wavelength and redshift. According to the model of \cite{Inoue2014}, the mean IGM transmission for LyC radiation at $\lambda \sim 750$ {\AA} for a source at  $z=2.2$ is $T_{\rm IGM}\approx40$ \%. We convolve the IGM transmission as a function of observed wavelength for a source at $z=2.2$ with the normalised transmission of the $NUV$ filter, see Fig. $\ref{fig:filter}$. This results in a bandpass-averaged $T_{\rm IGM, NUV} = 40.4$\%.

Combining equations 2 and 3 results in:
\begin{equation}
f_{\rm esc} = \frac{1-f_{\rm dust}}{(1+\alpha \frac{L_{\rm H\alpha}}{L_{NUV}})}
\end{equation}
where we define $\alpha = \epsilon \, c_{\rm H\alpha}^{-1} \, T_{\rm IGM, NUV}$. Combining our assumed values, we estimate $\alpha =8.09$. We note that $\epsilon$ and c$_{\rm H\alpha}$ are relatively insensitive to systematic uncertainties, while f$_{\rm dust}$ and T$_{\rm IGM}$ are highly uncertain for individual sources.

In addition to the absolute escape fraction of ionizing radiation, it is common to define the relative escape fraction of LyC photons to UV ($\sim 1500$ {\AA}) photons, since these are most commonly observed in high redshift galaxies. Following \cite{Steidel2001}, the relative escape fraction, f$_{\rm esc}^{rel}$, is defined as:

\begin{figure*}
\begin{tabular}{ccc}
\includegraphics[width=5.6cm]{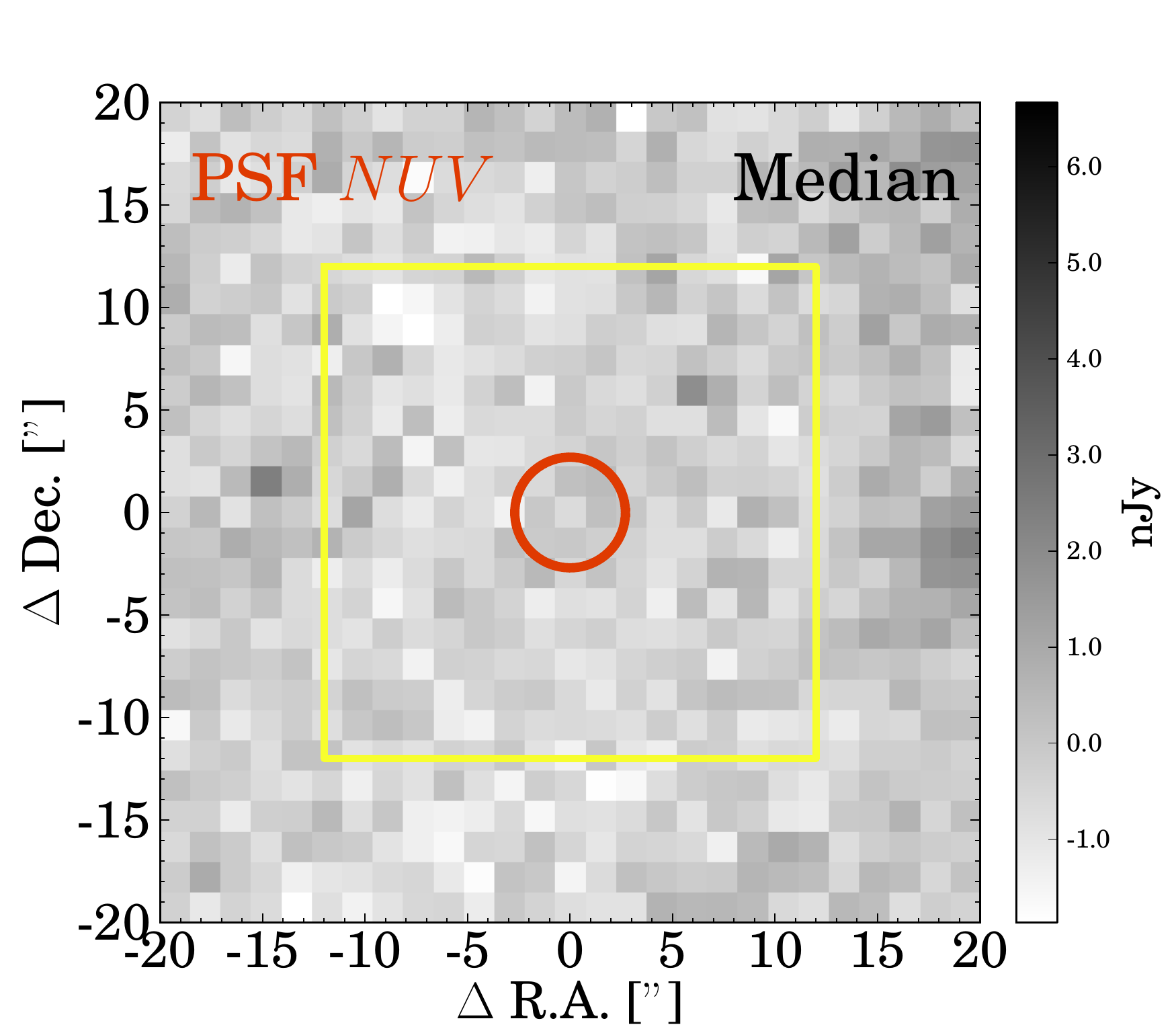}&
\includegraphics[width=5.6cm]{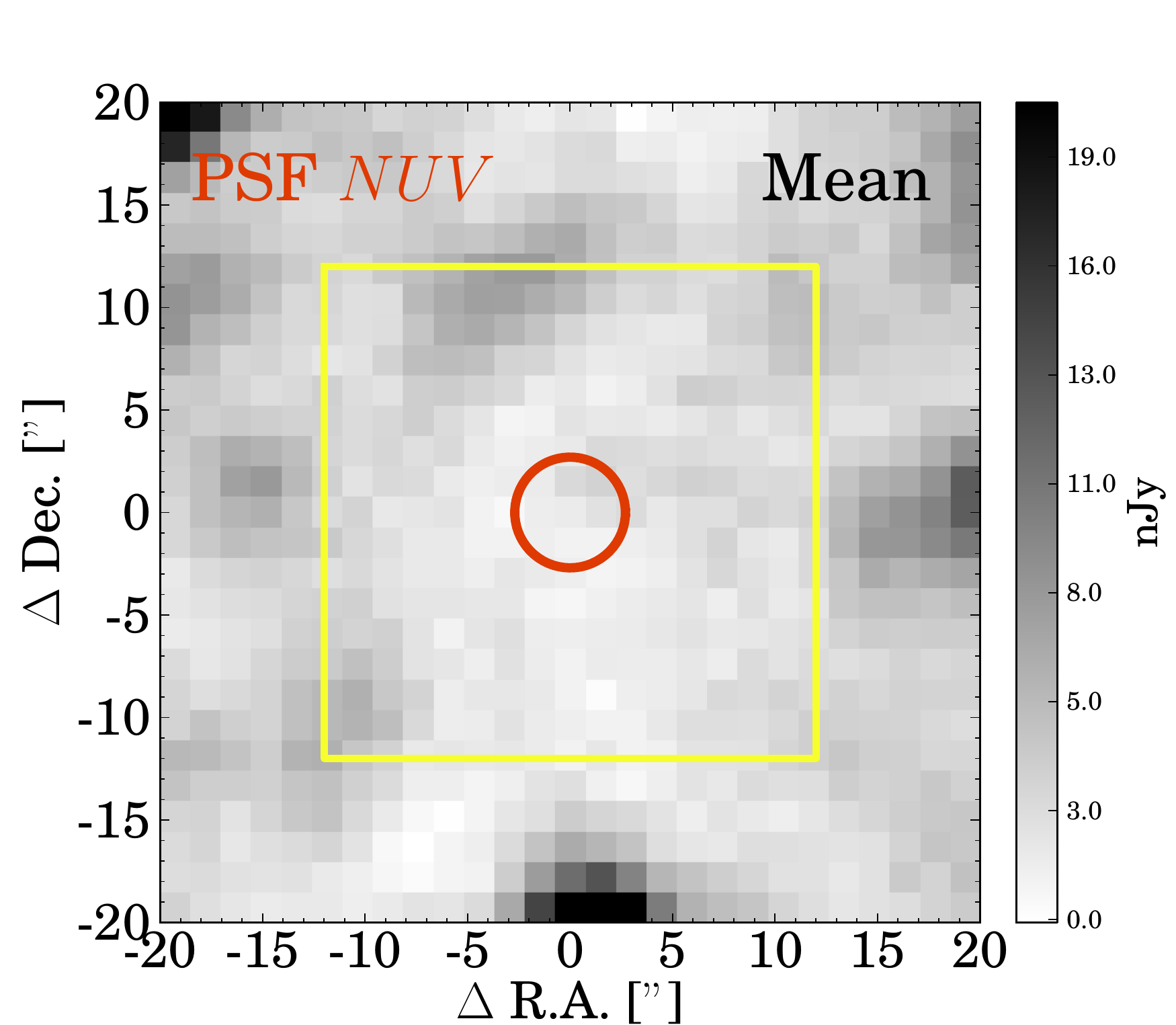}&
\includegraphics[width=5.6cm]{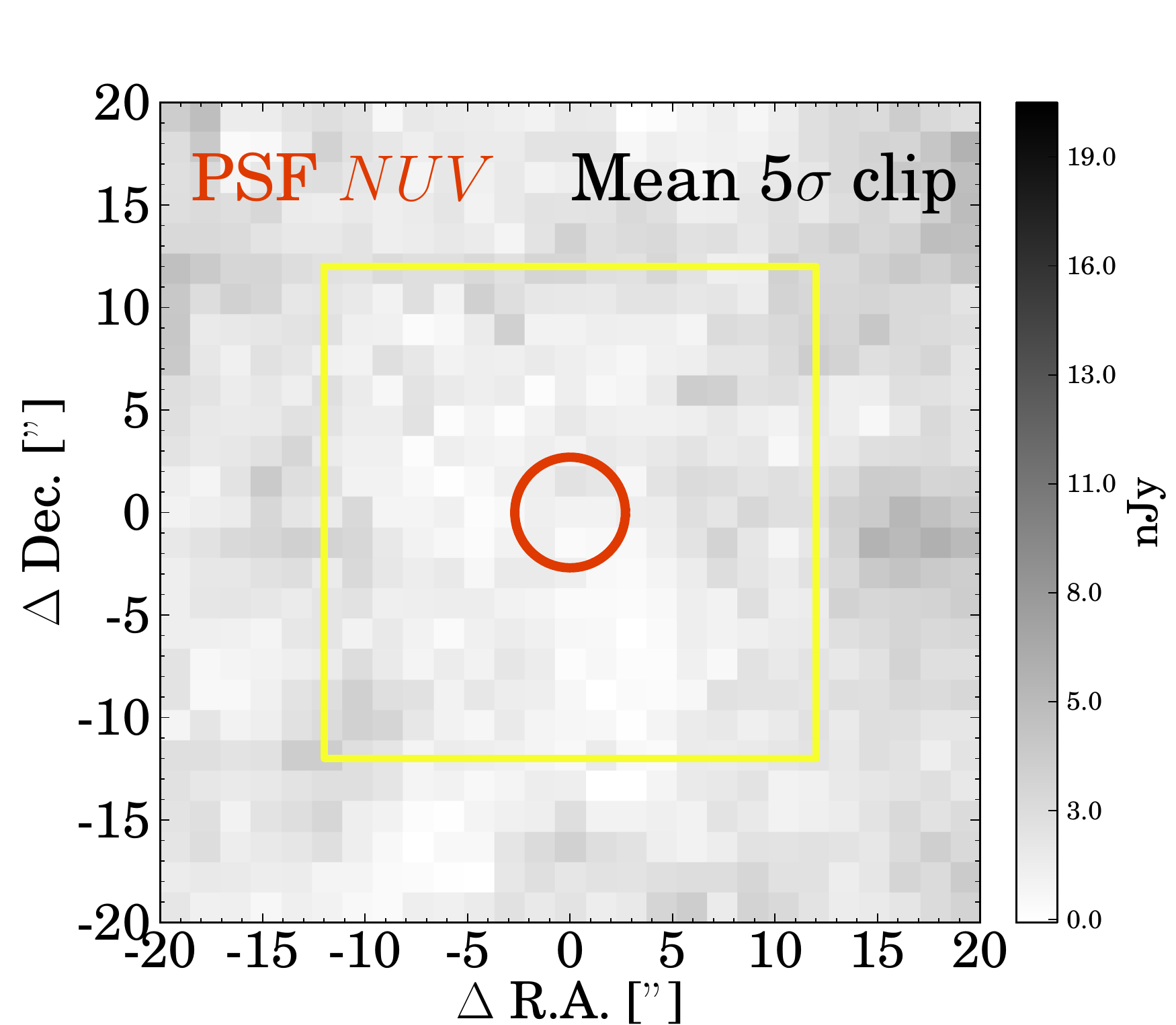}\\

\end{tabular}
\caption{\small{$20''\times20''$ thumbnail images of the $NUV$ stack for {\sc Clean}, star-forming HAEs in COSMOS, for three different stacking methods. The red circle shows the PSF-FWHM of $NUV$ on the central position. The yellow box is the box which is used to measure the depth of the stack. Note that the range of the color-bar of the median stack is different than the color-bar of the mean stacks because the median stack is deeper. }}
\label{fig:stacks_appendix}
\end{figure*}

\begin{equation}
 f_{\rm esc}^{rel} = f_{\rm esc} e^{\tau_{dust, UV}} = \frac{(L_{UV}/L_{NUV})_{int}}{(L_{UV}/L_{NUV})_{obs}} \,T_{\rm IGM, NUV}^{-1}
\end{equation}

In this equation, L$_{UV}$ is the luminosity in the observed $V$ band, $e^{\tau_{dust, UV}}$ is the correction for dust (see \S 2.1.3) and we adopt an intrinsic ratio of $(L_{UV}/L_{NUV})_{int}$ = 5 \citep[e.g.][]{Siana2007}. The relative escape fraction can be related to the absolute escape fraction when the dust attenuation for L$_{UV}$, $A_{UV}$, is known: $f_{\rm esc} =  f_{\rm esc}^{rel} \times 10^{-0.4 A_{UV}} $.

\begin{table*}
\begin{center}
\begin{small}
\caption{Stacked measurements for subsamples of HAEs and LAEs at $z=2.2$. \# indicates the number of objects in each subsample. We further show the general characteristics of the subsample with observed H$\alpha$ luminosity (corrected for {\sc [Nii]} contribution, see \S 2.1.2), the H$\alpha$ extinction with the E$(B-V)$ value and a Calzetti law, the median stellar mass and UV slope ($\beta$) inferred from $V-R$ colours. The $NUV$ column shows the limits on the $NUV$ magnitude. L$_{1500}$ is the rest-frame 1500 {\AA} luminosity obtained from the $V$ band. The absolute f$_{\rm esc}$ is measured from H$\alpha$ and the $NUV$ as described in \S 4.1. f$_{\rm esc, rel}$ is the relative escape fraction of ionizing photons to UV photons and is measured from $NUV$ and L$_{1500}$. Note that with a Calzetti law A$_{UV} = 3.1$A$_{\rm H\alpha}$. {\sc Clean} subsamples are samples without foreground/neighbouring source within the $NUV$ PSF (2.7$''$).} 

\begin{tabular}{ lrrrrrrrrr }
$\bf Subsample$ & $\bf \#$ &$\bf L_{H\alpha, obs}$ & $\bf A_{H\alpha}$ &\bf $\beta$ & \bf M$_{\rm star}$ & $\bf NUV$ & \bf L$_{1500}$ & $\bf f_{\rm esc}$ & $\bf f_{\rm esc}^{rel}$ \\ 
 & & erg s$^{-1}$ & mag & & log$_{10}$(M$_{\odot}$) & 1$\sigma$ AB & erg s$^{-1}$Hz$^{-1}$ & \% & \% \\ \hline
 {\bf Median stacking} & & & & & & & & & \\
COSMOS no AGN {\sc Clean} & 191 & $1.60\times10^{42}$ & 1.23 & -1.97 & 9.55 &  29.7 & 5.78$\times10^{28}$& $ <2.8$ & $<92.5$\\ 
 {\bf Mean stacking} & & & & & & & & & \\
COSMOS no AGN {\sc Clean} & & & & & & 27.9 &  & $ <11.7$ & $<465.4$ \\ 
--5$\sigma$ clip &  & & & & & 28.7 &  & $ <6.4$ & $<231.0$ \\ \hline

\end{tabular}
\label{tab:subsamples}
\end{small}
\end{center}
\end{table*}

\subsection{Individual $NUV$ detections}
By matching our sample of HAEs and LAEs with the public {\it GALEX} EM cleaned catalogue \citep[e.g.][]{Zamojski2007,EMphot}, we find that 33 HAEs and 5 LAEs have a detection with $NUV<26$ within a separation of 1$''$. However, most of these matches are identified as spurious,  foreground sources or significantly contaminated inside the large PSF-FWHM of $NUV$ imaging (see Appendix A). Yet, seven of these matches (of which five are AGN) are in the {\sc Clean} subsample without a clear foreground source and could thus potentially be LyC leakers. Because it is known that foreground contamination has been a major problem in studies of LyC leakage at $z\sim3$ \citep[e.g.][]{Mostardi2015,Siana2015}, we can only confirm the reality of these candidate LyC leakers with high resolution UV imaging with HST. We list the individual detections in Appendix A, but caution the reader that any further investigation requires these candidates to be confirmed first.

\subsection{Stacks of HAEs}
The majority of our sources are undetected in the $NUV$ imaging. Therefore, in order to constrain f$_{\rm esc}$ for typical star-forming galaxies, we stack $NUV$ thumbnails of our full sample of HAEs in COSMOS and also stack various subsets. We create thumbnails of $40''\times40''$ centered on the position of the NB$_K$ (H$\alpha$) detection and stack these by either median or mean combining the counts in each pixel. While median stacking results in optimal noise properties and is not dominated by outliers, it assumes that the underlying population is uniform, which is likely not the case. Mean stacking is much more sensitive to outliers (such as for example luminous AGN), but would give a more meaningful result as it gives the average f$_{\rm esc}$, which is the important quantity in assessing the ionizing photon output of the entire galaxy population. 

We measure the depth by randomly placing 100,000 empty apertures with a radius of $0.67\times$PSF-FWHM (similar to e.g. \citealt{Cowie2009,Rutkowski2015}) in a box of  $24''\times24''$ around the centre of the thumbnail (see Fig. $\ref{fig:stacks_appendix}$) and quote the 1$\sigma$ standard deviation as the depth. Apertures with a detection of $NUV<26$ AB magnitude are masked (this is particularly important for mean stacking). Counts are converted to AB magnitudes with the photometric zero-point of 20.08 \citep{Cowie2009}. For mean stacking, we experiment with an iterative 5$\sigma$ clipping method in order to have the background not dominated by a few luminous sources. To do this, we compute the standard deviation of the counts of the stacked sample in each pixel and ignore 5$\sigma$ outliers in computing the mean value of each pixel. This is iterated five times, although we note that most of the mean values already converge after a single iteration.
 
By stacking only sources from the {\sc Clean} sample and by removing X-ray AGN, the limiting $NUV$ magnitude of the stack of {\sc Clean} HAEs is $NUV \approx 29.7$ AB (see Table $\ref{tab:subsamples}$), which translates into an upper limit of f$_{\rm esc} < 2.8$ \%. Mean stacking gives shallower constraints f$_{\rm esc} < 11.7$ \%)because the noise does not decrease as rapidly by stacking more sources, possibly because of a contribution from faint background or companion sources below the detection limit. This is improved somewhat by our iterative 5$\sigma$ clipping (f$_{\rm esc} < 6.4$ \%), which effectively masks out the contribution from bright pixels. We show the stacked thumbnails of this sample in Fig. $\ref{fig:stacks_appendix}$.

The median (mean) upper limit on the relative escape fraction, f$_{\rm esc, rel}$, is much higher ($<92.5 (231)$ \%). However, if we correct for the dust attenuation with the \cite{Calzetti2000} law, we find A$_{UV} \approx 3.8$ and a dust corrected inferred escape fraction of $<2.8 (7.0)$ \%, in good agreement with our direct estimate from H$\alpha$, although we note that the additional uncertainty due to this dust correction is large. 

We have experimented by stacking subsets of galaxies in bins of stellar mass, SFR and UV magnitude or LAEs, but all result in a non-detection in the $NUV$, all with weaker upper limits than the stack of {\sc Clean} HAEs. 

\subsubsection{Systematic uncertainty due to the dust correction}
In this sub-section, we investigate how sensitive our results are to the method used to correct for dust, which is the most important systematic uncertainty. In Table $\ref{tab:subsamples}$, we have used the SED inferred value of E$(B-V)$ to infer A$_{\rm H\alpha}$:  A$_{\rm H\alpha} = E(B-V)\times k_{\rm H\alpha}$, where $k_{\rm H\alpha}=3.3277$ following \cite{Calzetti2000}, which results in A$_{\rm H\alpha} = 1.23$. However, it is also possible to infer A$_{\rm H\alpha}$ from a relation with the UV slope \citep[e.g.][]{Meurer1999}, such that A$_{\rm H\alpha} = 0.641 (\beta+2.23)$, for $\beta > -2.23$ and A$_{\rm H\alpha} = 0$ for $\beta < -2.23$. Finally, we also use the relation between A$_{\rm H\alpha}$ and stellar mass from \cite{GarnBest2010}, which is: A$_{\rm H\alpha} = 0.91+0.77 X+0.11 X^2-0.09 X^3$, where $X = $ log$_{10}$(M$_{\rm star}$/$10^{10}$ M$_{\odot}$). Note that we assume a \cite{Calzetti2000} dust law in all these prescriptions.

It is immediately clear that there is a large systematic uncertainty in the dust correction, as for our full sample of HAEs we infer A$_{\rm H\alpha} = 0.70$ with the \cite{GarnBest2010} prescription and A$_{\rm H\alpha} =0.19$ following \cite{Meurer1999}, meaning that the systematic uncertainty due to dust can be as large as a factor 3. Thus, these different dust corrections result in different upper limits on f$_{\rm esc}$. For the {\sc Clean}, star-forming HAE sample, the upper limit on f$_{\rm esc}$ from median stacking increases to $f_{\rm esc} <4.4\,(6.6)$ \%, using the attenuation based on stellar mass ($\beta$). With a simple 1 magnitude of extinction for H$\alpha$,  f$_{\rm esc} < 3.4$ \% and without correcting for dust results in f$_{\rm esc} < 7.7$ \%.

In addition to the uncertainty in the dust correction of the H$\alpha$ luminosity, another uncertainty in our method is the f$_{\rm dust}$ parameter introduced in Eq. 2. The dust attenuation curve at wavelengths below 912 {\AA} is highly uncertain, such that our estimate of f$_{\rm dust}$ is uncertain as well. However, because our limits on f$_{\rm esc}$ from the median stack are low, the results do not change significantly by altering f$_{\rm dust}$: if f$_{\rm dust}=0.75 (0.25)$, we find f$_{\rm esc}<1.4 (4.1)$ \%. This means that as long as the limit is low, our result is not very sensitive to the exact value of f$_{\rm dust}$.

\begin{table*}
\begin{center}
\begin{small}
\caption{Measurements of $\langle$f$_{\rm esc} \rangle$, the escape fraction of ionizing photons averaged over the galaxy population at $z\approx2-5$. Constraints on the IGM emissivity from absorption studies by \citet{BeckerBolton2013} have been used to infer the global escape fraction. For $z=2.2$, we have used the H$\alpha$ luminosity function from \citet{Sobral2013}. We have used the analytical formula from \citet{MadauHaardt2015} to estimate the contribution from quasars to the ionizing emissivity, which assumes that f$_{\rm esc, quasars} = 100$ \%. At $z=3.8$ and $z=4.9$ we have used the SFR function from \citet{Smit2016}.} 
\begin{tabular}{ lrr }
\bf Sample & \bf Method & \bf $\langle$f$_{\rm esc} \rangle$\\ \hline
\bf This paper & & \\
HAEs $z=2.2$ & full SFR integration, A$_{\rm H\alpha} = 1.0$ & $4.4^{+7.1}_{-2.0}$ \% \\
HAEs $z=2.2$ & SFR $> 3$ M$_{\odot}$/yr, A$_{\rm H\alpha} = 1.0$ & $6.7^{+10.8}_{-3.1}$ \%\\
HAEs $z=2.2$ & full SFR integration, A$_{\rm H\alpha} = 0.7$ & $5.9^{+9.3}_{-2.6}$ \% \\
\bf HAEs $z=2.2$ & \bf final estimate: full integration, A$_{\rm H\alpha} = 0.7$, conservative systematic errors & \bf $5.9^{+14.5}_{-4.2}$ \% \\ HAEs $z=2.2$ & full SFR integration, A$_{\rm H\alpha} = 1.0$, QSO contribution & $0.5^{+3.6}_{-0.5}$ \% \\
\\
LBGs $z=3.8$ & full SFR integration, H$\alpha$ from {\it Spitzer}/IRAC & $2.7^{+7.2}_{-2.3}$ \% \\
LBGs $z=3.8$ & full SFR integration, H$\alpha$ from {\it Spitzer}/IRAC, QSO contribution & $0.0^{+3.0}_{-0.0}$ \% \\

LBGs $z=4.9$ & full SFR integration, H$\alpha$ from {\it Spitzer}/IRAC & $6.0^{+13.9}_{-5.2}$ \% \\
LBGs $z=4.9$ & full SFR integration, H$\alpha$ from {\it Spitzer}/IRAC, QSO contribution & $2.1^{+6.2}_{-2.1}$ \% \\ \hline

\bf Literature & & \\
\citet{Cristiani2016} $z=3.8$ & integrated LBG LF + contribution from QSOs & $5.3^{+2.7}_{-1.2}$ \% \\ \hline
\end{tabular}
\label{tab:global_fesc}
\end{small}
\end{center}
\end{table*} 

\section{Constraining f$_{\rm esc}$ of HAEs from the ionizing background}
\label{sec:5}
In addition to constraining f$_{\rm esc}$ directly, we can obtain an indirect measurement of f$_{\rm esc}$ by using the ionizing emissivity, measured from quasar absorption studies, as a constraint. The emissivity is defined as the number of escaping ionizing photons per second per comoving volume: 
\begin{equation}
\dot{N}_{ion} = \langle {\rm f}_{\rm esc} \rangle \times \Phi({\rm H}\alpha)\times c^{-1}_{\rm H\alpha}
\end{equation}
Here, $\dot{N}_{ion}$ is in s$^{-1}$ Mpc$^{-3}$, $\langle$f$_{\rm esc} \rangle$ is the escape fraction averaged over the entire galaxy population, $\Phi(\rm H\alpha)$ is the H$\alpha$ luminosity density in erg s$^{-1}$ Mpc$^{-3}$ and $c_{\rm H\alpha}$ is the recombination coefficient as in Eq. 2. 

We first check whether our derived emissivity using our upper limit on f$_{\rm esc}$ for HAEs is consistent with published measurements of the emissivity. The H$\alpha$ luminosity density is measured in \cite{Sobral2013} as the full integral of the H$\alpha$ luminosity function, with a global dust correction of A$_{\rm H\alpha} = 1.0$. Using the mean limit on f$_{\rm esc}$ for our {\sc Clean} sample of HAEs (so f$_{\rm esc}\leq6.4$ \%), we find that $\dot{N}_{ion} \leq 1.3^{+0.2}_{-0.2}\times10^{51}$ s$^{-1}$ Mpc$^{-3}$, where the errors come from the uncertainty in the H$\alpha$ LF. We note that these numbers are relatively independent on the dust correction method because while a smaller dust attenuation would decrease the H$\alpha$ luminosity density, it would also raise the upper limit on the escape fraction, thus almost cancelling out. These upper limits on $\dot{N}_{ion}$ are consistent with the measured emissivity at $z=2.4$ of \cite{BeckerBolton2013}, who measured $\dot{N}_{ion} = 0.90^{+1.60}_{-0.52}\times10^{51}$ s$^{-1}$ Mpc$^{-3}$ (combined systematic and measurement errors) using the latest measurements of the IGM temperature and opacity to Ly$\alpha$ and LyC photons.
 
Now, by isolating $\langle$f$_{\rm esc} \rangle$ in Eq. 6, we can estimate the globally averaged escape fraction. If we assume that there is no evolution in the emissivity from \cite{BeckerBolton2013} between $z=2.2$ and $z=2.4$ and that the H$\alpha$ luminosity function captures all sources of ionizing photons, we find that $\langle$f$_{\rm esc} \rangle = 4.4^{+7.1}_{-2.0}$ \% for A$_{\rm H\alpha} = 1.0$. There are a number of systematic uncertainties that we will address now and add to the error bars of our final estimate. These uncertainties are: i) integration limit of the H$\alpha$ LF, ii) the dust attenuation to L(H$\alpha$), iii) the conversion from L(H$\alpha$) to ionizing numbers, and iv) the {\sc [Nii]} correction to the observed H$\alpha$ luminosity.

Integrating the H$\alpha$ LF only to SFR $\approx 3$ M$_{\odot}$ yr$^{-1}$, we find $\langle$f$_{\rm esc} \rangle = 6.7^{+10.8}_{-3.1}$ \%, such that the systematic uncertainty is of order 50 \%. If A$_{\rm H\alpha} = 0.7$, which is the median value when we correct for dust using stellar mass, and which may be more representative of fainter H$\alpha$ emitters (as faint sources are expected to have less dust), the escape fraction is somewhat higher, with $\langle$f$_{\rm esc} \rangle = 5.9^{+9.3}_{-2.6}$ \%. These numbers are summarised in Table $\ref{tab:global_fesc}$. The uncertainty in c$_{\rm H\alpha}$ is relatively small, as c$_{\rm H\alpha}$ depends only modestly on the density and the temperature. For example, in the case of a temperature of T = $30 000$ K, c$_{\rm H\alpha}$ decreases only by $\approx10$\% \citep{Schaerer2002}. This adds a 10 \% uncertainty in the escape fraction. As explained in \S 2.1.2, there is an uncertainty in the measured H$\alpha$ luminosity due to the contribution of the {\sc [Nii]} doublet to the observed narrow-band flux, for which we correct using a relation with observed EW. By comparing this method with the method from \cite{Erb2006}, which is based on the observed mass-metallicity relation of a sample of LBGs at $z\sim2$, we find that the inferred H$\alpha$ luminosity density would conservatively be 10 \% higher, such that this correction adds another 10 \% systematic uncertainty in the escape fraction.

For our final estimate of $\langle$f$_{\rm esc} \rangle$ we use the dust correction based on stellar mass, fully integrate the H$\alpha$ luminosity function and add a 10 \% error in quadrate for the systematic uncertainty in each of the parameters as described above, 50 \% due to the uncertain integration limits and add a 40 \% error due to the systematics in the dust attenuation. This results in $\langle$f$_{\rm esc} \rangle = 5.9^{+14.5}_{-4.2}$ \% at $z=2.2$.

An additional contribution to the ionizing emissivity from rarer sources than sources with number densities $<10^{-5}$ Mpc$^{-3}$ such as quasars, would lower the escape fraction for HAEs. While \cite{MadauHaardt2015} argue that the ionizing budget at $z\approx2-3$ is dominated by quasars, this measurement may be overestimated by assuming quasars have a 100 \% escape fraction. Recently, \cite{Micheva2016} obtained a much lower emissivity (up to a factor of 10) from quasars by directly measuring f$_{\rm esc}$ for a sample of $z\sim3$ AGN. Using a large sample of quasars at $z=3.6-4.0$, \cite{Cristiani2016}, measure a mean $\langle$f$_{\rm esc, quasar} \rangle \approx 70$ \%, which means that quasars do not dominate the ionizing background at $z\approx4$. When we include a quasar contribution from \cite{MadauHaardt2015} in the most conservative way (meaning that we assume f$_{\rm esc}$ = 100 \% for quasars), we find that $\langle$f$_{\rm esc} \rangle = 0.5^{+3.6}_{-0.5}$ \%. If the escape fraction for quasars is 70 \%, $\langle$f$_{\rm esc} \rangle = 1.6^{+5.4}_{-1.3}$ \%, such that a non-zero contribution from star-forming galaxies is not ruled out.

We note that, these measurements of $\langle$f$_{\rm esc} \rangle$ contain significantly less (systematic) uncertainties than measurements based on the integral of the UV luminosity function \citep[e.g.][]{BeckerBolton2013,Khaire2016}. This is because: i) UV selected galaxy samples do not necessarily span the entire range of SFGs \citep[e.g.][]{Oteo2015} and might thus miss dusty star-forming galaxies and ii) there are additional uncertainties in converting non-ionizing UV luminosity to intrinsic LyC luminosity (in particular the dust corrections in $\xi_{ion}$ and uncertainties in the detailed SED models in $(L_{UV}/L_{NUV})_{int}$). An issue is that H$\alpha$ is very challenging to observe at $z\gtrsim2.8$ and that a potential spectroscopic follow-up study of UV selected galaxies with the {\it JWST} might yield biased results. 

\begin{figure}
\includegraphics[width=8.6cm]{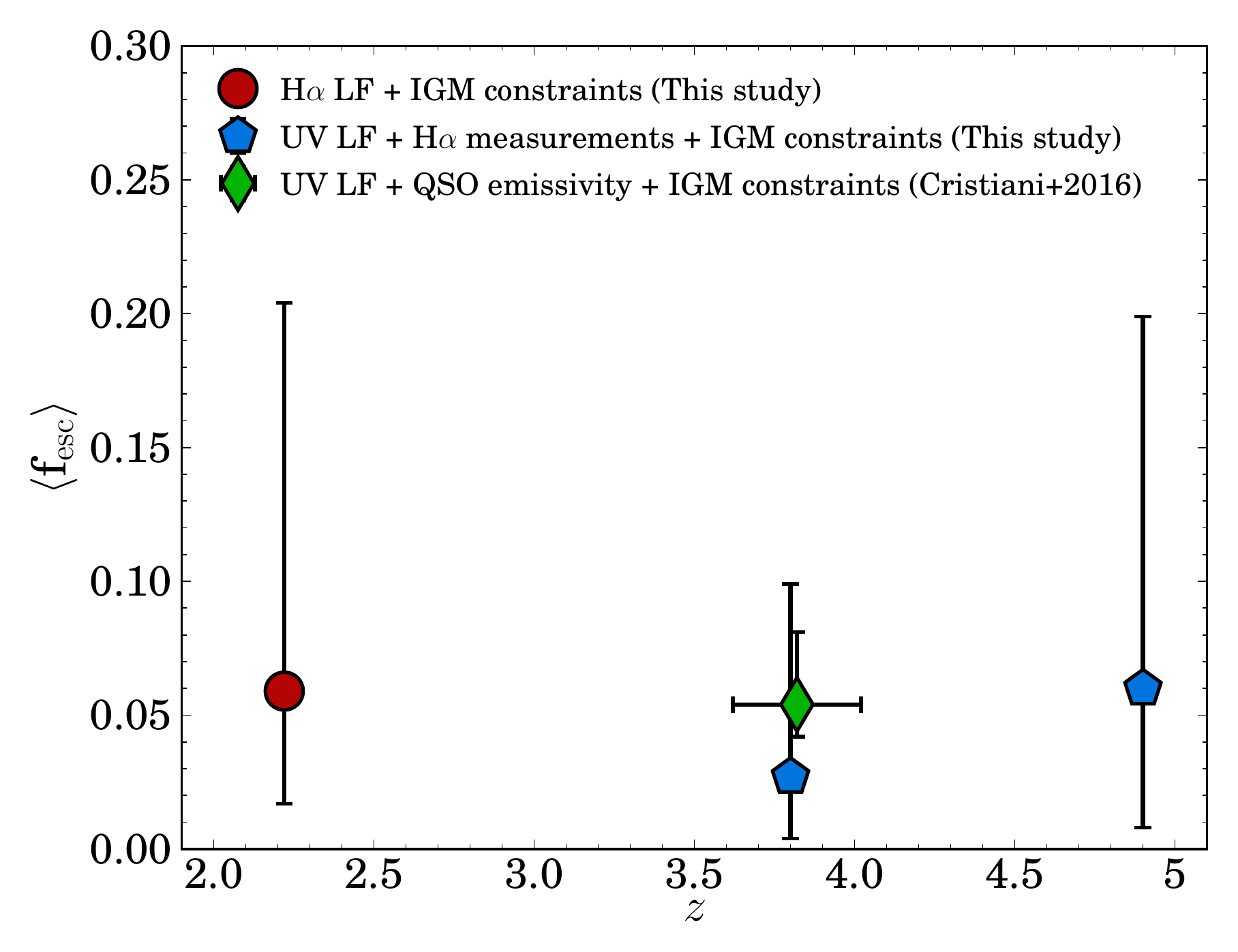}
\caption{\small{Evolution of the globally averaged $\langle$f$_{\rm esc}\rangle$, which is obtained by forcing the emissivity of the integrated H$\alpha$ ($z=2.2$) and UV ($z\approx4-5$) LF to be equal to the emissivity measured by IGM absorption models from \citealt{BeckerBolton2013}. The $z\approx4-5$ results are based on a UV luminosity function which is then corrected to a SFR function with H$\alpha$ measurements from {\it Spitzer}/IRAC, which implicitly means using a value of $\xi_{ion}$ (SFR functions are presented in \citealt{Smit2016}, but see also \citealt{Bouwens2015xi}). The error bars of red and blue symbols include estimates of the systematic uncertainties. The green diamond shows the estimated value by \citealt{Cristiani2016}, who combined IGM constraints with a UV LBG and the emissivity of QSOs at $z=3.6-4.0$.}} 
\label{fig:fesc_evolution}
\end{figure}

\subsection{Redshift evolution}
Using the methodology described in \S 5, we also compute the average f$_{\rm esc}$ at $z=3.8$ and $z=4.9$ by using the SFR functions of \cite{Smit2016}, which are derived from UV luminosity functions, a \cite{Meurer1999} dust correction and a general offset to correct for the difference between SFR(UV) and SFR(H$\alpha$), estimated from {\it Spitzer}/IRAC photometry. This offset is implicitly related to the value of $\xi_{ion}$ from \cite{Bouwens2015xi}, which is estimated from the same measurements. We combine these SFR functions, converted to the H$\alpha$ luminosity function as in \S 2.1.2, with the IGM emissivity from \cite{BeckerBolton2013} at $z=4.0$ and $z=4.75$, respectively. Similarly to the H$\alpha$ luminosity density, we use the analytical integral of the Schechter function. Similarly as at $z=2.2$, we conservatively increase the error bars by a factor $\sqrt 2$ to take systematic uncertainties into account. This results in $\langle$f$_{\rm esc} \rangle = 2.7^{+7.2}_{-2.3}$ \% and $\langle$f$_{\rm esc} \rangle = 6.0^{+13.9}_{-5.2}$ \% at $z\approx 4$ and $z\approx 5$, respectively, see Table $\ref{tab:global_fesc}$. When including a (maximum) quasar contribution from \cite{MadauHaardt2015} as described above, we find $\langle$f$_{\rm esc} \rangle = 0.0^{+3.0}_{-0.0}$ \% at $z\approx4$ and $\langle$f$_{\rm esc} \rangle = 2.1^{+6.2}_{-2.1}$ \%. 

As illustrated in Fig. $\ref{fig:fesc_evolution}$, the global escape fraction is low at $z\approx2-5$. While dust has been corrected for with different methods at $z=2.2$ and $z\approx4-5$, we note that the differences between different dust correction methods are not expected to be very large at $z\approx4-5$. This is because higher redshift galaxies typically have lower mass, which results in a higher agreement between dust correction methods based on either M$_{\rm star}$ or $\beta$.  One potentially important caveat is that our computation assumes that the H$\alpha$ and UV luminosity functions include all sources of ionizing photons in addition to quasars. An additional contribution of ionizing photons from galaxies which have potentially been missed by a UV selection (for example sub-mm galaxies) would decrease the global f$_{\rm esc}$. Such a bias is likely more important at $z\approx3-5$ than $z\approx2$ because the $z\approx2$ sample is selected with H$\alpha$ which is able to recover sub-mm galaxies. Even under current uncertainties, we rule out a globally averaged $\langle$f$_{\rm esc} \rangle > 20$ \% at redshifts lower than $z\approx5$. 

These indirectly derived escape fractions of $\sim 4$ \% at $z\approx2-5$ are consistent with recently published upper limits from \cite{Sandberg2015b} at $z=2.2$ and similar to strict upper limits on f$_{\rm esc}$ at $z\sim1$ measured by \cite{Rutkowski2015}, see also \cite{Cowie2009,Bridge2010}. Recently, \cite{Cristiani2016} estimated that galaxies have on average $\langle$f$_{\rm esc} \rangle = 5.3^{+2.7}_{-1.2}$ \% at $z\approx4$ by combining IGM constraints with the UV luminosity function from \cite{Bouwens2011} and by including the contribution from quasars to the total emissivity. This result is still consistent within the error-bars with our estimate using the \cite{MadauHaardt2015} quasar contribution and \cite{Smit2016} SFR function. Part of this is because we use a different conversion from UV luminosity to the number of produced ionizing photons based on H$\alpha$ estimates with {\it Spitzer}/IRAC, and because our computation assumes f$_{\rm esc, quasars} = 100$\%, while \cite{Cristiani2016} uses f$_{\rm esc, quasars} \approx 70$\%.

Furthermore, our results are also consistent with observations from \cite{Chen2007} who find a mean escape fraction of $2\pm2$ \% averaged over galaxy viewing angles using spectroscopy of the afterglow of a sample of $\gamma$-Ray bursts at $z>2$. \cite{Grazian2016} measures a strict median upper limit of f$_{\rm esc}^{rel} < 2$ \% at $z=3.3$, although this limit is for relatively luminous Lyman-break galaxies and not for the entire population of SFGs. This would potentially indicate that the majority of LyC photons escape from galaxies with lower luminosity, or galaxies missed by a Lyman-break selection, i.e. \cite{Cooke2014} or that they come from just a sub-set of the population, and thus the median f$_{\rm esc}$ can even be close to zero. \cite{Khaire2016} finds that f$_{\rm esc}$ must evolve from $\approx5-20$ \% between $z=3-5$, which is allowed within the errors. However, we note that they assume that the number of produced ionizing photons per unit UV luminosity does not evolve with redshift. In \S 6.5 we find that there is evolution of this number by roughly a factor 1.5, such that the required evolution of f$_{\rm esc}$ would only be a factor $\approx 3$. While our results indicate little to no evolution in the average escape fraction up to $z\approx5$, this does not rule out an increasing f$_{\rm esc}$ at $z>5$, where theoretical models expect an evolving f$_{\rm esc}$ \citep[e.g.][]{Kuhlen2012,FerraraLoeb2013,Mitra2013,Khaire2016,Sharma2016,Price2016}, see also a recent observational claim of evolving f$_{\rm esc}$ with redshift \citep{Smith2016}. 

Finally, we stress that a low $\langle$f$_{\rm esc} \rangle$ is not inconsistent with the recent detection of the high f$_{\rm esc}$ of above 50 \% from a galaxy at $z\approx3$ \citep{deBarros2016,Vanzella2016}, which may simply reflect that there is a broad distribution of escape fractions. For example, if only a small fraction ($<5$ \%) of galaxies are LyC leakers with f$_{\rm esc} \approx 75$ \%, the average f$_{\rm esc}$ over the galaxy population is $\approx 4$ \%, consistent with the indirect measurement, even if f$_{\rm esc} = 0$ for all other galaxies. Such a scenario would be the case if the escape of LyC photons is a very stochastic process, for example if it is highly direction or time dependent. This can be tested with deeper LyC limits on individual galaxies over a complete selection of star-forming galaxies.

\begin{figure}
\includegraphics[width=8.5cm]{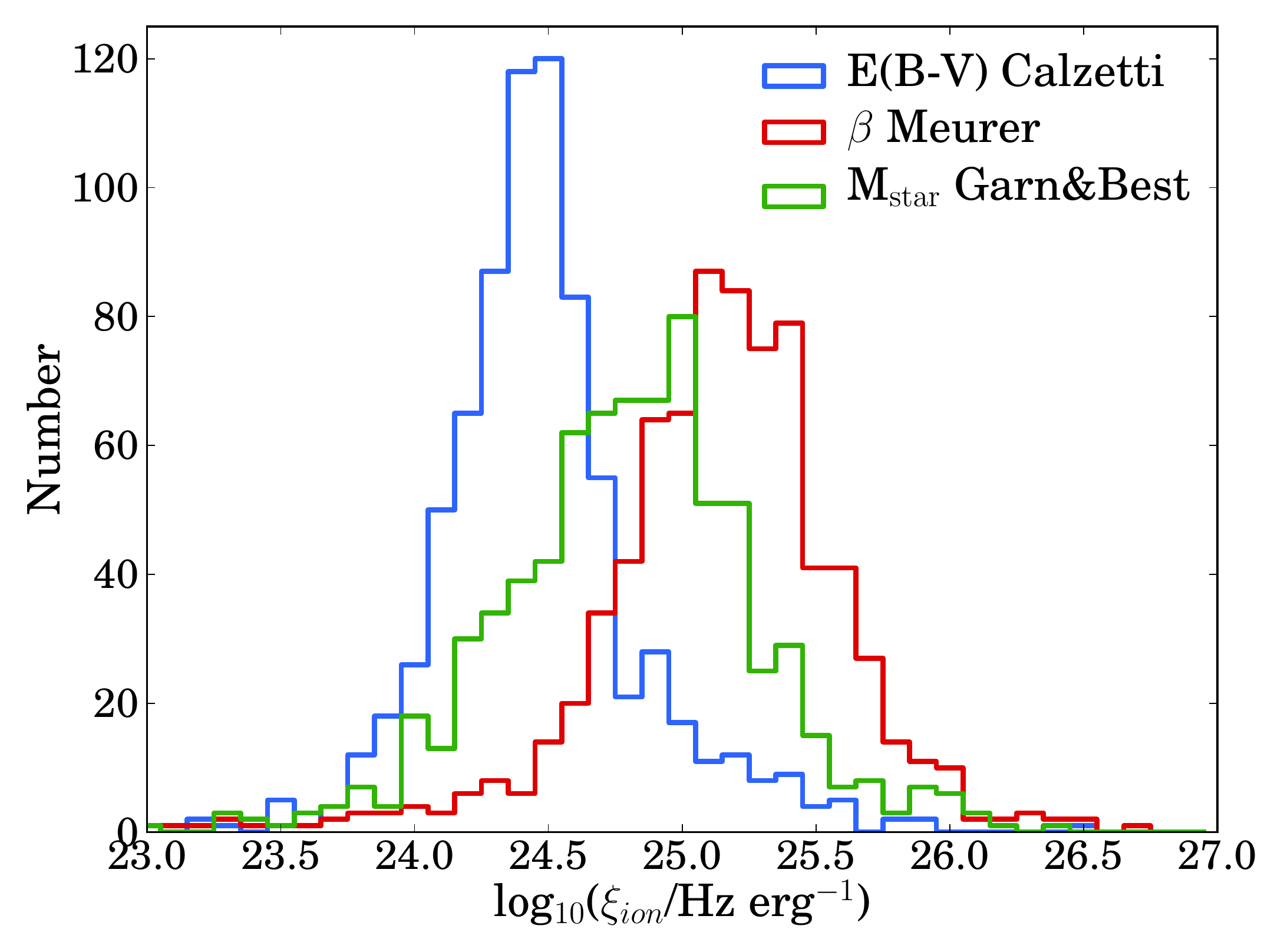}
\caption{\small{Histogram of the values of $\xi_{ion}$ for HAEs with three different methods to correct for dust attenuation. The blue histogram shows values of $\xi_{ion}$ when dust is corrected with the E$(B-V)$ value from the SED in combination with a Calzetti law (see \S 2.1). The red histogram is corrected for dust with the \citealt{Meurer1999} prescription based on the UV slope and the green histogram is corrected for dust with the prescription from \citealt{GarnBest2010} based on a relation between dust attenuation and stellar mass. As can be seen, the measured values of $\xi_{ion}$ differ significantly, with the highest values found when correcting for dust with the UV slope. When the nebular attenuation is higher than the stellar attenuation, $\xi_{ion}$ would shift to higher values.}} 
\label{fig:hist_xion}
\end{figure}

\begin{figure*}
\includegraphics[width=17.7cm]{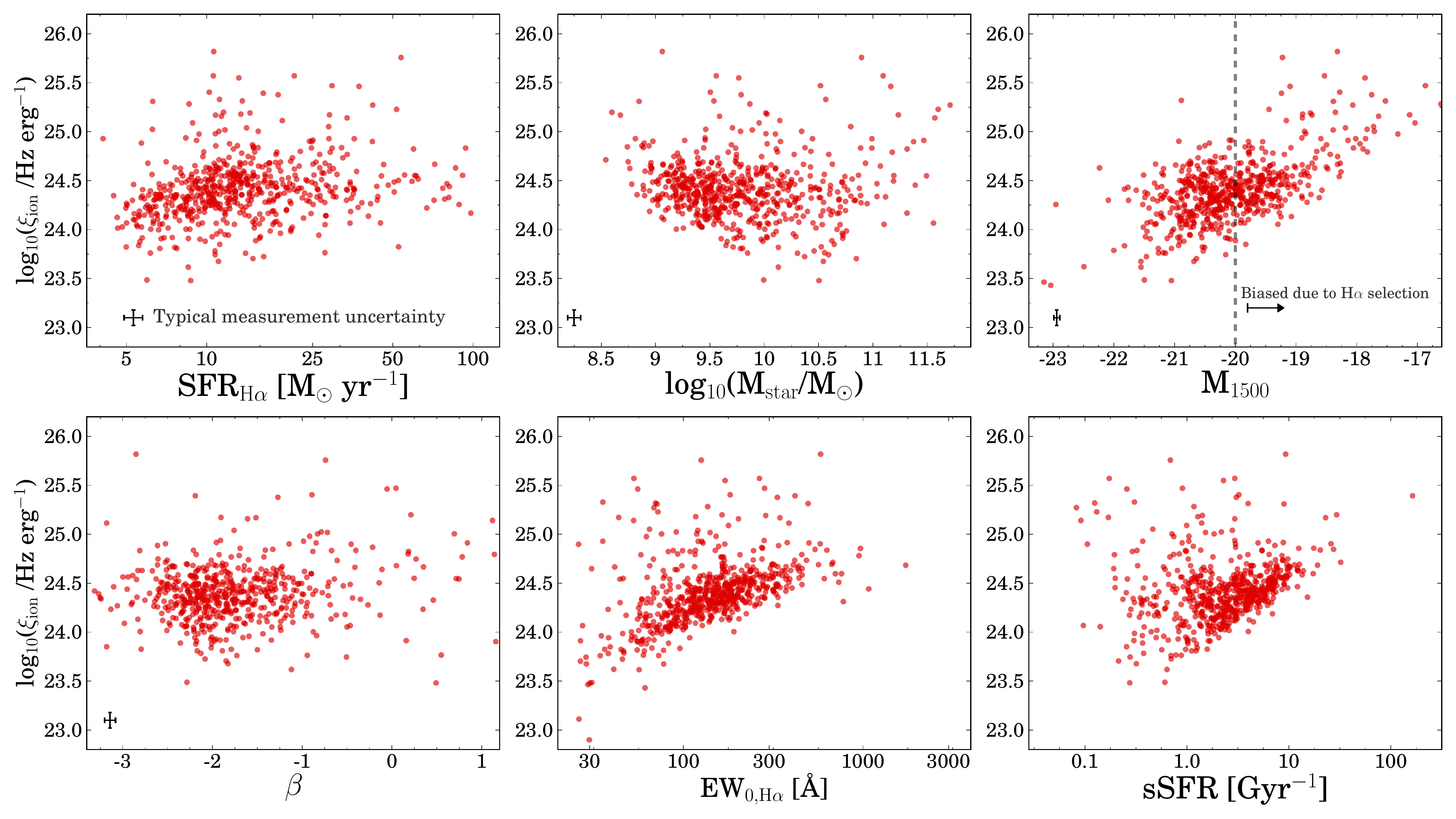}
\caption{\small{Correlations between $\xi_{ion}$ and galaxy properties for HAEs, when dust is corrected using the SED fitted E$(B-V)$ values. $\xi_{ion}$ does not clearly correlate with SFR(H$\alpha$), M$_{\rm star}$ or $\beta$. A correlation between $\xi_{ion}$ and M$_{1500}$ is expected of similar strength as seen, based on the definition of $\xi_{ion}$. $\xi_{ion}$ increases strongly with H$\alpha$ EW and sSFR. High values of $\xi_{ion}$ at low sSFRs are mostly due to the dust correction. }}
\label{fig:xion_properties}
\end{figure*} 


\begin{table}
\begin{center}
\begin{small}
\caption{Ionizing properties of HAEs and LAEs for various methods to correct for dust attenuations and different subsets. We show the median stellar mass of each subsample. Errors on $\xi_{ion}$ are computed as $\sigma_{\xi_{ion}}/\sqrt{N}$, where $\sigma_{\xi_{ion}}$ is the median measurement error of $\xi_{ion}$ and $N$ the number of sources. For the \citet{Bouwens2015xi} measurements, we show only dust corrections with a \citet{Calzetti2000} curve. The subsample of `low mass' HAEs has M$_{\rm star} = 10^{9.0-9.4}$ M$_{\odot}$. `UV faint' HAEs have $M_{1500}>-19$.} 
\begin{tabular}{ lrrr }
\bf Sample & \bf <M$_{\rm star}$>& \bf log$_{10}$ $\xi_{ion}$ &\bf Dust\\ 
& log$_{10}$ M$_{\odot}$ & Hz erg$^{-1}$ & \\ \hline
This paper & & & \\
HAEs $z=2.2$ & 9.8 & $24.39\pm0.04$ & E$(B-V)$ \\
 &  & $25.11\pm0.04$ & $\beta$  \\
 &  & $24.77\pm0.04$ & M$_{\rm star}$  \\ 
   &  & $25.41\pm0.05$ & No dust  \\
  &  & $24.57\pm0.04$ & A$_{\rm H\alpha} = 1$ \\

Low mass & 9.2  & $24.49\pm0.06$ & E$(B-V)$ \\
 &  & $25.22\pm0.06$ & $\beta$  \\
 &   & $24.99\pm0.06$ & M$_{\rm star}$  \\
UV faint & 10.2 & $24.93\pm0.07$ & E$(B-V)$ \\
&  & $25.39\pm0.07$ & $\beta$ \\
& & $25.24\pm0.07$ & M$_{\rm star}$ \\

LAEs $z=2.2$ & 8.5 & $24.84\pm0.09$ & E$(B-V)$ \\
 &  & $25.37\pm0.09$ & $\beta$\\
 &  & $25.14\pm0.09$ & M$_{\rm star}$  \\
  &  & $25.39\pm0.09$ & No dust  \\ \hline

\citet{Bouwens2015xi} & & & \\
 LBGs $z=3.8-5.0$ & 9.2 & $25.27\pm0.03$ & $\beta$  \\
 LBGs $z=5.1-5.4$ & 9.2 & $25.44\pm0.12$ & $\beta$  \\ \hline
\end{tabular}
\label{tab:Xion_subsets}
\end{small}
\end{center}
\end{table} 

\section{The ionizing properties of star-forming galaxies at $z=2.2$}
\label{sec:6}
\subsection{How to measure $\xi_{ion}$?}
The number of ionizing photons produced per unit UV luminosity, $\xi_{ion}$, is used to convert the observed UV luminosity of high-redshift galaxies to the number of produced ionizing photons. It can thus be interpreted as the production efficiency of ionizing photons. $\xi_{ion}$ is defined as:
\begin{equation}
 \xi_{ion} = Q_{\rm ion} / L_{UV, \rm int}
 \end{equation}
As described in the previous section, $Q_{\rm ion}$ (in s$^{-1}$) can be measured directly from the dust-corrected H$\alpha$ luminosity by rewriting Eq. 2 and assuming f$_{\rm esc} = 0$. $L_{UV, \rm int}$ (in erg s$^{-1}$ Hz$^{-1}$) is obtained by correcting the observed UV magnitudes for dust attenuation. With a \cite{Calzetti2000} attenuation curve A$_{UV} = 3.1$A$_{\rm H\alpha}$.  

In our definition of $\xi_{ion}$, we assume that the escape fraction of ionizing photons is $\approx0$. Our direct constraint of f$_{\rm esc} \lesssim 6$\% and our indirect global measurement of f$_{\rm esc} \approx 5$ \% validate this assumption. If the average escape fraction is f$_{\rm esc} = 10$\%, $\xi_{ion}$ is higher by a factor 1.11 (so only 0.04 dex), such that $\xi_{ion}$ is relatively insensitive to the escape fraction as long as the escape fraction is low. We also note that the $\xi_{ion}$ measurements at $z\approx4-5$ from \cite{Bouwens2015xi} are validated by our finding that the global escape fraction at $z<5$ is consistent with being very low, $< 5$ \%.

\begin{figure*}
\includegraphics[width=17.7cm]{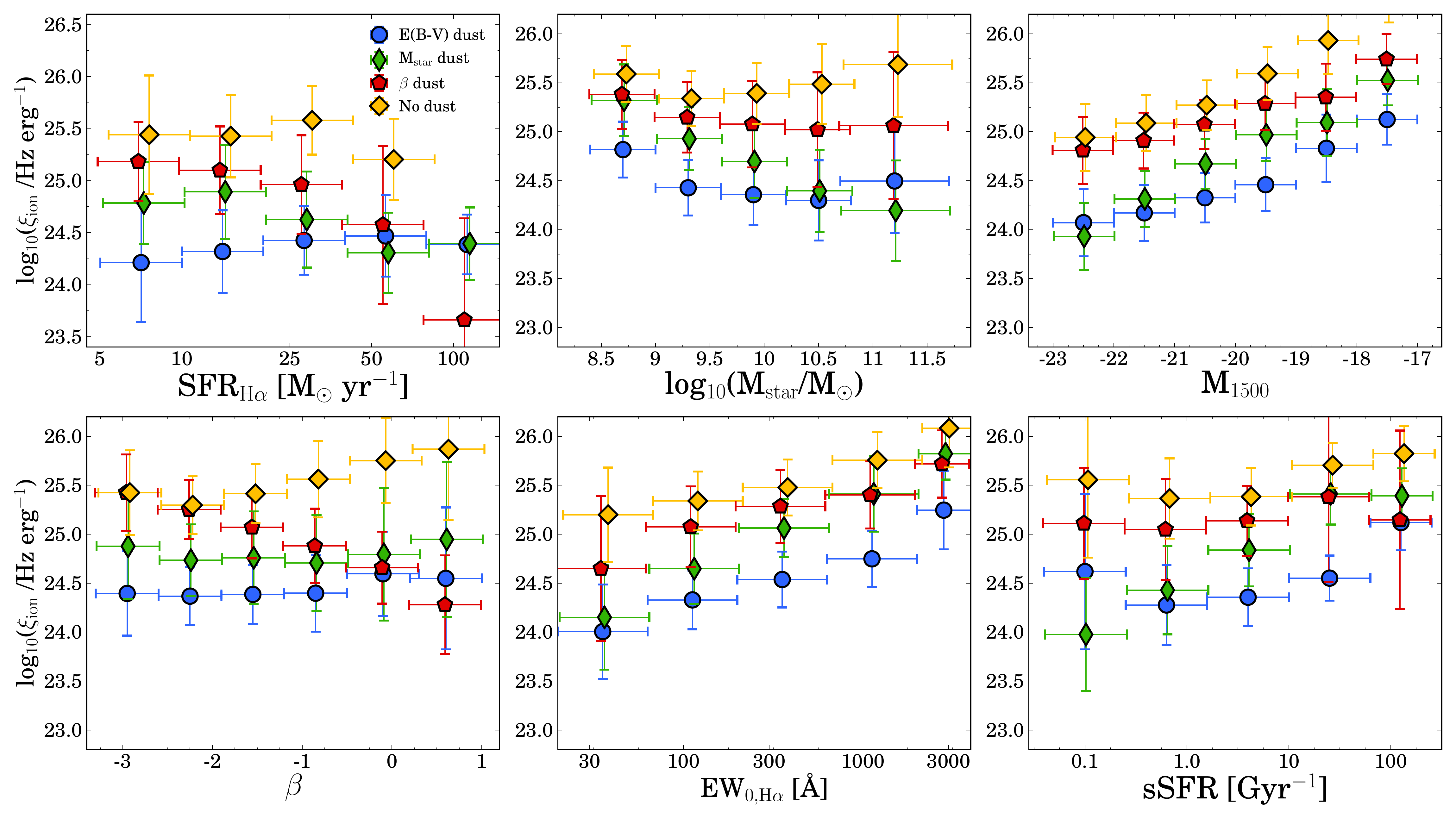}
\caption{\small{Correlations between $\xi_{ion}$ and galaxy properties for different methods to correct for dust attenuation. To facilitate the comparison, HAEs were binned on the x-axis. The value of $\xi_{ion}$ is the median value in each bin, while the vertical error is the standard deviation. Blue bins show the values where dust is corrected with the E$(B-V)$ value from the SED. The red bins are corrected for dust with the \citet{Meurer1999} prescription based on $\beta$ and the green bins are corrected for dust with the prescription from \citet{GarnBest2010} based on stellar mass. Yellow bins show the results where we assume that there is no dust.}}
\label{fig:xion_properties2}
\end{figure*} 

\subsection{$\xi_{ion}$ at $z=2.2$}
We show our measured values of $\xi_{ion}$ for HAEs in Fig. $\ref{fig:hist_xion}$ and Table $\ref{tab:Xion_subsets}$, where dust attenuation is corrected with three different methods based either on the E$(B-V)$ value of the SED fit, the UV slope $\beta$ or the stellar mass. It can be seen that the average value of $\xi_{ion}$ is very sensitive to the dust correction method, ranging from $\xi_{ion} = 10^{24.39\pm0.04}$ Hz erg$^{-1}$ for the SED method to $\xi_{ion} = 10^{25.11\pm0.04}$ Hz erg$^{-1}$ for the $\beta$ method. For the dust correction based on stellar mass the value lies in between, with $\xi_{ion} = 10^{24.85\pm0.04}$ Hz erg$^{-1}$. In the case of a higher nebular attenuation than the stellar attenuation, as for example by a factor $\approx 2$ as in the original \cite{Calzetti2000} prescription, $\xi_{ion}$ increases by 0.4 dex to $\xi_{ion} = 10^{24.79\pm0.04}$ Hz erg$^{-1}$ when correcting for dust with the SED fit.

We note that independent (stacking) measurements of the dust attenuation from {\it Herschel} and Balmer decrements at $z\sim1-2$ indicate that dust attenuations agree very well with the \cite{GarnBest2010} prescription \citep[e.g.][]{Sobral2012,Ibar2013,Buat2015,Pannella2015}, thus favouring the intermediate value of $\xi_{ion}$. Without correcting $\xi_{ion}$ for dust, we find $\xi_{ion} = 10^{25.41\pm0.05}$ Hz erg$^{-1}$. With 1 magnitude of extinction for H$\alpha$, as for example used in the conversion of the H$\alpha$ luminosity density to a SFR density in \cite{Sobral2013}, $\xi_{ion} = 10^{24.57\pm0.04}$ Hz erg$^{-1}$. 

Since individual H$\alpha$ measurements for LAEs are uncertain due to the difference in filter transmissions depending on the exact redshift (see \citealt{Matthee2016}), we only investigate $\xi_{ion}$ for our sample of LAEs in the stacks described in \cite{Sobral2015survey}. With stacking, we measure the median H$\alpha$ flux of LAEs convolved through the filter profile and the median UV luminosity by stacking $V$ band imaging. As seen in Table $\ref{tab:Xion_subsets}$, the median $\xi_{ion}$ is higher than the median $\xi_{ion}$ for HAEs for each dust correction. However, this difference disappears without correcting for dust. Therefore, the higher values of $\xi_{ion}$ for LAEs simply indicate that the median LAE has a bluer UV slope, lower stellar mass and lower E$(B-V)$ than the median HAE. More accurate dust measurements are required to investigate whether $\xi_{ion}$ is really higher for LAEs. We note that $\approx 40$ \% of the LAEs are undetected in the broad-bands and thus assigned a stellar mass of $10^8$ M$_{\odot}$ and E$(B-V) = 0.1$ when computing the median dust attenuation. Therefore, the $\xi_{ion}$ values for LAEs could be under-estimated if the real dust attenuation is even lower.

\subsection{Dependence on galaxy properties}
In this section we investigate how $\xi_{ion}$ depends on the galaxy properties that are defined in \S 2.1 and also check whether subsets of galaxies lie in a specific parameter space. As illustrated in Fig. $\ref{fig:xion_properties}$ (where we correct for dust with E$(B-V)$), we find that $\xi_{ion}$ does not depend strongly on SFR(H$\alpha$) with a Spearman correlation rank (R$_s$) of R$_s = 0.11$. Such a correlation would naively be expected if the H$\alpha$ SFRs are not related closely to UV SFRs, since $\xi_{ion} \propto L_{\rm H\alpha}/L_{1500} \propto$ SFR(H$\alpha$)/SFR(UV). However, for our sample of galaxies these SFRs are strongly correlated with only 0.3 dex of scatter, see also \cite{Oteo2015}, leading to a relatively constant $\xi_{ion}$ with SFR. 

For the same reason, we measure a relatively weak slope of $\approx 0.25$ when we fit a simple linear relation between log$_{10}$($\xi_{ion}$) and M$_{1500}$, instead of the naively expected value of $\xi_{ion} \propto 0.4 M_{1500}$. At M$_{1500} > - 20$, our H$\alpha$ selection is biased towards high values of H$\alpha$ relative to the UV, leading to a bias in high values of $\xi_{ion}$. For sources with M$_{1500} < -20$, we measure a slope of $\approx 0.2$. This means that $\xi_{ion}$ does not increase rapidly with decreasing UV luminosity. This is because H$\alpha$ luminosity and dust attenuation themselves are also related to M$_{1500}$. Indeed, we find that the H$\alpha$ luminosity anti-correlates with the UV magnitude and E$(B-V)$ increases for fainter UV magnitudes. 

The stellar mass and $\beta$ are not by definition directly related to $\xi_{ion}$. Therefore, a possible upturn of $\xi_{ion}$ at low masses (see the middle-top panel in Fig. $\ref{fig:xion_properties}$) may be a real physical effect, although we note that we are not mass-complete below M$_{\rm star} < 10^{10}$ M$_{\odot}$ and an H$\alpha$ selected sample of galaxies likely misses low-mass galaxies with lower values of $\xi_{ion}$. 

We find that the number of ionizing photons per unit UV luminosity is strongly related to the H$\alpha$ EW (with a slope of $\sim 0.6$ in log-log space), see Fig. $\ref{fig:xion_properties}$. Such a correlation is expected because of our definition of $\xi_{ion}$: i) the H$\alpha$ EW increases mildly with increasing H$\alpha$ (line-)luminosity and ii) the H$\alpha$ EW is weakly anti-related with the UV (continuum) luminosity, such that $\xi_{ion}$ increases relatively strongly with EW. Since there is a relation between H$\alpha$ EW and specific SFR (sSFR = SFR/M$_{\rm star}$, e.g. \citealt{Fumagalli2012}), we also find that $\xi_{ion}$ increases strongly with increasing sSFR, see Fig. $\ref{fig:xion_properties}$. 

In Fig. $\ref{fig:xion_properties2}$ we show the same correlations as discussed above, but now compare the results for different methods to correct for dust. For comparison, we only show the median $\xi_{ion}$ in bins of the property on the x-axis. The vertical error on the bins is the standard deviation of the values of $\xi_{ion}$ in the bin. As $\xi_{ion}$ depends on the dust correction, we find that $\xi_{ion}$ correlates with the galaxy property that was used to correct for dust in the case of $\beta$ (red symbols) and M$_{\rm star}$ (green symbols). Specific SFR depends on stellar mass, so we also find the strongest correlation between sSFR and $\xi_{ion}$ when $\xi_{ion}$ is corrected for dust with the \cite{GarnBest2010} prescription. We only find a relation between $\xi_{ion}$ and $\beta$ when dust is corrected with the \cite{Meurer1999} prescription. For UV magnitude only the normalisation of $\xi_{ion}$ changes with the dust correction method. 

It is more interesting to look at correlations between $\xi_{ion}$ and galaxy properties which are not directly related to the computation of $\xi_{ion}$ or the dust correction. Hence, we note that irrespective of the dust correction method, $\xi_{ion}$ appears to be somewhat higher for lower mass galaxies (although this is likely a selection effect as discussed above). Irrespective of the dust correction method, $\xi_{ion}$ increases with increasing H$\alpha$ EW and fainter M$_{1500}$, where the particular dust correction method used only sets the normalisation. We return to this relation between $\xi_{ion}$ and H$\alpha$ EW in \S 6.5.

\begin{figure*}
\includegraphics[width=16cm]{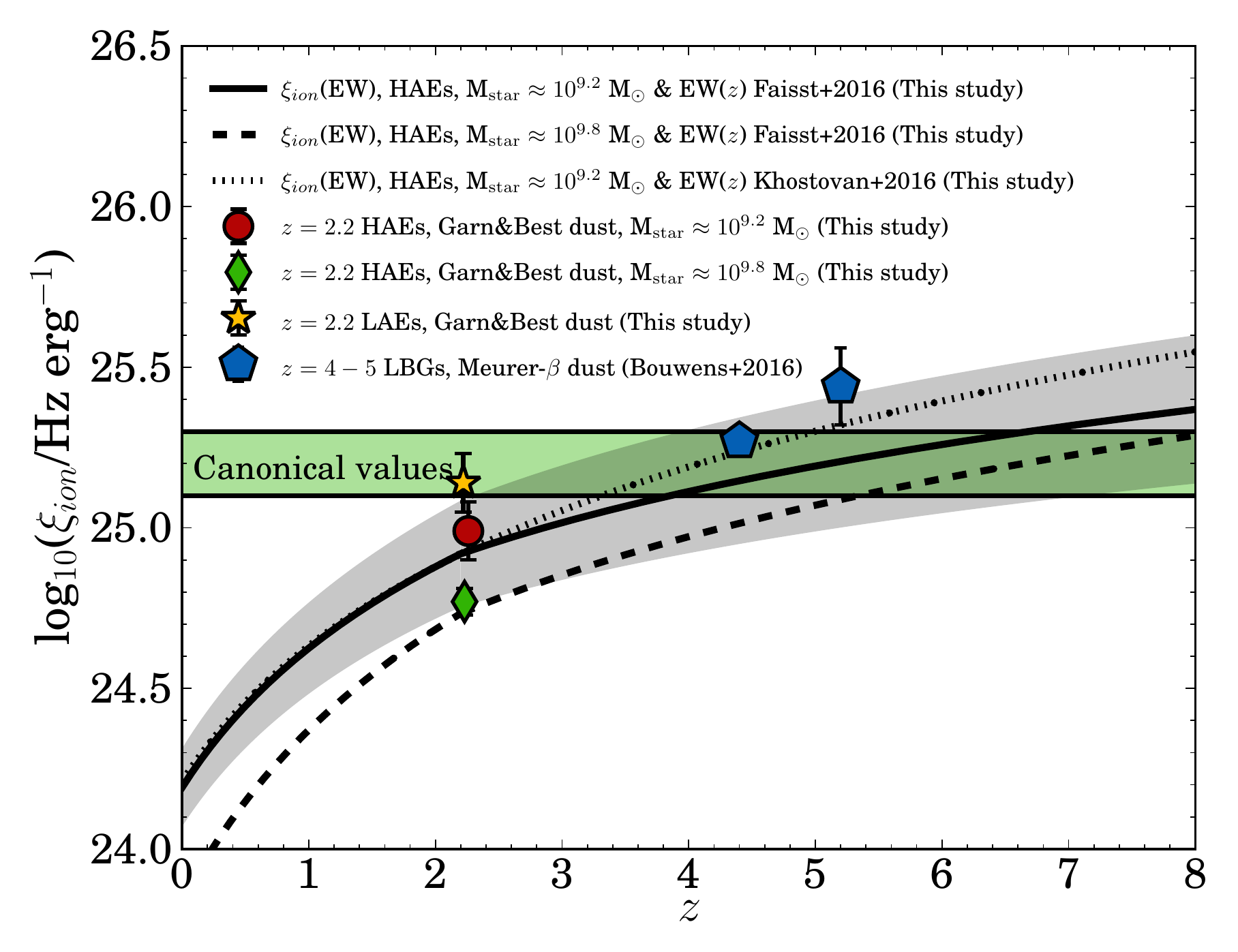}
\caption{\small{Inferred evolution of $\xi_{ion}$ (corrected for dust with M$_{\rm star}$) with redshift based on our observed trend between $\xi_{ion}$ and H$\alpha$ EW, for different stellar masses (compare the solid with the dashed line) and EW($z$) evolutions (compare the solid with the dotted line). The grey shaded region indicates the errors on the redshift evolution of $\xi_{ion}$. The normalisation of $\xi_{ion}$ is higher for lower mass galaxies or LAEs. The green region shows the typically assumed values. The estimated evolution of $\xi_{ion}$ with redshift is consistent with the typically assumed values of $\xi_{ion}$ in the reionization era and with recent measurements at $z=4-5$. }} 
\label{fig:xion_evolution}
\end{figure*} 

\subsection{Redshift evolution of $\xi_{ion}$}
Because of its dependency on galaxy properties, it is possible that $\xi_{ion}$ evolves with redshift. In fact, such an evolution is expected as more evolved galaxies (particularly with declining star formation histories) have a relatively stronger UV luminosity than H$\alpha$ and a higher dust content, likely leading to a lower $\xi_{ion}$ at $z=2.2$ than at $z>6$. 

By comparing our measurement of $\xi_{ion}$ with those from \cite{Bouwens2015xi} at $z=4-5$, we already find such an evolution (see Table $\ref{tab:Xion_subsets}$), although we note that the samples of galaxies are selected differently and that there are many other differences, such as the dust attenuation, typical stellar mass and the H$\alpha$ measurement. 
If we mimic a Lyman-break selected sample by only selecting HAEs with E$(B-V) < 0.3$ (typical for UV selected galaxies, e.g. \citealt{Steidel2011}), we find that $\xi_{ion}$ increases by (maximally) 0.1 dex, such that this does likely not explain the difference in $\xi_{ion}$ at $z=2.2$ and $z\approx4-5$ of $\approx 0.5$ dex. Furthermore, our H$\alpha$ selection is likely biased towards high values of $\xi_{ion}$ for $M_{1500} > -20$, which mitigates the difference on the median $\xi_{ion}$. If we select only low mass galaxies such that the median stellar mass resembles that of \cite{Bouwens2015xi}, the difference is only $\approx 0.2\pm0.1$ dex, which still would suggest evolution. 

We estimate the redshift evolution of $\xi_{ion}$ by combining the relation between $\xi_{ion}$ and H$\alpha$ EW with the redshift evolution of the H$\alpha$ EW. Several studies have recently noted that the H$\alpha$ EW (and related sSFR) increases with increasing redshift \citep[e.g.][]{Fumagalli2012,Sobral2014,Smit2014,Marmol2015,Faisst2016,Khostovan2016}. Furthermore, the EW is mildly dependent on stellar mass as EW $\sim \rm M_{\rm star}^{-0.25}$ \citep{Sobral2014,Marmol2015}. In order to estimate the $\xi_{ion}$ using the H$\alpha$ EW evolution, we: 

{\bf i)} Select a subset of our HAEs with stellar mass between $10^{9-9.4}$ M$_{\odot}$, with a median of M$_{\rm star} \approx 10^{9.2}$ M$_{\odot}$, which is similar to the mass of the sample from \cite{Bouwens2015xi}, see \cite{Smit2016}, 

{\bf ii)} Fit a linear trend between log$_{10}$(EW) and log$_{10}(\xi_{ion})$ (with the \cite{GarnBest2010} prescription to correct for dust attenuation). We note that the trend between EW and $\xi_{ion}$ will be steepened if dust is corrected with a  prescription based on stellar mass (since H$\alpha$ EW anti-correlates with stellar mass, see also Table $\ref{tab:xionfit}$). However, this is validated by several independent observations from either Herschel or Balmer decrements which confirm that dust attenuation increases with stellar mass at a wide range of redshifts \citep{Dominguez2013,Buat2015,Koyama2015,Pannella2015,SobralDUSQ}.

\begin{table}
\begin{center}
\begin{small}
\caption{Fit parameters for log$_{10}\ \xi_{ion} = a + b $ log$_{10}\ $EW(H$\alpha$) for different selections and dust corrections} 
\begin{tabular}{ lrrrr }
\bf Sample & \bf <M$_{\rm star}$>& \bf $a$ &\bf $b$& \bf Dust \\ 
& log$_{10}$ M$_{\odot}$ && & \\ \hline
All HAEs & 9.8 & 23.12 & 0.59 & E$(B-V)$ \\
 &  & 23.66 & 0.64 & $\beta$  \\
 &  & 22.60 & 0.97 & M$_{\rm star}$  \\
  &  & 23.59 & 0.45 & A$_{\rm H\alpha} = 1$ \\

Low mass & 9.2  & 22.64 & 0.78 & E$(B-V)$ \\
 &  & 23.68 & 0.64 & $\beta$  \\
 &   & 23.19 & 0.77 & M$_{\rm star}$  \\
  &  & 22.77 & 0.75 & A$_{\rm H\alpha} = 1$ \\ \hline

\end{tabular}
\label{tab:xionfit}
\end{small}
\end{center}
\end{table} 

Using a simple least squares algorithm, we find: 
\begin{equation}
{\rm log}_{10}(\xi_{ion})=23.19^{+0.09}_{-0.09} + 0.77^{+0.04}_{-0.04}\times {\rm log}_{10}(\rm EW)
\end{equation} 

{\bf iii)} Combine the trend between H$\alpha$ EW and redshift with the trend between $\xi_{ion}$ and H$\alpha$ EW. We use the redshift evolution of the H$\alpha$ EW from \cite{Faisst2016}, which has been inferred from fitting SEDs, and measured up to $z\approx6$. In this parametrisation, the slope changes from EW$\approx (1+z)^{1.87}$ at $z<2.2$ to EW$\approx (1+z)^{1.3}$ at $z>2.2$. Below $z<2.2$, this trend is fully consistent with the EW evolution from HiZELS \citep{Sobral2014}, which is measured with narrow-band imaging. Although HiZELS does not have H$\alpha$ emitters at $z>2.2$, the EW evolution of [O{\sc iii}]+H$\beta$ is found to flatten at $z>2.2$ as well \citep{Khostovan2016}. We note that we assume that the slope of the H$\alpha$ EW evolution with redshift does not vary strongly for stellar masses between $10^{9.2}$ M$_{\odot}$ and $10^{9.8}$ M$_{\odot}$, since the following equations are measured at stellar mass $\approx 10^{9.6}$ M$_{\odot}$ \citep{Faisst2016}, hence: 
\begin{equation}
{\rm EW}(z) = 
\begin{cases}
20\times(1+z)^{1.87},&  z<2.2\\
37.4\times(1+z)^{1.3},    &z\geq2.2
\end{cases}
\end{equation}

\noindent This results in:
\begin{equation}
 {\rm log}_{10}(\xi_{ion}(z)) = 
 \begin{cases}
 24.19+1.44\times {\rm log}_{10}(1+z) ,&  z<2.2\\
 24.40+1.00\times {\rm log}_{10}(1+z),    &z\geq2.2 
 \end{cases}
 \end{equation}
where $\xi_{ion}$ is in Hz erg$^{-1}$. The error on the normalisation is 0.09 Hz erg$^{-1}$ and the error on the slope is 0.18.
For our typical mass of M$_{\rm star} = 10^{9.8}$ M$_{\odot}$, the normalisation is roughly 0.2 dex lower and the slope a factor $\approx 1.1$ higher compared to the fit at lower stellar masses. This is due to a slightly different relation between $\xi_{ion}$ and EW (see Table $\ref{tab:xionfit}$). The evolving $\xi_{ion}$ is consistent with the typically assumed value of $\xi_{ion} = 10^{25.2\pm0.1}$ Hz erg$^{-1}$ \citep[e.g.][]{Robertson2013} at $z\approx2.5-12$ within the 1$\sigma$ error bars.  
 
We show the inferred evolution of $\xi_{ion}$ with redshift in Fig. $\ref{fig:xion_evolution}$. The solid and dashed line use the EW($z$) evolution from \cite{Faisst2016}, while the dotted line uses the \cite{Khostovan2016} parametrisation. The grey shaded region indicates the errors on the redshift evolution of $\xi_{ion}$. Due to the anti-correlation between EW and stellar mass, galaxies with a lower stellar mass have a higher $\xi_{ion}$ (which is then even strengthened by a higher dust attenuation at high masses). 

Relatively independent of the dust correction (as discussed in Fig. $\ref{fig:xion_evolution_variation}$), the median $\xi_{ion}$ increases $\approx 0.2$ dex at fixed stellar mass between $z=2.2$ and $z=4.5$. This can easily explain the 0.2 dex difference between our measurement at $z=2.2$ and the \cite{Bouwens2015xi} measurements at $z=4-5$ (see Fig. $\ref{fig:xion_evolution}$), such that it is plausible that $\xi_{ion}$ evolves to higher values in the reionization epoch, of roughly $\xi_{ion} \approx 10^{25.4}$ Hz erg$^{-1}$ at $z\approx8$. Interestingly, LAEs at $z=2.2$ already have $\xi_{ion}$ similar to the canonical value in the reionization era.

\section{Implications for reionization}
\label{sec:7}
The product of $f_{\rm esc} \xi_{ion}$ is an important parameter in assessing whether galaxies have provided the photons to reionize the Universe, because these convert the (non-ionizing) UV luminosity density (obtained from integrating the dust-corrected UV luminosity function) to the ionizing emissivity. The typical adopted values are $\xi_{ion} \approx10^{25.2-25.3}$ Hz erg$^{-1}$ and f$_{\rm esc} \approx 0.1-0.2$ \citep[e.g.][]{Robertson2015}, such that the product is $f_{\rm esc} \xi_{ion} \approx 10^{24.2-24.6}$ Hz erg$^{-1}$. This is significantly higher than our upper limit of $f_{\rm esc} \xi_{ion} \lesssim 10^{23.5}$ Hz erg$^{-1}$ (using $\langle$f$_{\rm esc}\rangle$ and $\xi_{ion}$ where dust is corrected with M$_{\rm star}$, see \S 5 and \S 6). However, as shown in \S 6.5, we expect $\xi_{ion} \approx 10^{25.4}$ Hz erg$^{-1}$ in the reionization era due to the dependency of $\xi_{ion}$ on EW(H$\alpha$), such that escape fractions of $f_{\rm esc}\approx 7$ \% would suffice for $f_{\rm esc} \xi_{ion} = 10^{24.2}$ Hz erg$^{-1}$. \cite{BeckerBolton2013} find an evolution in the product of $f_{\rm esc} \xi_{ion}$ of a factor 4 between $z=3-5$ (similar to \citealt{HaardtMadau2012}), which is consistent with our measurements. 

Recently, \cite{Faisst2016b} inferred that f$_{\rm esc}$ may evolve with redshift by combining a relation between f$_{\rm esc}$ and the {\sc [Oiii]/[Oii]} ratio with the inferred redshift evolution of the {\sc [Oiii]/[Oii]} ratio. This redshift evolution is estimated from local analogs to high redshift galaxies selected on H$\alpha$ EW, such that the redshift evolution of f$_{\rm esc}$ is implicitly coupled to the evolution of H$\alpha$ EW as in our model of $\xi_{ion}(z)$. \cite{Faisst2016b} estimate that f$_{\rm esc}$ evolves from $\approx 2$ \% at $z=2$ to $\approx 5$ \% at $z=5$, which is consistent with our measurements of $\langle$f$_{\rm esc}\rangle$ (see Fig. $\ref{fig:fesc_evolution}$). With this evolving escape fraction, galaxies can provide sufficient amounts of photons to reionize the Universe, consistent with the most recent CMB constraints \cite{Planck2016}. This calculation assumes $\xi_{ion} = 10^{25.4}$ Hz erg$^{-1}$, which is the same value our model predicts for $\xi_{ion}$ in the reionization era.

In addition to understanding whether galaxies have reionized the Universe, it is perhaps more interesting to understand which galaxies have been the most important to do so. For example, \cite{Sharma2016} argue that the distribution of escape fractions in galaxies is likely very bimodal and dependent on the SFR surface density, which could mean that LyC photons preferentially escape from bright galaxies. Such a scenario may agree better with a late and rapid reionization process such as favoured by the new low optical depth measurement from \cite{Planck2016}. We note that the apparent discrepancy between the f$_{\rm esc}$ upper limit from median stacking (f$_{\rm esc}< 2.8$ \%) and the average f$_{\rm esc}$ from the integrated luminosity density combined with IGM constraints ($\langle$f$_{\rm esc} \rangle = 5.9$ \%) can be understood in a scenario where the average f$_{\rm esc}$ is driven by a few galaxies with high f$_{\rm esc}$, or by a scenario where f$_{\rm esc}$ is higher for galaxies below the H$\alpha$ detection threshold (which corresponds to SFR$>4$ M$_{\odot}$ yr$^{-1}$), contrarily to a scenario where each typical HAE has an escape fraction of $\approx5-6$ \%. 

\cite{DijkstraGronke2016} argue a connection between the escape of Ly$\alpha$ photons and LyC photons, such that LAEs could potentially be important contributors to the photon budget in the reionization era (particularly since we find that $\xi_{ion}$ is higher for LAEs than for more normal star-forming galaxies at $z=2.2$). Hence, LAEs may have been important contributors of the photons that reionized the Universe.

To make progress we need a detailed understanding of the physical processes which drive f$_{\rm esc}$, for which a significant sample of directly detected LyC leakers at a range of redshifts and galaxy properties is required. It is challenging to measure f$_{\rm esc}$ directly at $z>3$ (and practically impossible at $z>5$) due to the increasing optical depth of the IGM with redshift, such that indirect methods to estimate f$_{\rm esc}$ may be more successful \citep[e.g.][]{Jones2013,Zackrisson2013,Verhamme2015}. However, the validity of these methods remains to be evaluated \citep[i.e.][]{Vasei2016}.

\section{Conclusions}
\label{sec:8}
We have studied the production and escape of ionizing photons (LyC, $\lambda_0 < 912$ {\AA}) for a large sample of H$\alpha$ selected galaxies at $z=2.2$. Thanks to the joint coverage of the rest-frame LyC, UV and H$\alpha$ (and, in some cases, Ly$\alpha$), we have been able to reliably estimate the intrinsic LyC luminosity and the number of ionizing photons per unit UV luminosity ($\xi_{ion}$), for which we (indirectly) constrained the escape fraction of ionizing photons (f$_{\rm esc}$). Our results are:
\begin{enumerate}
\item We have stacked the $NUV$ thumbnails for all HAEs and subsets of galaxies in order to obtain constraints on f$_{\rm esc}$. None of the stacks shows a direct detection of LyC flux, allowing us to place a median (mean) upper limit of f$_{\rm esc} < 2.8\, (6.4)$ \% for the stack of star-forming HAEs (\S 4.3). A low escape fraction validates our method to estimate $\xi_{ion}$, the production efficiency of ionizing photons.
\item Combining the IGM emissivity measurements from \cite{BeckerBolton2013} with the integrated H$\alpha$ luminosity function from \cite{Sobral2013} at $z=2.2$, we find a globally averaged $\langle$f$_{\rm esc} \rangle = 5.9^{+14.5}_{-4.2}$ \% at $z=2.2$ (\S 5), where the errors include conservative estimates of the systematic uncertainties. Combined with recent estimates of the QSO emissivity at $z\approx2.2$, we can not fully rule out a non-zero contribution from star-forming galaxies to the ionizing emissivity. We speculate that the apparent discrepancy between the f$_{\rm esc}$ upper limit from median stacking and $\langle$f$_{\rm esc} \rangle$ can be understood in a scenario where the average f$_{\rm esc}$ is driven by a few galaxies with high f$_{\rm esc}$, or by a scenario where f$_{\rm esc}$ is higher for galaxies below the H$\alpha$ detection threshold (SFR$>4$ M$_{\odot}$ yr$^{-1}$).
\item Applying a similar analysis to published data at $z\approx4-5$ results in a relatively constant f$_{\rm esc}$ with redshift (see Table $\ref{tab:global_fesc}$ and Fig. $\ref{fig:fesc_evolution}$). We rule out $\langle$f$_{\rm esc}\rangle > 20$ \% at redshifts lower than $z\approx5$. An additional contribution of ionizing photons from rare quasars strengthens this constraint. 
\item We find that $\xi_{ion}$ increases strongly with increasing sSFR and H$\alpha$ EW and decreasing UV luminosity, independently on the dust correction method. We find no significant correlations between $\xi_{ion}$ and SFR(H$\alpha$), $\beta$ or M$_{\rm star}$. On average, LAEs have a higher $\xi_{ion}$ than HAEs, a consequence of LAEs having typically bluer UV slopes, lower masses and lower values of E$(B-V)$ (\S 6) -- properties which are typical for galaxies at the highest redshift.
\item The median $\xi_{ion}$ of HAEs at $z=2.2$ is $\xi_{ion} \approx 10^{24.77\pm0.04}$ Hz erg$^{-1}$, which is $\approx0.4$ dex lower than the typically assumed values in the reionization era or recent measurements at $z\sim4-5$ \citep{Bouwens2015xi}, see Table $\ref{tab:Xion_subsets}$. Only half of this difference is explained by the lower stellar mass and dust attenuation of the galaxies in the \cite{Bouwens2015xi} sample. 
\item For LAEs at $z=2.2$ we find a higher $\xi_{ion} =10^{25.14\pm0.09}$ Hz erg$^{-1}$, already similar to the typical value assumed in the reionization era. This difference is driven by the fact that LAEs are typically less massive and bluer and thus have less dust than HAEs.
\item By combining our trend between $\xi_{ion}$ and H$\alpha$ EW with the redshift evolution of H$\alpha$ EW, we find that $\xi_{ion}$ increases with $\approx 0.2$ dex between $z=2.2$ and $z=4-5$, resulting in perfect agreement with the results from \cite{Bouwens2015xi}. Extrapolating this trend leads to a median value of $\xi_{ion} \approx 10^{25.4}$ Hz erg$^{-1}$ at $z\sim8$, slightly higher than the typically assumed value in the reionization epoch (\S 7), such that a relatively low global f$_{\rm esc}$ (consistent with our global estimates at $z\approx2-5$) would suffice to provide the photons to reionize the Universe. 
\end{enumerate}

\section*{Acknowledgments}
We thank the referee for the many helpful and constructive comments which have significantly improved this paper. JM acknowledges the support of a Huygens PhD fellowship from Leiden University. DS acknowledges financial support from the Netherlands Organisation for Scientific research (NWO) through a Veni fellowship and from FCT through a FCT Investigator Starting Grant and Start-up Grant (IF/01154/2012/CP0189/CT0010). PNB is grateful for support from the UK STFC via grant ST/M001229/1. IO acknowledges support from the European Research Council in the form of the Advanced Investigator Programme, 321302, {\sc cosmicism}. The authors thank Andreas Faisst, Michael Rutkowski and Andreas Sandberg for answering questions related to this work and Daniel Schaerer and Mark Dijkstra for discussions.
We acknowledge the work that has been done by both the COSMOS team in assembling such large, state-of-the-art multi-wavelength data-set, as this has been crucial for the results presented in this paper.
We have benefited greatly from the public available programming language {\sc Python}, including the {\sc numpy, matplotlib, pyfits, scipy} \citep{SCIPY,MATPLOTLIB,NUMPY} and {\sc astropy} \citep{ASTROPY} packages, the astronomical imaging tools {\sc SExtractor} and {\sc Swarp} \citep{Bertin1996,Bertin2010} and the {\sc Topcat} analysis program \citep{Topcat}.




\bibliographystyle{mnras}

\bibliography{bib_LAEevo.bib}


\appendix
\section{Individual $NUV$ detections}
We search for individual galaxies possibly leaking LyC photons by matching our {\sc Clean} galaxy sample with the public {\it GALEX} EM cleaned catalogue \citep[e.g.][]{Zamojski2007,EMphot}, which is $U$ band detected. In total, we find 19 matches between {\sc Clean} HAEs and {\it GALEX} sources with $NUV<26$ within 1$''$ (33 matches when using all HAEs), and 9 matches between LAEs and {\it GALEX} sources (four out of these 9 are also in the HAE sample and we will discuss these as HAEs). By visual inspection of the {\it HST}/ACS F814W and CFHT/$U$ band imaging, we mark 8/19 HAEs and 2/5 LAEs as reliable $NUV$ detections. The 14 matches that we discarded were either unreliable detections in $NUV$ (9 times, caused by local variations in the depth, such that the detections are at 2$\sigma$ level) or a fake source in $NUV$ (5 times, caused by artefacts of bright objects). We note however that in most of the remaining 10 $NUV$ detections (8 HAEs, 2 LAEs) the $NUV$ photometry is blended with a source at a distance of $\approx$4$''$, see Fig. $\ref{fig:thumbnails_appendix}$.  

In order to get a first order estimate of the contamination from neighbouring sources to the $NUV$ flux, we perform the following simulation. First, we simulate the $NUV$ flux of the candidate LyC leakers and all sources within 10$''$ by placing Moffat flux distributions with the PSF-FWHM of $NUV$ imaging and $\beta = 3$. These flux distributions are normalised by the $U$ band magnitude of each source, since the catalog that we use to measure $NUV$ imaging uses $U$ band imaging as a prior. We then measure the fraction of the flux that is coming from neighbouring sources within an aperture with radius $0.67\times$FWHM centred at the position of the $NUV$ detection of the candidate LyC leaker. We find that contamination for most candidates is significant, and remove three candidates for which the estimated contamination is larger than $>50$ \%. The remaining candidates have contaminations ranging from 0-39 \% and we subtract this contamination from the measured $NUV$ flux when estimating their escape fractions. We estimate the uncertainty in our contamination estimate due to variations in the PSF and in the flux normalisation (due to $NUV-U$ colours) as follows: we first simulate the contamination with a gaussian PSF and Moffat PSFs with increasing $\beta$ up to $\beta =7$ and also by correcting the $U$ band magnitude prior with the observed $U-B$ and $NUV-U$ colours. We then estimate the systematic uncertainty by measuring the standard deviation of the contamination rates estimated with the different simulations. For sources with little contamination, the systematic uncertainty in the contamination estimate is of the order 5 \%. 

We test whether the $NUV$ detections for these sources could arise solely from flux at $\lambda_0 > 912$ {\AA} in the far red wing of the $NUV$ filter (see \S 3.2). For each galaxy, we obtain the best-fit {\sc Starburst99} model by matching the H$\alpha$ EW, as H$\alpha$ EW is most strongly related to the SED shape around 900 {\AA}. We redshift this model to a redshift of 2.22 and normalise the SED to reproduce the $V$ band magnitude (we assume zero dust attenuation, which is a conservative assumption for this analysis, see below) and convolve the model with the mean IGM transmission at $z=2.22$. Then, we measure the predicted $NUV$ magnitude in the case that the flux is only non-zero at $\lambda_0 > 912$ {\AA}. We find that, in all cases, this magnitude is too faint to explain the NUV detections, ranging from $NUV=30.1-32.5$, well below the detection limits. In the presence of dust, the attenuation at $\lambda \sim 912-930$ {\AA} is stronger than at $\lambda \sim 1600$ {\AA} \citep[e.g.][]{Reddy2016}, such that the predicted $NUV$ magnitude would be even fainter. We test the robustness of this estimate by varying the SED models (lowering the H$\alpha$ EWs), neglecting the IGM absorption or by perturbing the redshift between $z=2.20-2.24$, but find that this changes the result only by up to 1 magnitude if all effects are combined. For ID 1139 and 7801 we also test simple AGN power-law models (f$_{\lambda} \propto \lambda^{\beta}$) with UV slopes ranging from -2.0 to -2.7, but find that pure non-ionizing flux can not explain the $NUV$ photometry. Therefore, it is unlikely that the $NUV$ detections arise purely from flux at $\lambda_0 > 912$ {\AA}, just because the filter transmission at these wavelengths is very low, and the wavelength range constitutes only a fraction of the full filter width.

\begin{table*}
\begin{center}
\begin{small}
\caption{Candidate LyC leakers among the HAE/LAE sample. ID numbers of HAEs refer to the IDs in the HiZELS catalog (\citealt{Sobral2013}). IDs indicated with a * are X-Ray AGN. The coordinates correspond to the peak of H$\alpha$/Ly$\alpha$ emission. The redshift is either spectroscopic ($^s$), photometric ($^p$) or from a dual-narrow band emission-line confirmation ($^d$). The $NUV$ contamination fraction is estimated as described in the text. f$_{\rm esc}$ is corrected for contamination from nearby sources to the $NUV$ flux. Because of the absence of H$\alpha$ measurements for LAEs, we do not estimate the SFR(H$\alpha$) or f$_{\rm esc}$.} 
\begin{tabular}{ lrrrrrrrrr }
\bf ID & \bf R.A. & \bf Dec. & \bf Redshift & \bf M$_{\rm star}$& \bf SFR(H$\alpha$) &\bf M$_{1500}$ &$\bf NUV$ & \bf NUV contamination &\bf f$_{\rm esc}$ \\ 
 & (J2000) & (J2000) & & log$_{10}$(M$_{\odot}$) & M$_{\odot}$ yr$^{-1}$ & mag & mag & & \% \\ \hline
  1139* & 10:00:55.39 & +01:59:55.39 & 2.219$^s$ & 10.1& 34.8 & -21.6 & 25.9 & 0.0  &  30\\
  1872 & 10:01:56.39 & +02:17:36.65 & 2.22$\pm0.02^p$ &  9.4 &  9.2 & -21.0 & 25.7 & 0.14 &  43\\
  1993 & 10:02:08.70 & +02:21:19.88 & 2.22$\pm0.01^d$  &9.6 & 8.2 & -21.3 & 24.6& 0.39 &  45\\
  2258 & 10:01:29.69 & +02:24:28.50 & 2.22$\pm0.02^p$ & 10.3 & 7.3 & -21.0 & 25.1 & 0.21 &  43\\
  7801* & 10:02:08.55 & +01:45:53.60 & 2.215$^s$ & 10.4 & 43.3 & -23.5 & 24.9 & 0.05 &  37\\ \hline
  C8* & 09:59:34.82 & +02:02:49.94 &2.182$^s$ & 10.9 & & -22.5 & 24.6 & 0.03 & \\ 
  C10* & 09:59:05.14 & +02:15:29.86 & 2.222$^s$ & 10.6 & & -23.5 & 23.7 & 0.03 & \\
\end{tabular}
\label{tab:LyC_candidates}
\end{small}
\end{center}
\end{table*}

For the five candidate LyC leakers with H$\alpha$ measurements, we measure escape fractions ranging from $\approx 35-46$ \%, see Table. $\ref{tab:LyC_candidates}$, although we note that these escape fractions are still uncertain due to i) possible underestimated foreground contamination from sources not detected in $U$ (or not detected as individual source due to blending) or with very blue $NUV-U$ colours, ii) uncertain dust attenuation of the H$\alpha$ luminosity, iii) underestimated contribution from flux at $\lambda_0 > 912$ {\AA} due to different SED shapes than expected or (photometric) redshift errors. Observations with higher spatial resolution and detailed spectroscopy are required in order to confirm whether these 7 candidates are really leaking LyC photons and at what rate.

Four isolated LyC leaker candidates (including two LAEs) are X-Ray AGN, and all have been spectroscopically confirmed at $z=2.2$ \citep{Lilly2009,Civano2012}. HiZELS-ID 1993 is detected in two other narrow-bands than the H$\alpha$ narrow-band: Ly$\alpha$ (EW$_{0, \rm Ly\alpha} = 67$ {\AA}) and {\sc [Oiii]} (EW$_{0,\rm [OIII]} > 100$ {\AA}), and is thus known to be at $z=2.22\pm0.01$ very robustly. ID 1872 and 2258 are selected as HAE at $z=2.2$ based on their photometry (see \citealt{Sobral2013}), such that it is possible that they are interlopers (with the second most likely emission-line being {\sc [Oiii]} at $z\sim3.3$, but other rarer possibilities such as Paschen series lines at $z<1$). We show thumbnail images of our candidate LyC leakers in the $NUV$, $F814W$ and $U$ bands in Fig. $\ref{fig:thumbnails_appendix}$ and Fig. $\ref{fig:thumbnails_appendix2}$.

\begin{figure*}
\begin{tabular}{ccc}
\includegraphics[width=5.5cm]{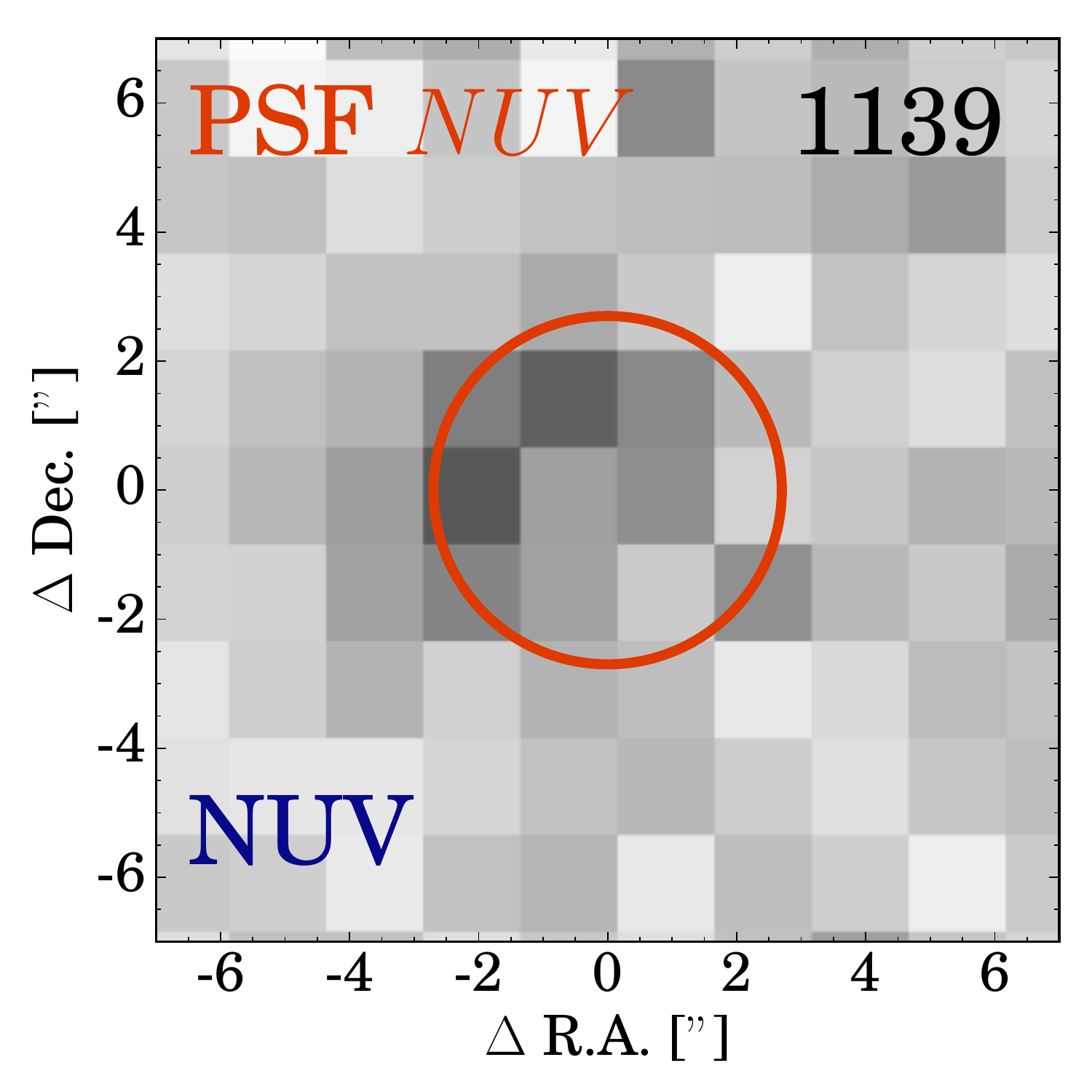}&
\includegraphics[width=5.5cm]{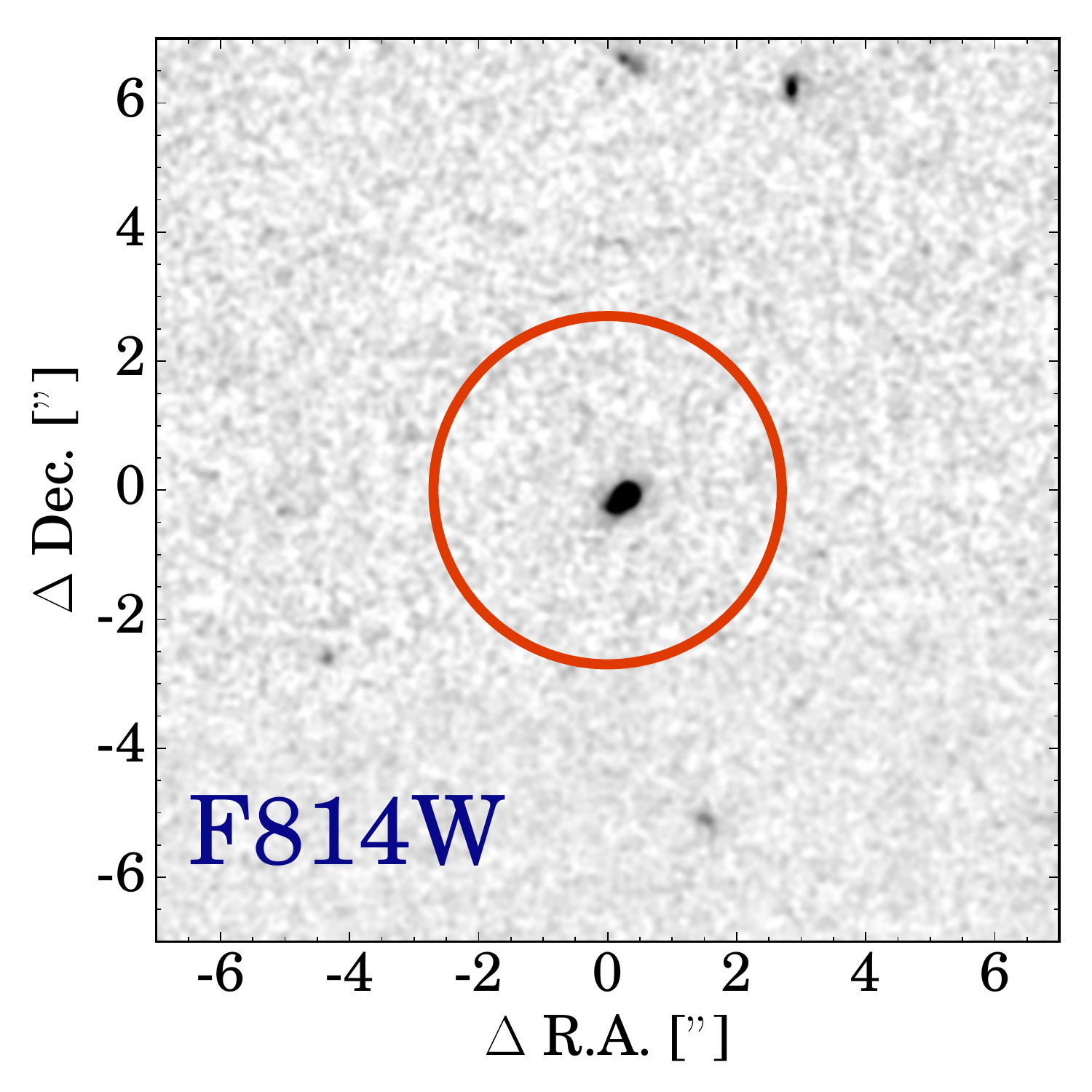}&
\includegraphics[width=5.5cm]{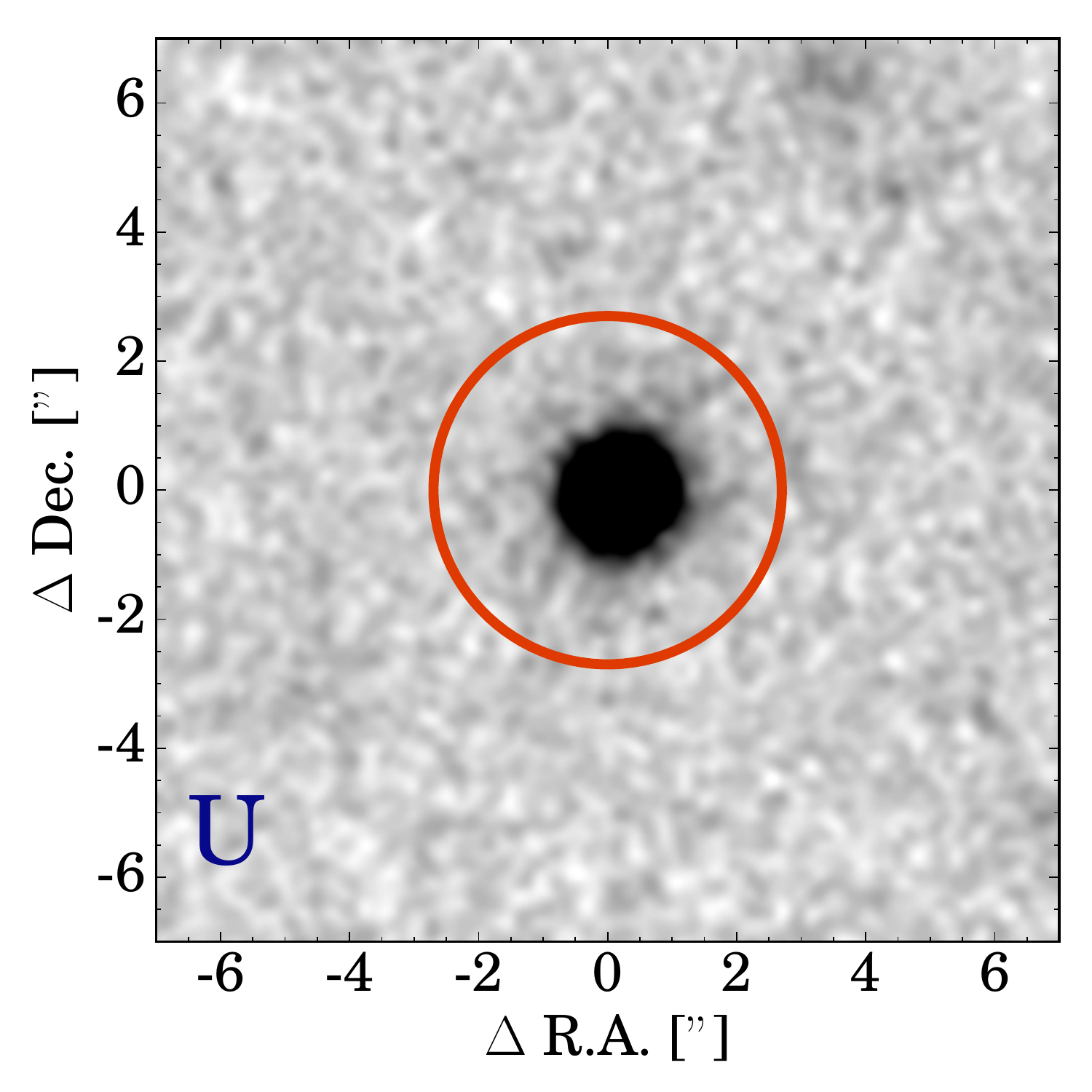}\\
\includegraphics[width=5.5cm]{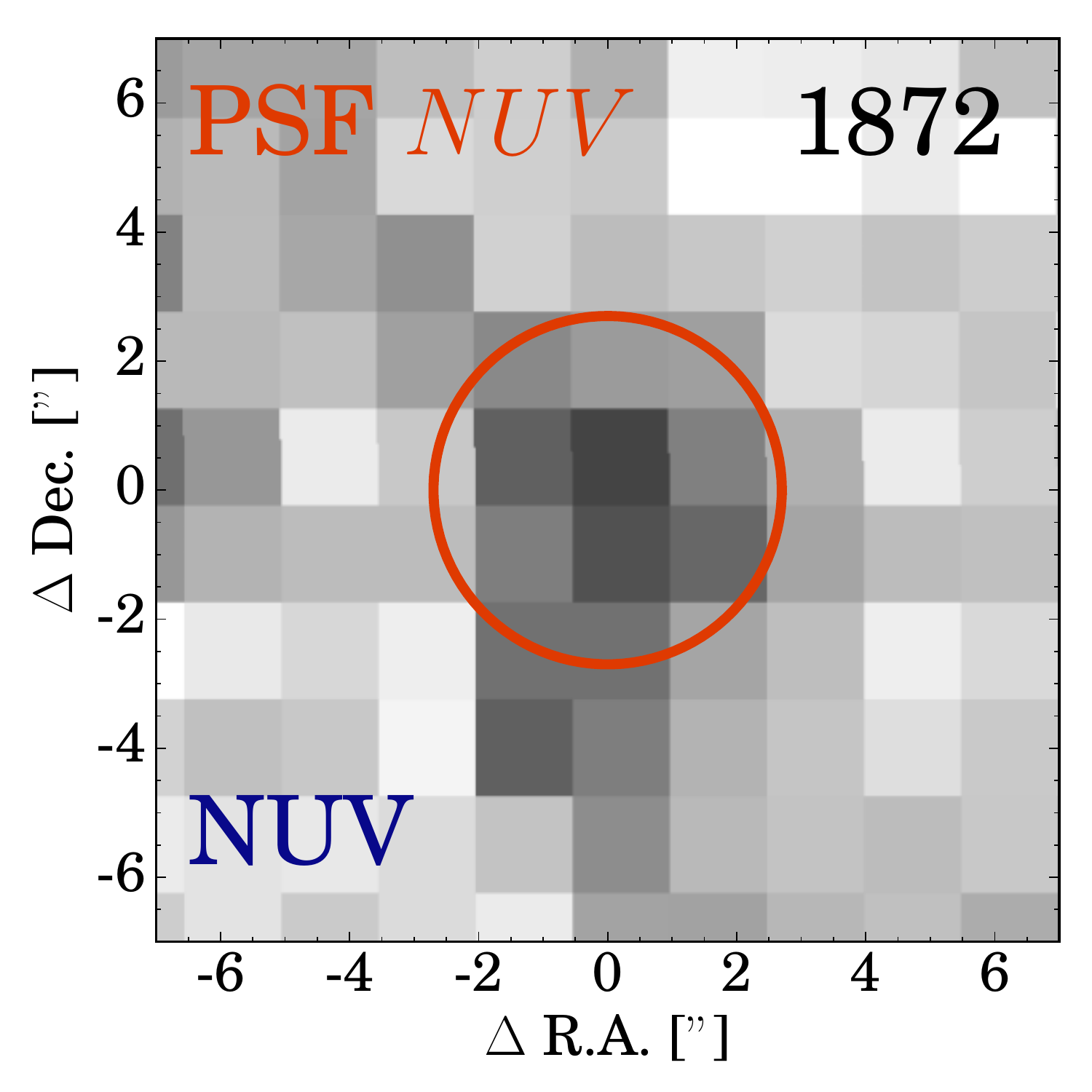}&
\includegraphics[width=5.5cm]{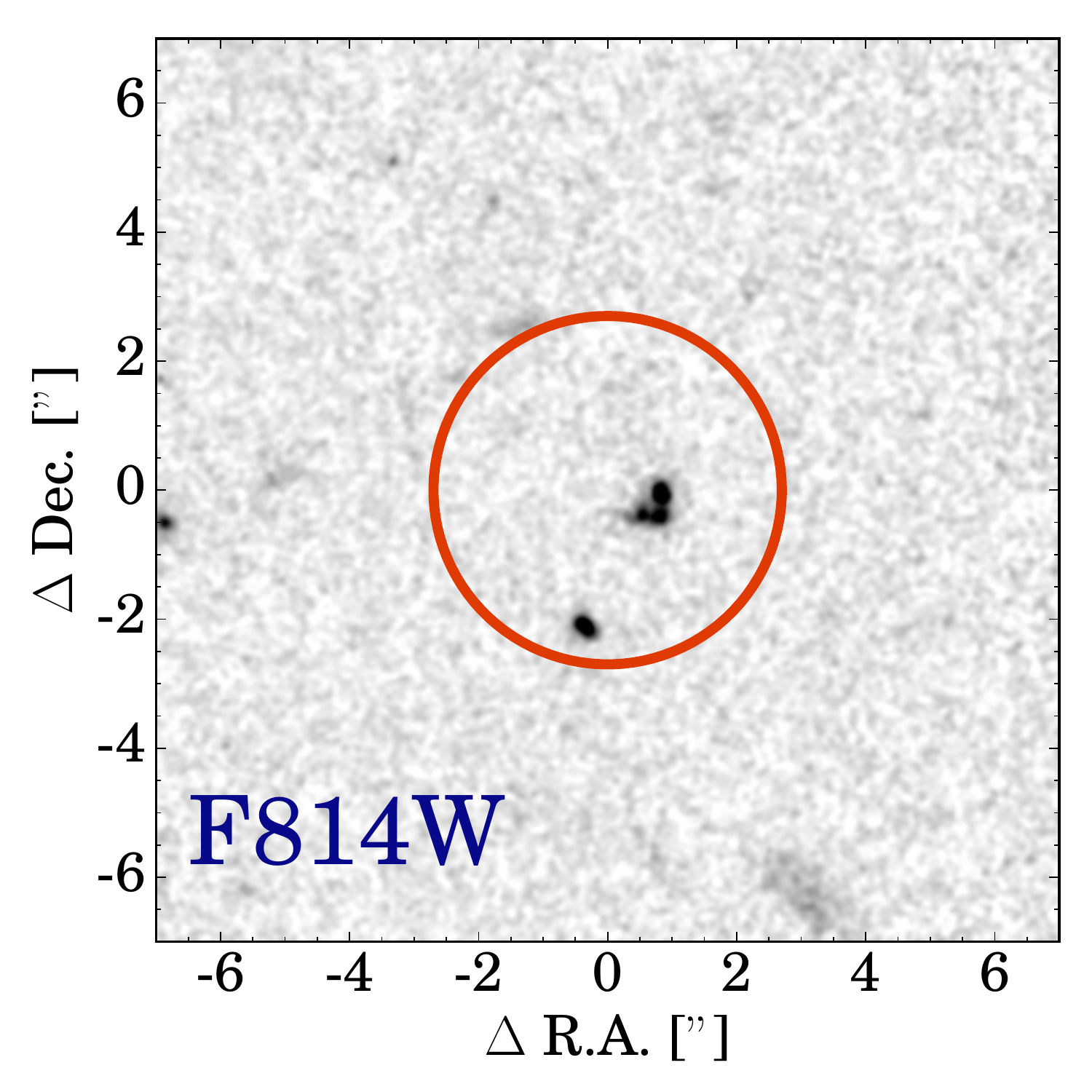}&
\includegraphics[width=5.5cm]{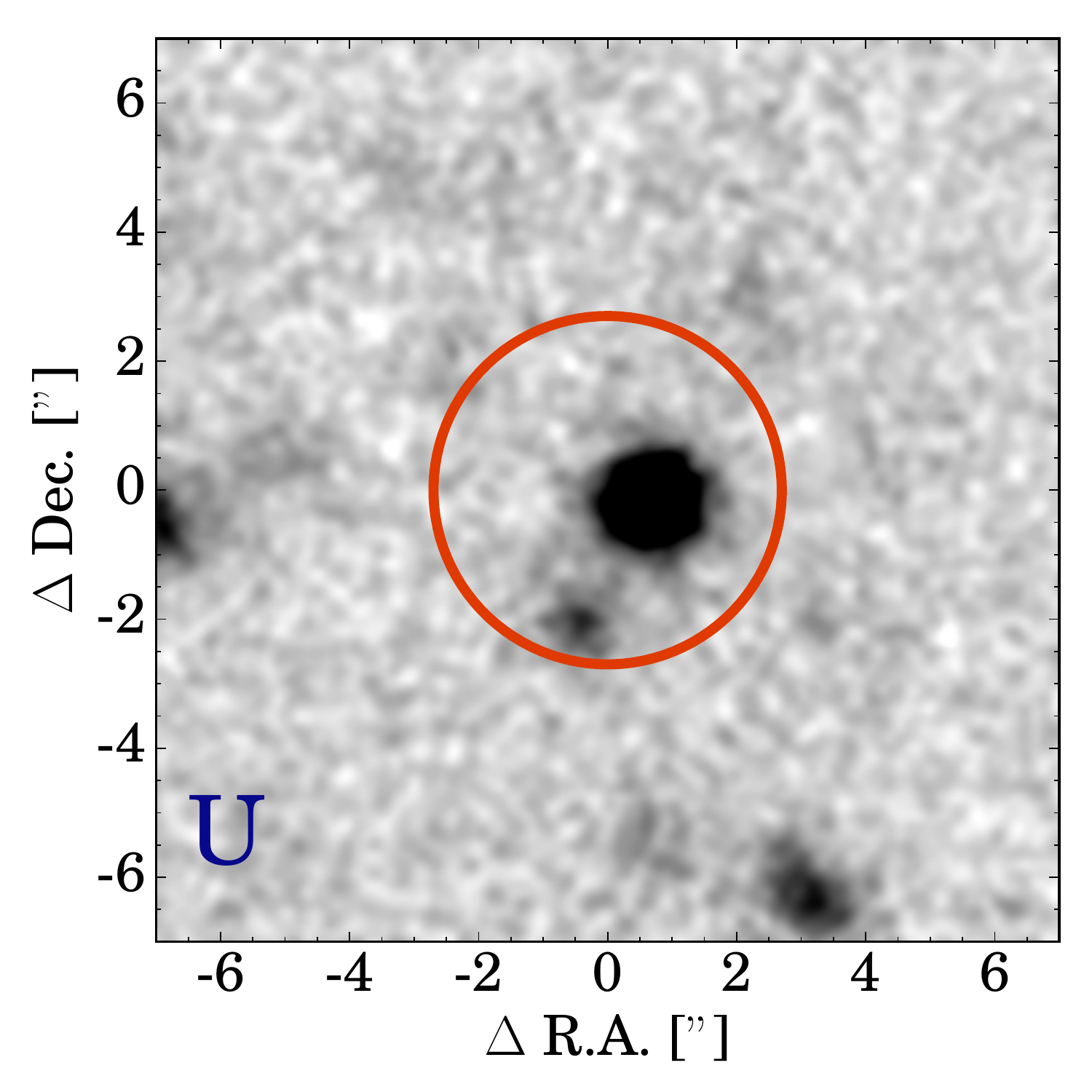}\\
\includegraphics[width=5.5cm]{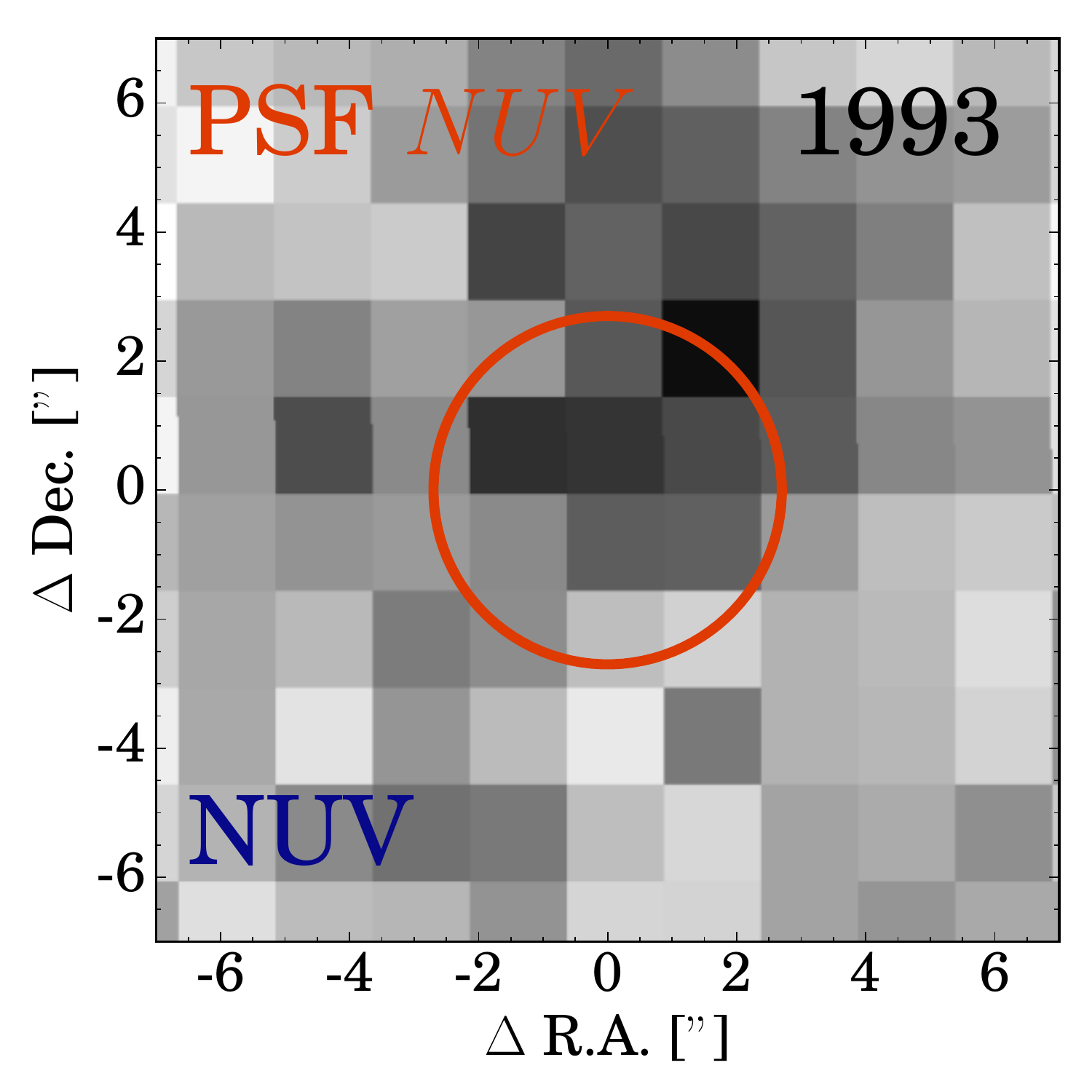}&
\includegraphics[width=5.5cm]{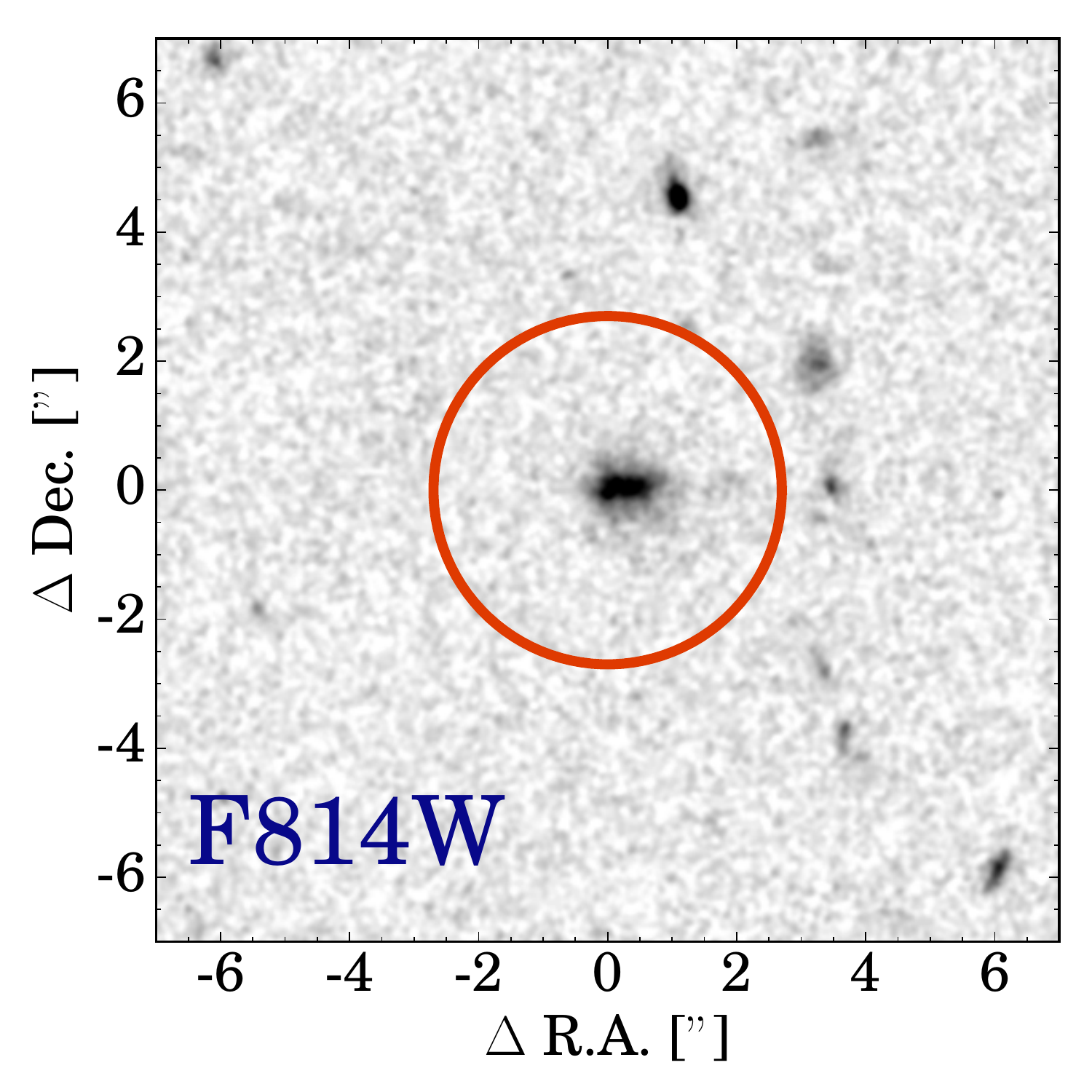}&
\includegraphics[width=5.5cm]{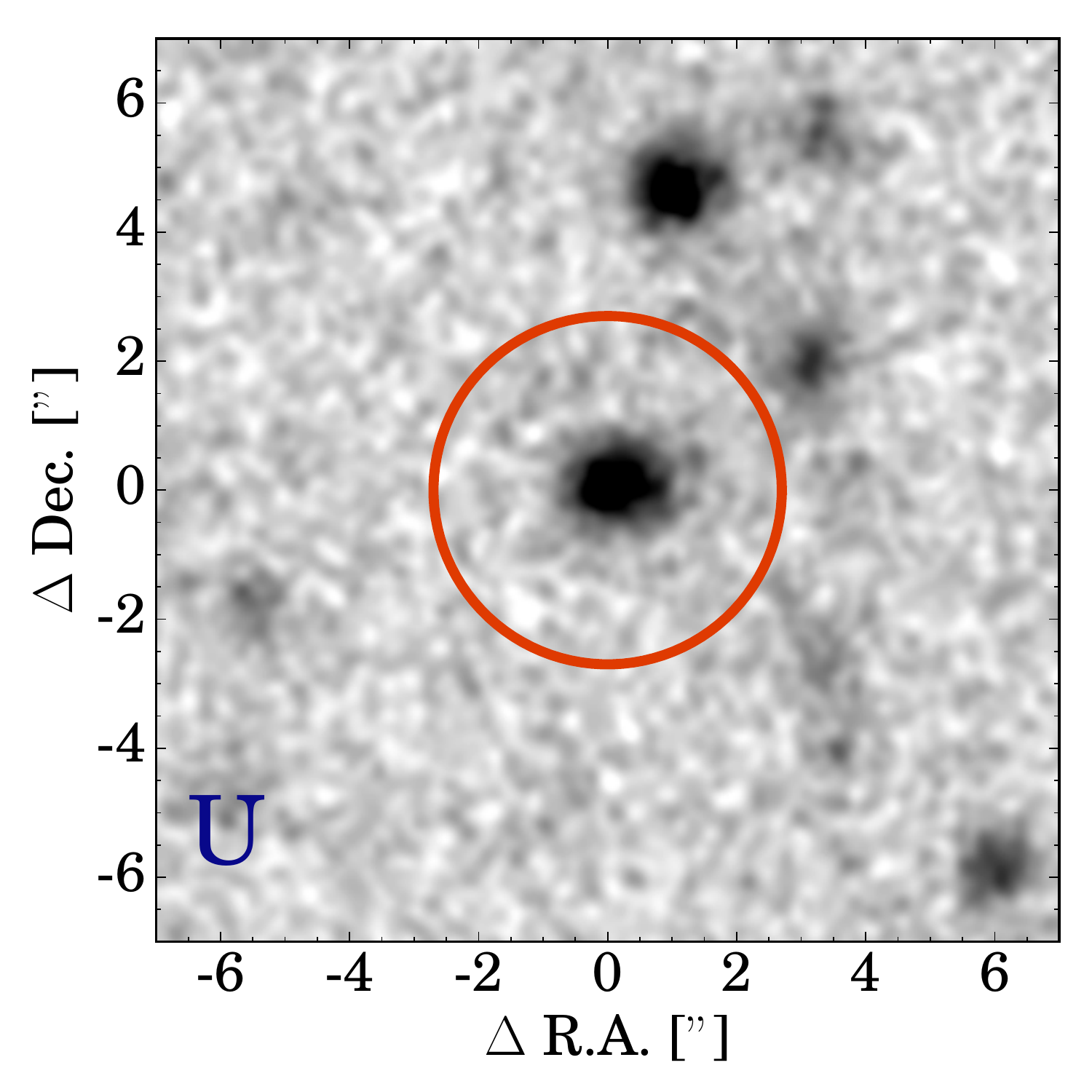}\\
\end{tabular}
\caption{\small{$15\times15''$ thumbnail images in $NUV$, $F814W$ and $U$ of candidate LyC leaking H$\alpha$ and Ly$\alpha$ emitters at $z=2.2$, centered on the positions of the HAE/LAE. The images are annotated with the IDs of the galaxies in the HiZELS catalogue (\citealt{Sobral2013}). Ly$\alpha$ emitters are identified with a ``C". IDs 1139, 1993 and 7801 are detected in both H$\alpha$ and Ly$\alpha$. IDs 1139, 7801, C-8 and C-10 are X-ray AGN. All other sources than the central source seen in thumbnails have photometric redshifts of $<1.5$.}}
\label{fig:thumbnails_appendix}
\end{figure*}

\begin{figure*}
\begin{tabular}{ccc}
\includegraphics[width=5.5cm]{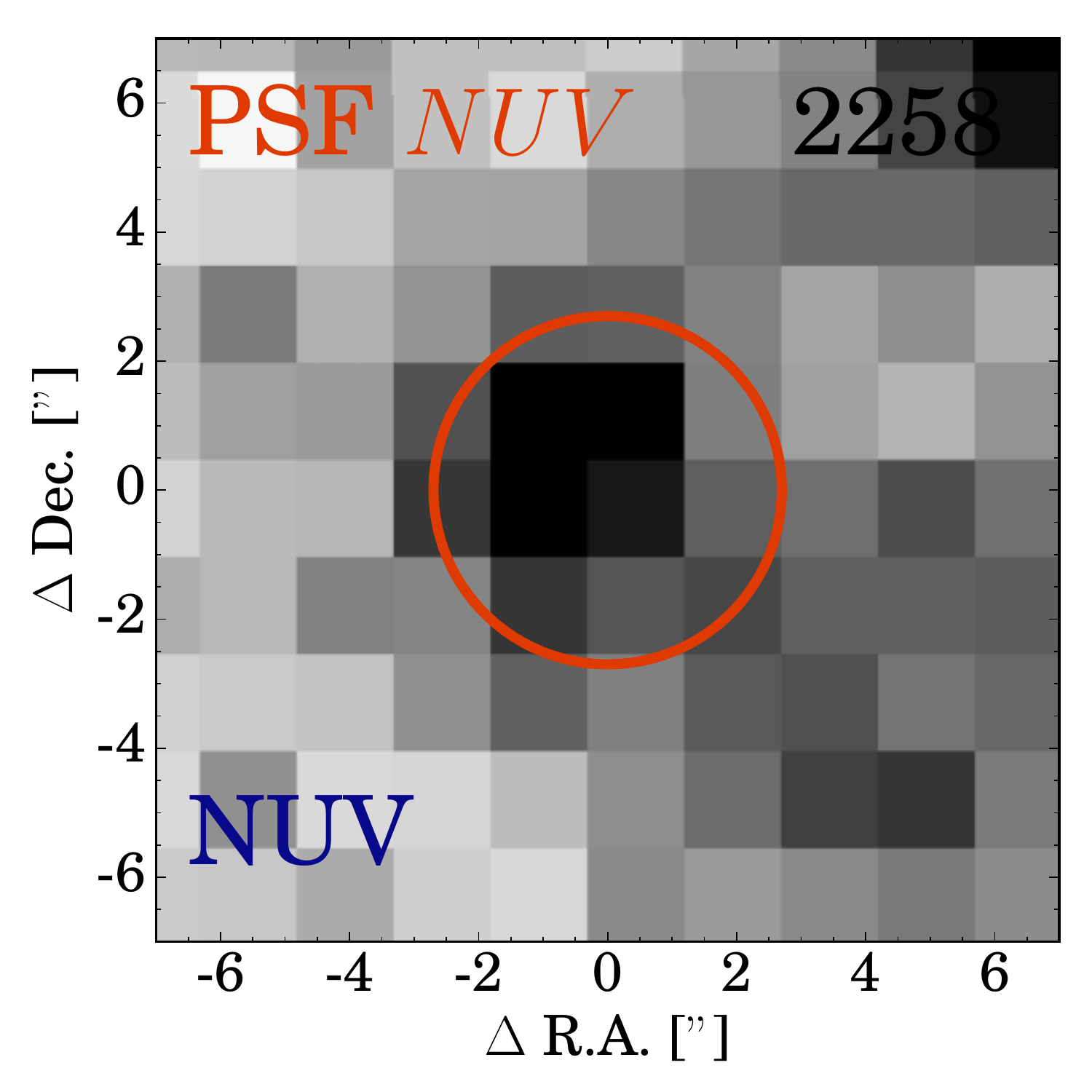}&
\includegraphics[width=5.5cm]{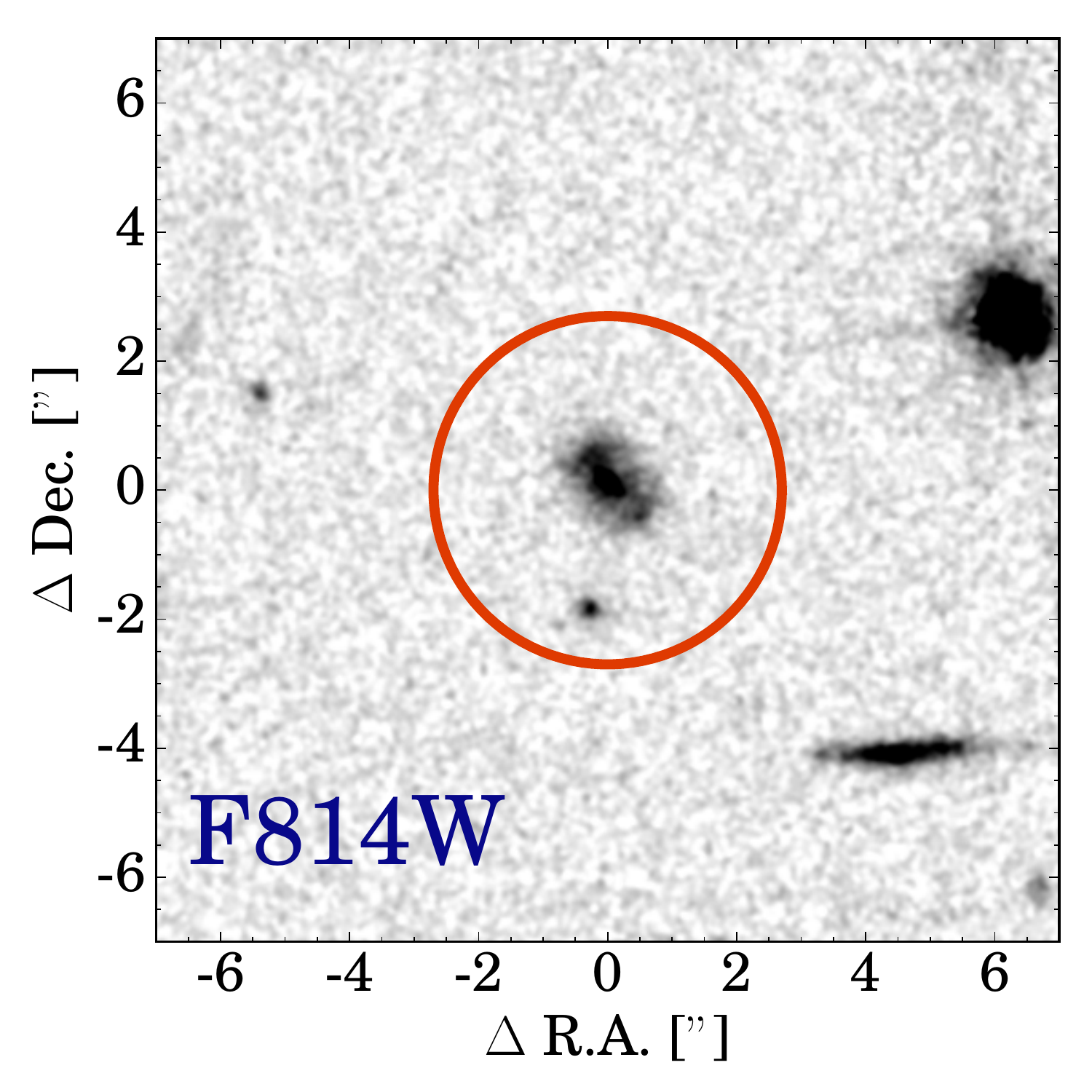}&
\includegraphics[width=5.5cm]{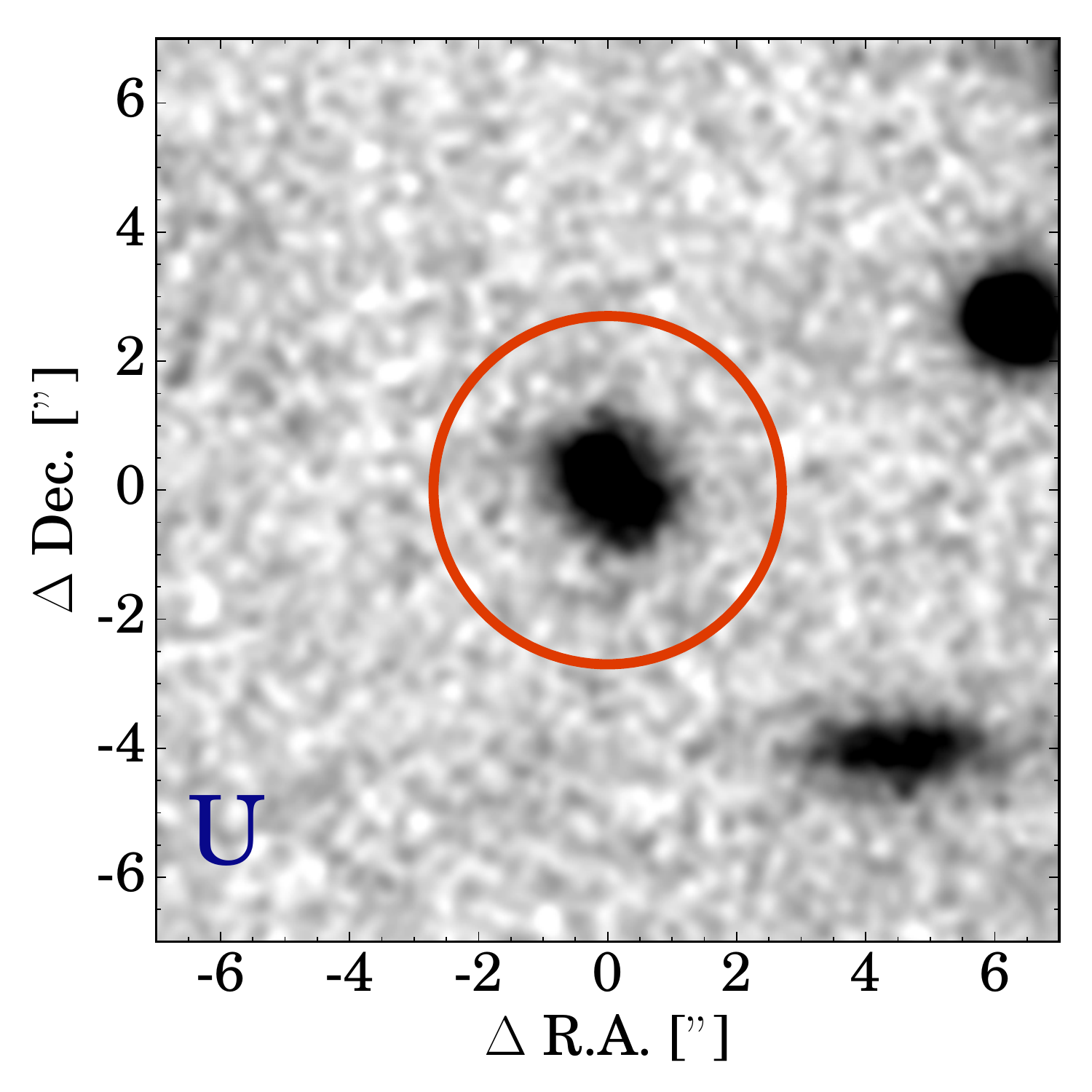}\\
\includegraphics[width=5.5cm]{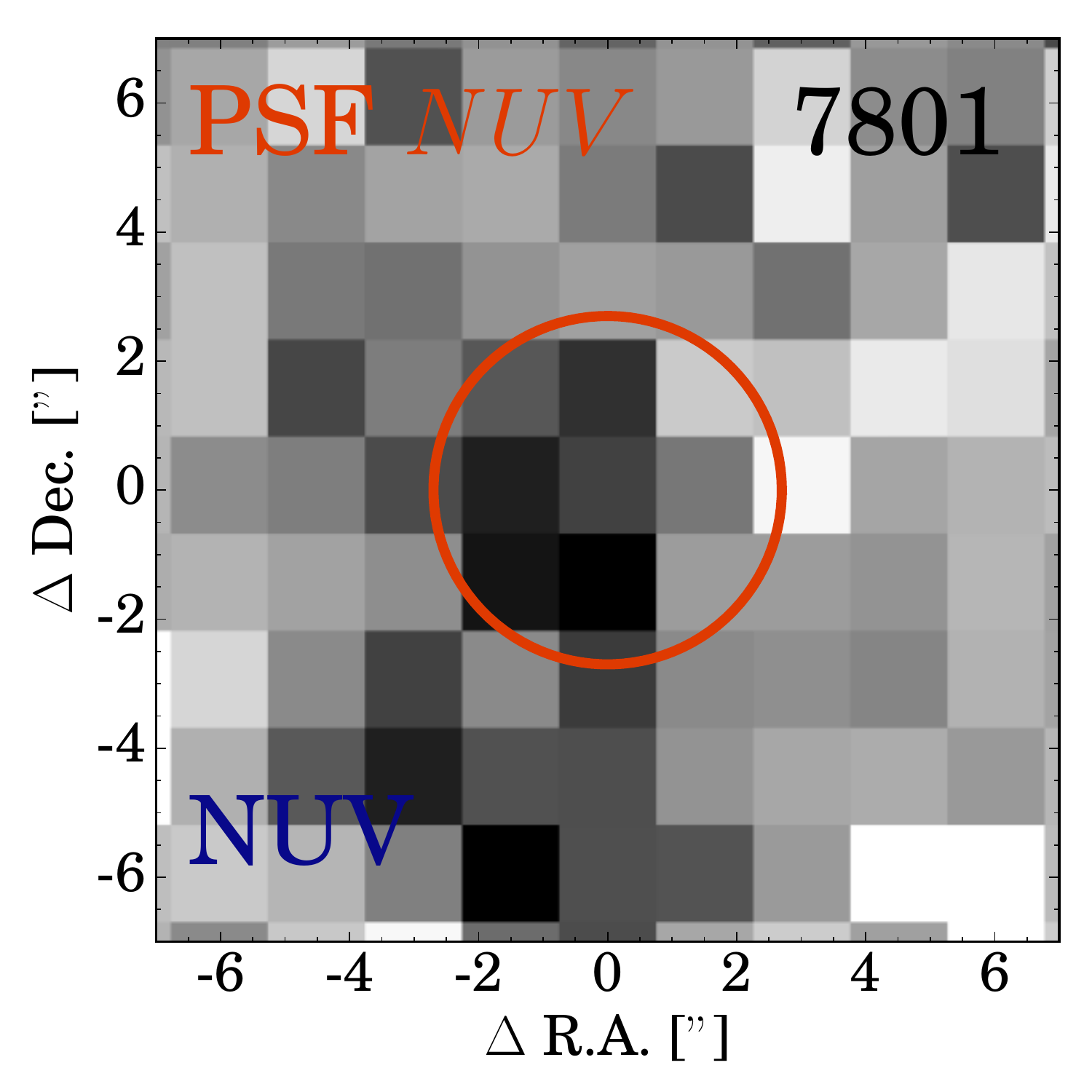}&
\includegraphics[width=5.5cm]{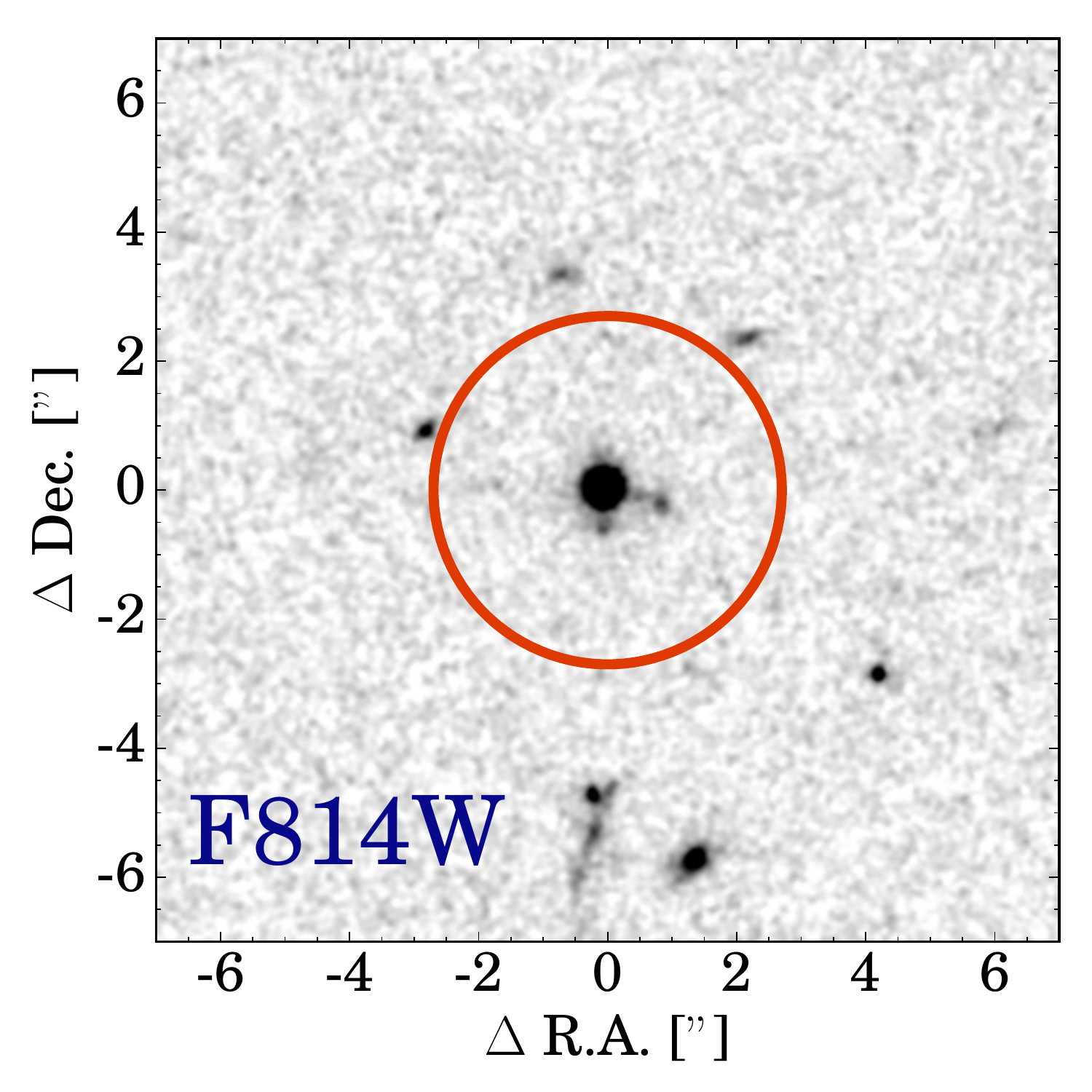}&
\includegraphics[width=5.5cm]{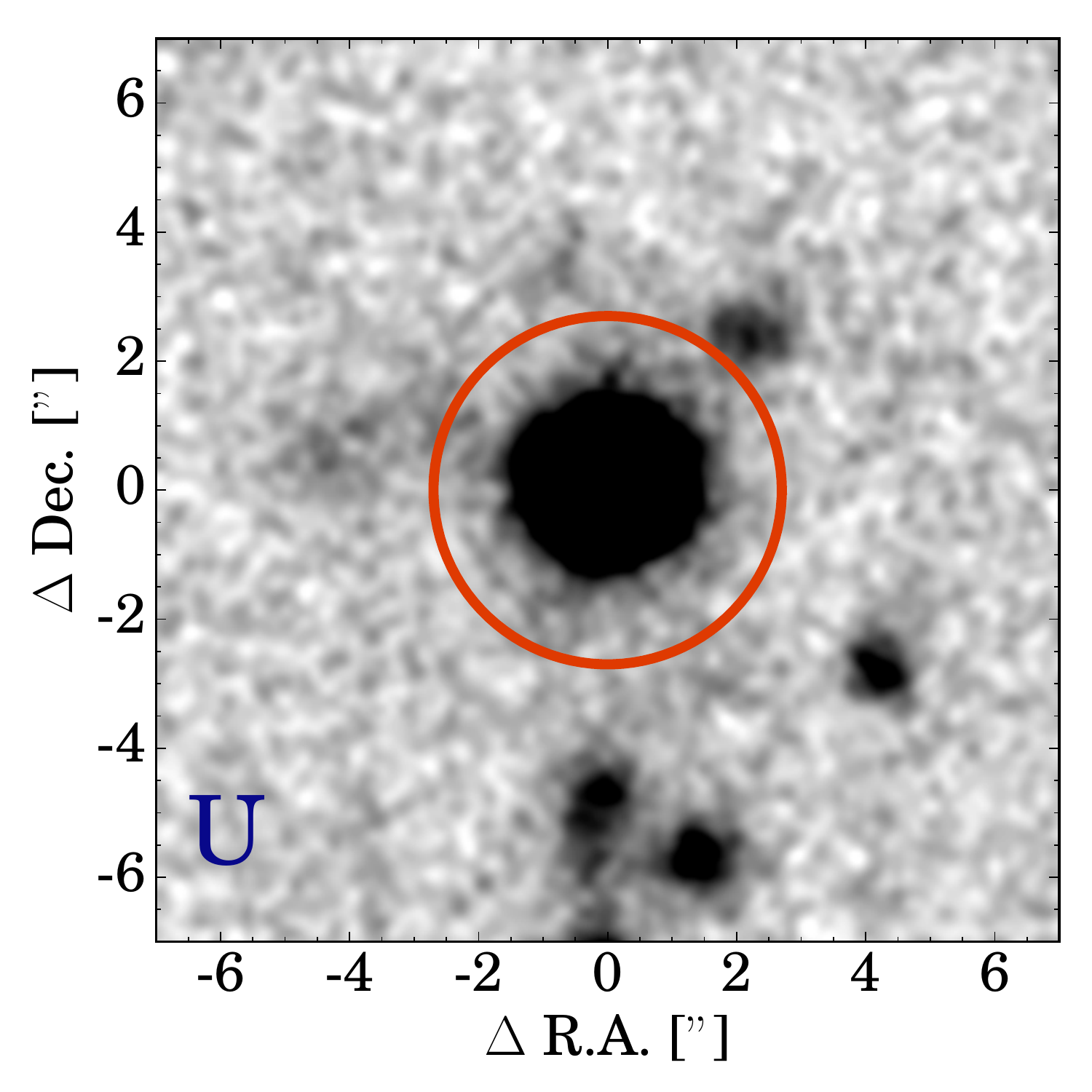}\\
\includegraphics[width=5.5cm]{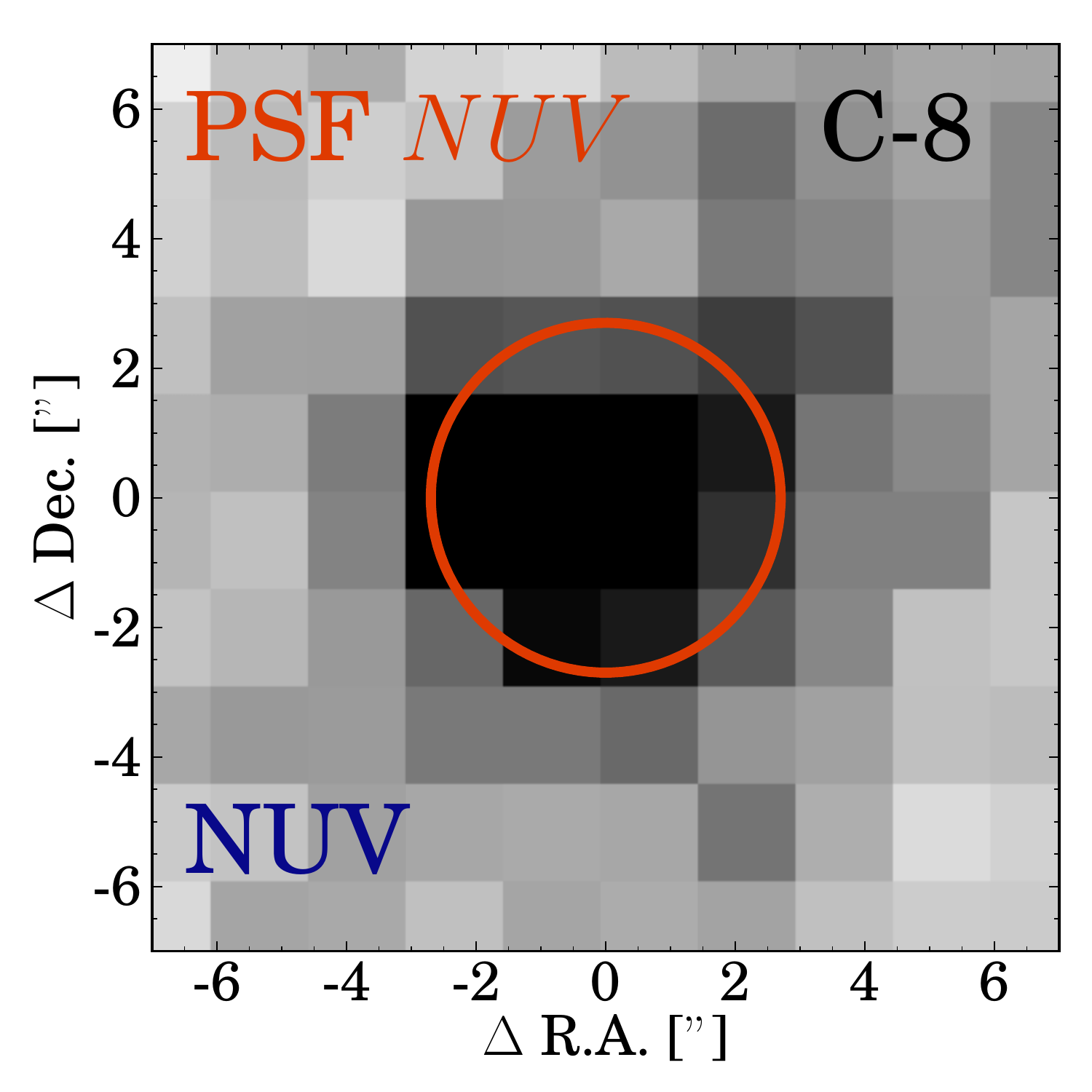}&
\includegraphics[width=5.5cm]{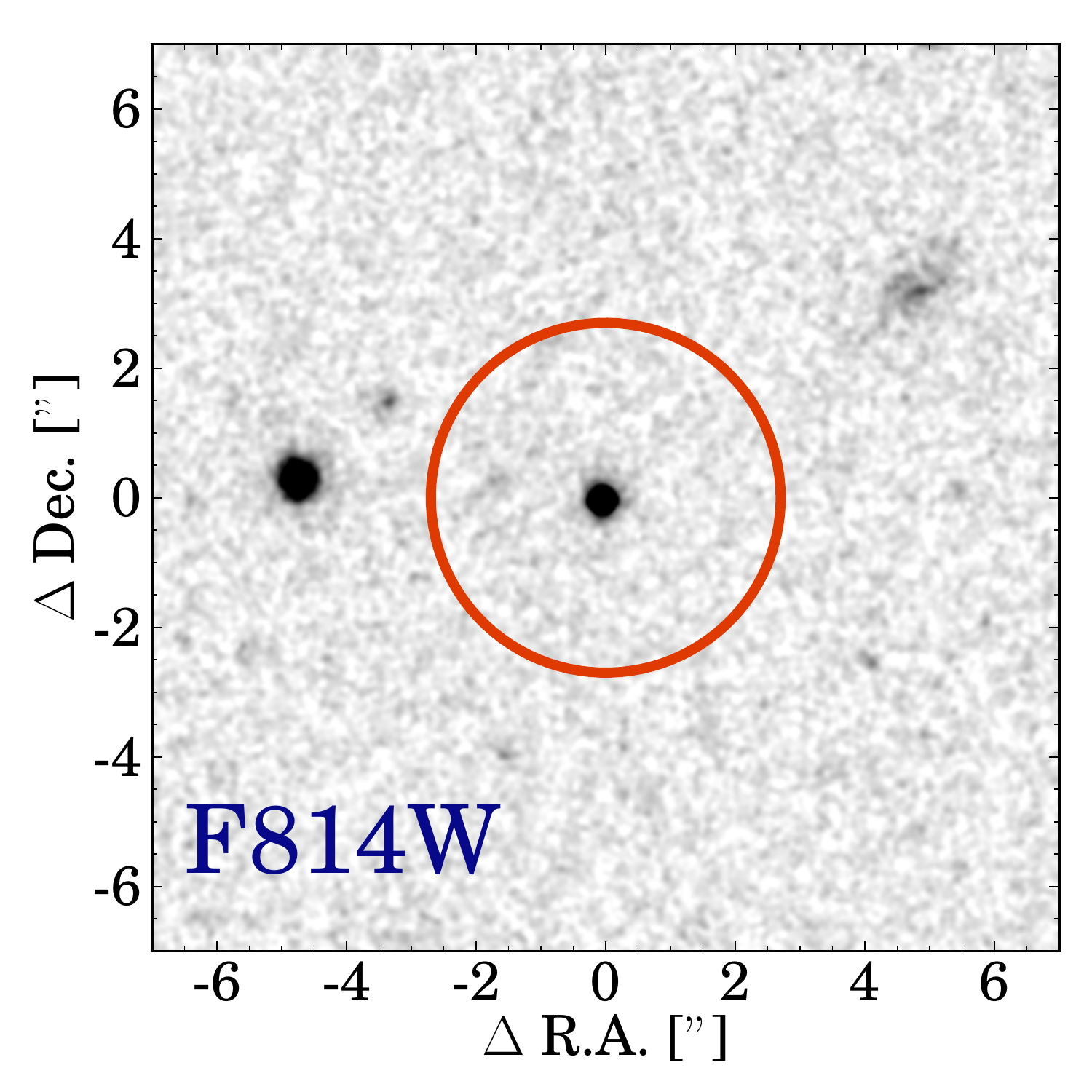}&
\includegraphics[width=5.5cm]{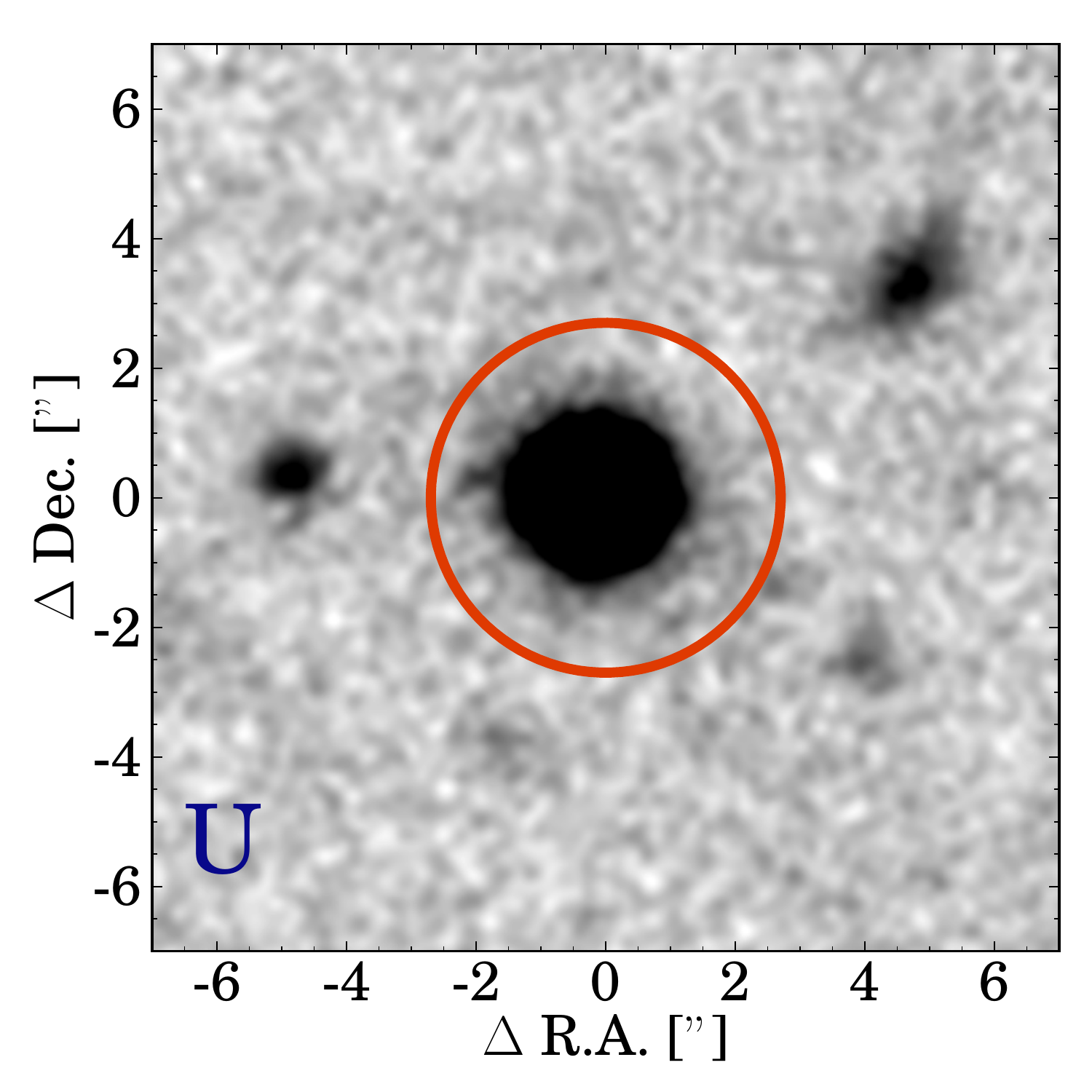}\\
\includegraphics[width=5.5cm]{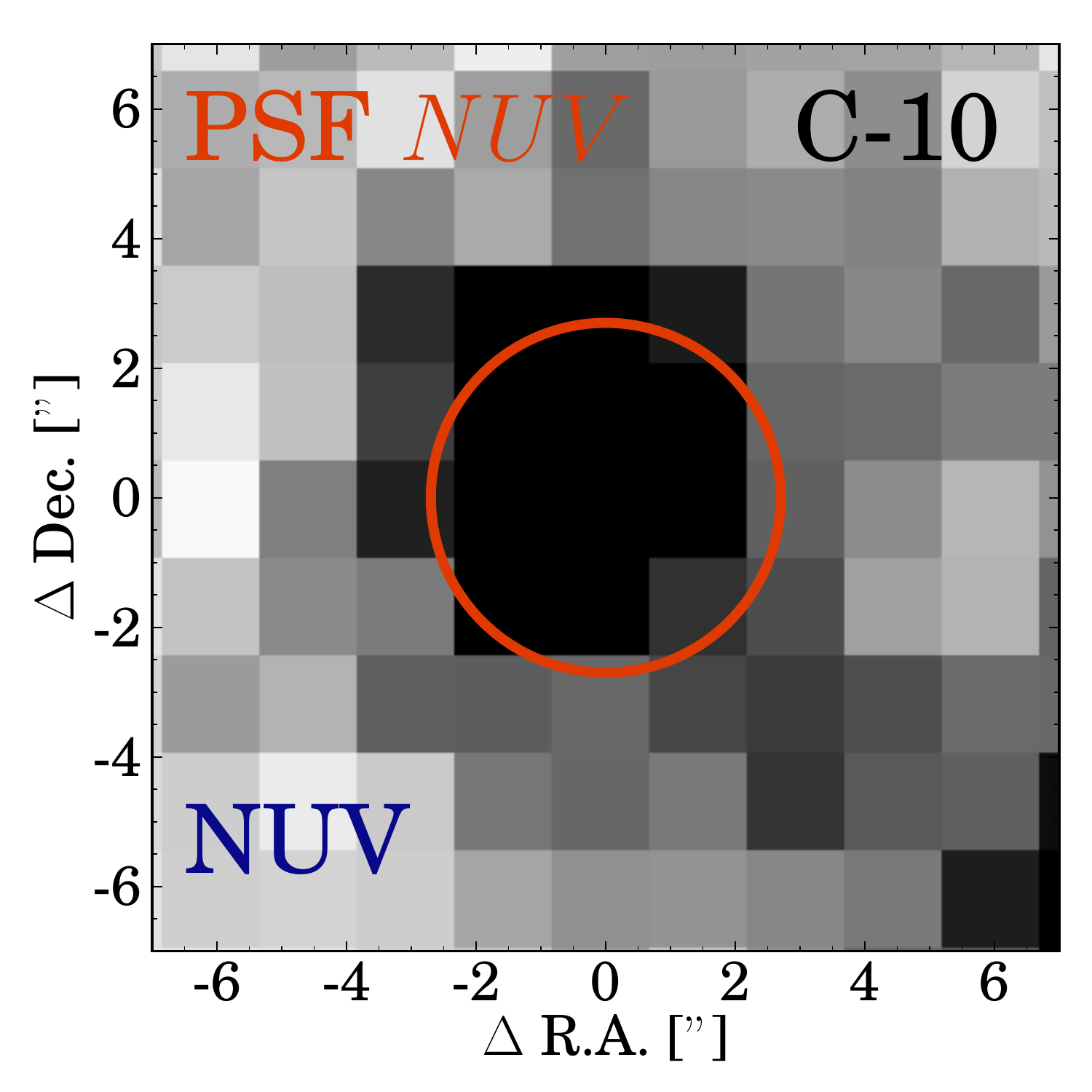}&
\includegraphics[width=5.5cm]{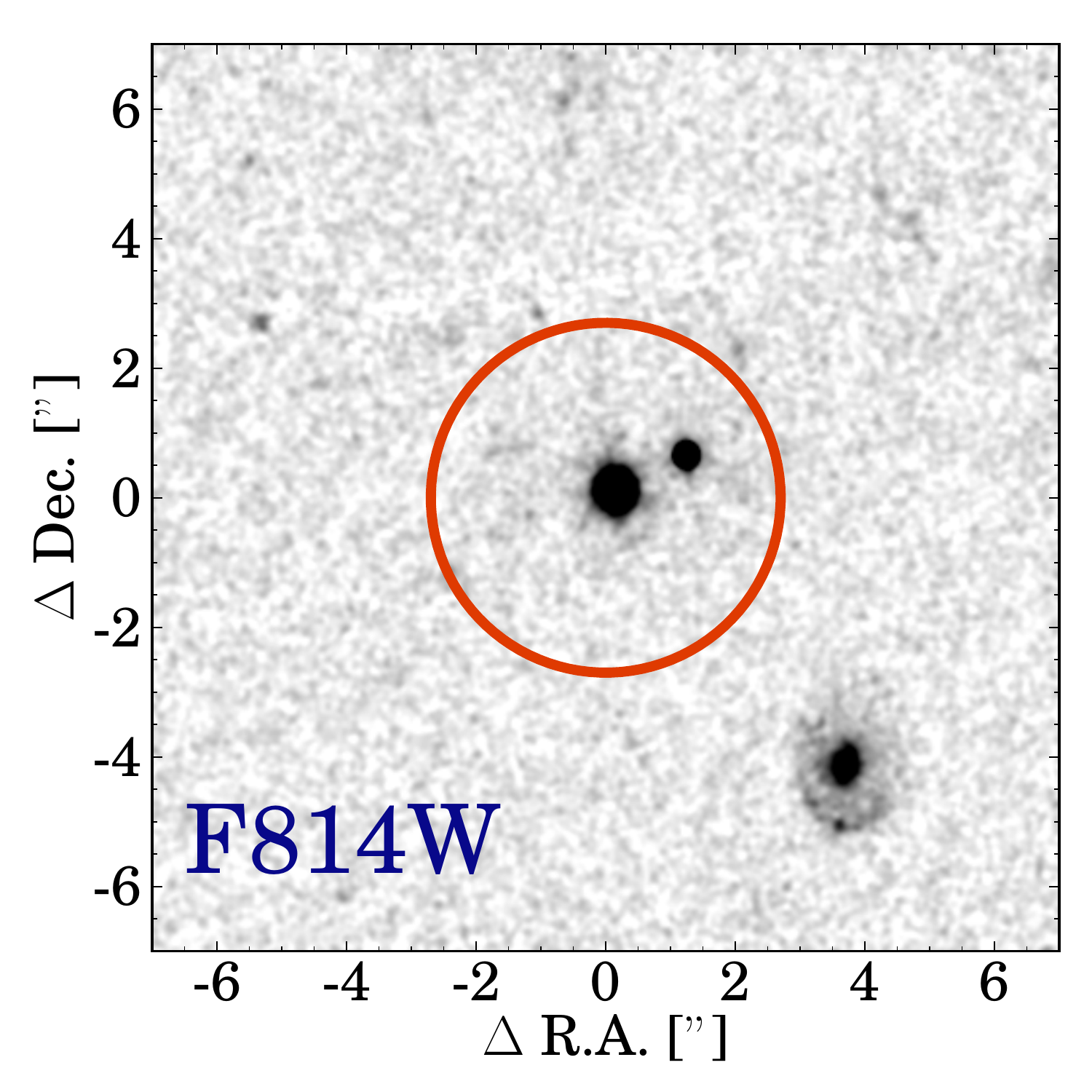}&
\includegraphics[width=5.5cm]{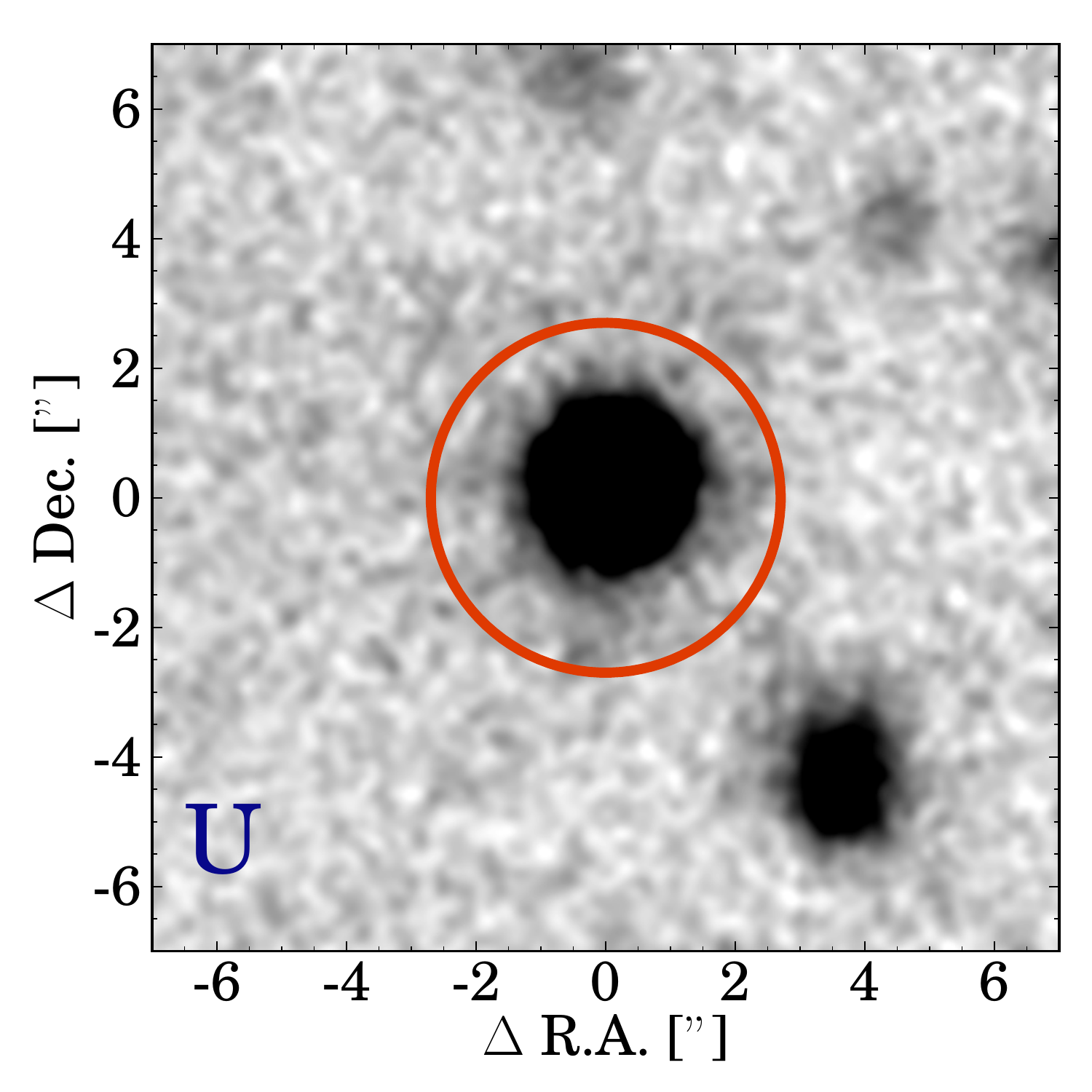}\\
\end{tabular}
\caption{\small{Continued from Fig. $\ref{fig:thumbnails_appendix}$.}}
\label{fig:thumbnails_appendix2}
\end{figure*}

\section{Redshift evolution of $\xi_{ion}$ with different dust corrections}
\begin{figure}
\includegraphics[width=8.6cm]{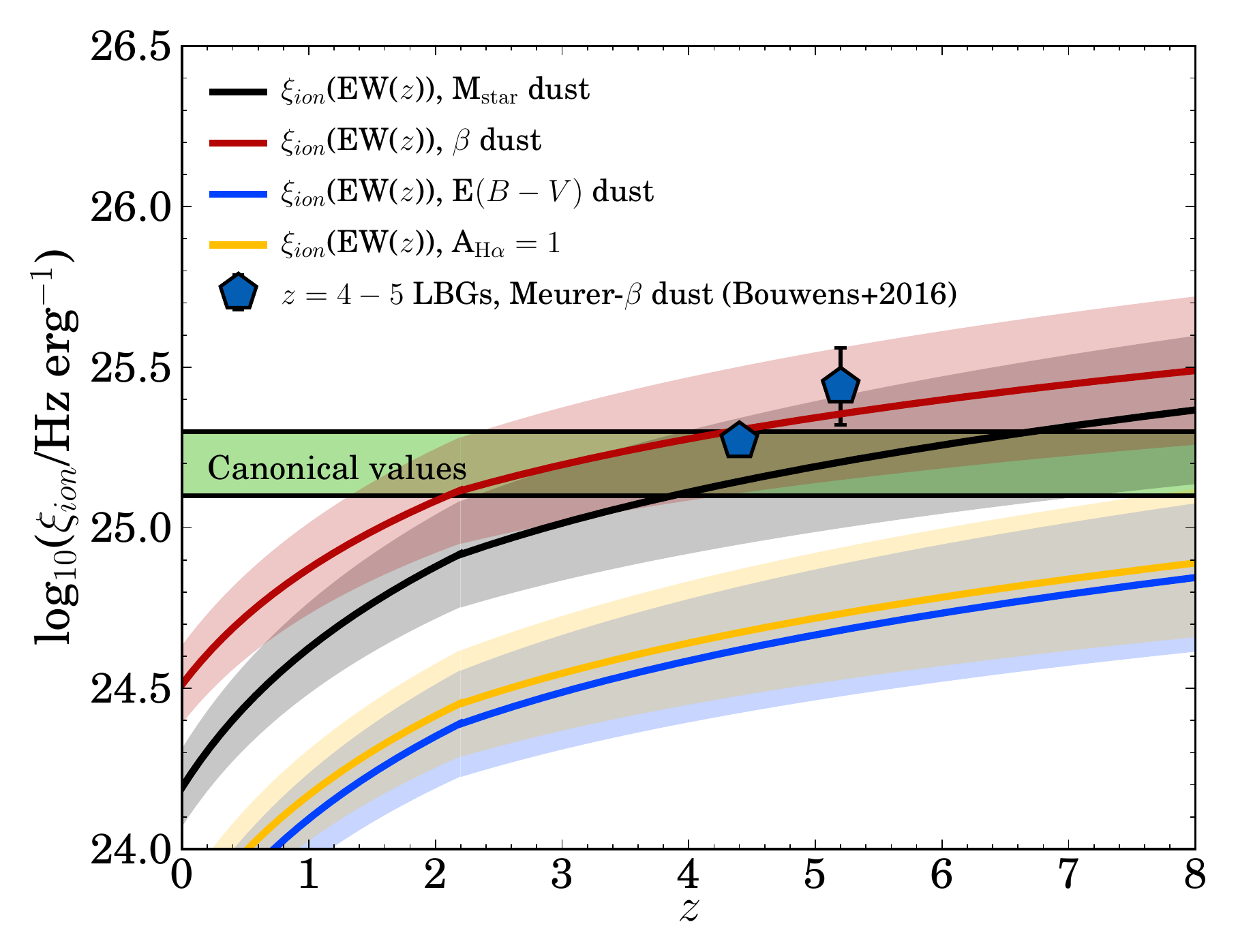}
\caption{\small{Inferred evolution of $\xi_{ion}$ with redshift based on the EW(H$\alpha$) evolution from \citet{Faisst2016} and our observed trend between $\xi_{ion}$ and H$\alpha$ EW for HAEs with M$_{\rm star} \sim 10^{9.2}$ M$_{\odot}$, for different methods to correct for dust. The black line shows the results when correcting for dust with M$_{\rm star}$, the red line shows dust corrected with $\beta$, the blue line shows dust corrected with the E$(B-V)$ values from SED fitting and the yellow line shows the results when we apply a global correction of A$_{\rm H\alpha} = 1$. The shaded regions indicate the errors on the redshift evolution of $\xi_{ion}$.}}   
\label{fig:xion_evolution_variation}
\end{figure}

In Fig. $\ref{fig:xion_evolution_variation}$ we show the inferred redshift evolution of $\xi_{ion}$ when we apply different methods to correct $\xi_{ion}$ for dust. Most of the differences are caused by a varying normalisation of $\xi_{ion}$, since we find that the slope of the fit between $\xi_{ion}$ and H$\alpha$ EW varies only mildly for various dust correction methods, see Table $\ref{tab:xionfit}$. However, we note again that most independent (stacking) observations from Balmer decrements and {\it Herschel} prefer dust attenuations similar to the dust attenuation we use when correcting for dust with stellar mass. 

\bsp	
\label{lastpage}
\end{document}